\documentclass[12pt]{article}
\usepackage{amsfonts}
\begin{document}
\newfont{\elevenmib}{cmmib10 scaled\magstep1}%
\renewcommand{\theequation}{\arabic{section}.\arabic{equation}}
\newcommand{\tabtopsp}[1]{\vbox{\vbox to#1{}\vbox to12pt{}}}
\font\larl=cmr10 at 24pt
\font\oneeight=cmr10 at 18pt
\font\twozero=cmr10 at 20pt
\newcommand{\vT}{\vphantom{\mbox{\twozero I}}}
\newcommand{\vTm}{\vphantom{\mbox{\oneeight I}}}
\newcommand{\vTb}{\vphantom{\mbox{\larl I}}}

\newcommand{\preprint}{
             \begin{flushleft}
    \elevenmib Yukawa\, Institute\, Kyoto\\
             \end{flushleft}\vspace{-1.3cm}
             \begin{flushright}\normalsize  \sf
             YITP-02-23\\
            {\tt hep-th/0204039} \\ April 2002
             \end{flushright}}
\newcommand{\Title}[1]{{\baselineskip=26pt \begin{center}
             \Large   \bf #1 \\ \ \\ \end{center}}}
\hspace*{2.13cm}%
\hspace*{0.7cm}%
\newcommand{\Author}{\begin{center}\large \bf
            E. Corrigan${}^a$  and R.
Sasaki${}^b$ \end{center}}
\newcommand{\Address}{\begin{center}
             $^a$ Department of Mathematics,
         University of York,\\
  Heslington, York  YO10 5DD,  United Kingdom\\
      ${}^b$ Yukawa Institute for Theoretical Physics,\\
      Kyoto University, Kyoto 606-8502, Japan
       \end{center}}
\newcommand{\Accepted}[1]{\begin{center}{\large \sf #1}\\
             \vspace{1mm}{\small \sf Accepted for Publication}
             \end{center}}
\baselineskip=20pt

\preprint
\thispagestyle{empty}
\bigskip
\bigskip
\bigskip

\Title{Quantum vs Classical  Integrability\\ in Calogero-Moser
Systems}
\Author

\Address
\vspace{1cm}

\begin{abstract}
  Calogero-Moser systems are classical and quantum integrable multi-particle
dynamics defined for any root system $\Delta$. The {\em quantum\/}
Calogero systems having
$1/q^2$ potential and a confining $q^2$ potential and the Sutherland systems
with $1/\sin^2q$ potentials  have  ``integer"
energy spectra characterised by the root system $\Delta$.
Various quantities of the corresponding {\em classical\/} systems,
{\em e.g.\/} minimum energy, frequencies of small oscillations, the
eigenvalues of the classical Lax pair matrices, etc. at
the equilibrium point of the potential are investigated
analytically as well as numerically for all root systems.
To our surprise, most of these classical data are also ``integers",
or they appear to be ``quantised". To be more precise, these quantities
are polynomials of the coupling constant(s) with integer coefficients.
The close relationship between quantum and classical integrability
in Calogero-Moser systems
deserves fuller analytical treatment, which would lead to better
understanding of these systems and of integrable systems in general.
\end{abstract}
\bigskip
\bigskip
\bigskip

\section{Introduction}
\label{intro}
\setcounter{equation}{0}
The contrast and resemblance between classical and quantum mechanics
and/or field theory has been a good source of stimulus for theoretical
physicists since the inception of quantum theory at the beginning of
the twentieth century.
In spite of the well-publicised differences such as the instability
(stability) of the hydrogen atom in classical (quantum) mechanics,
the  photo-electric effect, and tunneling effects,
classical and quantum mechanics share many
common theoretical structures (in particular, the canonical formalism)
and under certain circumstances provide (almost) the same predictions, as
exemplified by the correspondence principle and Ehrenfest's theorem.

In this paper we discuss issues related with the
{\em quantum\/} and {\em classical integrability\/} in
Calogero-Moser systems \cite{Cal, Sut},
\cite{CalMo}, having a rational potential with
harmonic confining force (the Calogero systems) and/or
a trigonometric potential (the Sutherland systems).
This is a part of a program for establishing a
{\em quantum Liouville theorem\/} on completely integrable
systems.
As is well-known, 
a classical Hamiltonian system with
  finitely many degrees of freedom can be transformed into
action-angle variables by {\em quadrature\/} if a complete
set of involutive independent conserved quantities  can be obtained. It
is a good challenge to formulate a quantum counterpart of the
`transformation into the action-angle variables by quadrature'.
Calogero-Moser systems are expected to provide the
best materials in this quest.
They are known to be integrable at both quantum and classical levels, and the
integrability is deeply related to the invariance of the Hamiltonian
with respect to a finite (Coxeter, Weyl) reflection  group $G_{\Delta}$
based on the root system $\Delta$.

Calogero-Moser systems for any root systems were formulated by
Olshanetsky and Perelomov \cite{OP1}, who  provided Lax pairs
for the systems based on the classical root systems, {\em i.e.\/}
the $A$, $B$, $C$, $D$ and $BC$ type root systems.
A universal {\em classical\/} Lax pair applicable to
all the Calogero-Moser
systems
based on any root systems including the $E_{8}$ and the
non-crystallographic root systems was derived by Bordner-Corrigan-Sasaki
\cite{bcs2}
which unified various types of Lax pairs known at
that time
\cite{DHoker_Phong, bcs1}.
A universal {\em quantum\/} Lax pair applicable to all the Calogero-Moser
systems
based on any root systems and for degenerate potentials
  was derived by Bordner-Manton-Sasaki \cite{bms}
  which provided the basic tools
for the present paper.
These universal classical and quantum Lax pairs are very closely
related to each other and also to the Dunkl operators \cite{Dunk,bms},
another well-known tool for quantum systems.
For quantum systems, universal formulae for the discrete spectra and
the ground state wave functions as well
as the proof of lower triangularity of the Hamiltonian and the
creation-annihilation
operator formalism etc have been obtained by Khastgir-Pocklington-Sasaki
\cite{kps}
based on the universal quantum Lax pair.
In this respect, the works of Heckman and Opdam \cite{Heck,HeOp}
offer a different approach
based on Dunkl operators.

The quantum Calogero and Sutherland systems have ``integer" energy eigenvalues
characterised by the root system $\Delta$.
Various quantities of the corresponding {\em classical\/} systems,
for example, minimum energy, frequencies of small oscillations, the
eigenvalues of the classical Lax pair matrices, etc., at
the equilibrium point of the potential are investigated in the present paper.
Some of these problems were tackled by Calogero and his collaborators
\cite{calmat,calpere,ahmcal}, about a quarter of a century ago.
They showed, mainly for the $A$-type theories, that the eigenvalues of Lax
matrices at equilibrium are ``integers",
and that the equilibrium positions are
related to zeros of classical polynomials (Hermite, Laguerre), etc.
The present paper provides systematic answers, both analytical and numerical,
to these old problems and presents new results thanks to the universal Lax pair
\cite{bcs2,bms} which are applicable to all root systems.
To our surprise, most of the
classical data are  ``integers", and appear to be ``quantised".

The present paper is organised as follows.
In section two we recapitulate the basic ingredients of the
Calogero-Moser systems and the solution mechanisms,
the reflection operators and the
root systems,  the quantum and classical Hamiltonian and potentials (in
\S\ref{ham-pot}),  the discrete spectra (in \S\ref{spectra}),
classical Lax pairs
(in \S\ref{laxpairs}) in order to introduce notation.
In section three the properties of the classical equilibrium point and
its uniqueness, its
representation in terms of the Lax pairs, are discussed.
The importance of the
{\em pre-potential\/} $W$, which is the logarithm of the ground
state wave function
(\ref{expform}), is stressed. The formulation of the spin exchange models
\cite{halsha,fmp,ber,yam},
by `freezing the dynamical freedom at the equilibrium point'
\cite{is1} is explained.
Their   definition is also based on a root system $\Delta$ and a
set of  vectors ${\cal R}$.
The uniqueness of the equilibrium point and the minimality of the
classical potential
as well as the maximality of the pre-potential are proved universally.
The explanation of  the highly organised nature of the energy spectra
of the spin
exchange models \cite{halsha,is2} in terms of the Lax pairs at
equilibrium is one of the motivations of the present paper.
Sections four and five contain the main results---the classical data
of the Calogero systems (\S\ref{cdata1}), and of the Sutherland systems
(\S\ref{cdata2}). In \S\ref{ratminener} we show that the minimum
energies are ``integer-valued". A general `virial theorem' is derived
based on the classical potential and the pre-potential.
In \S\ref{ratdetequi} the determination of  the classical
equilibrium points is discussed.
For  $A_r$ and $B_r$ the equilibrium points are known to be
given by the zeros of the
Hermite  and Laguerre polynomials \cite{calmat, calpere}.
For the other root systems,  the equilibrium points
are determined numerically.
In \S\ref{rateiglax}  the Lax pair matrices ($L$ and $M$) at the
equilibrium points are shown to satisfy classical versions of
the creation-annihilation operator relations. As a consequence,
the eigenvalues of the $M$
matrix at  equilibrium are shown to be equally spaced. The
eigenvalue-multiplicity relation of the $M$ matrix at  equilibrium is
shown to be the same as the height-multiplicity relation of the
chosen set of vectors ${\cal R}$.
The eigenvalues of $L^+L^-$ at  equilibrium are also evaluated.
In \S\ref{trigminener} the minimum energy of the Sutherland system is shown
to be ``quantised" since it is identical with the ground state
energy of the quantum
system.
The equilibrium position of the $A_r$ Sutherland system is known to be
``equally-spaced" (\ref{eqspaced}). We show in \S\ref{bcrtrigWspec}
  that  the equilibrium positions of $BC_r$ and
$D_r$  Sutherland systems are given as zeros of Jacobi polynomials,
which is a new
analytical result. The equivalence to the classical problem of maximising the
van der Mond determinant is also noted.
The Jacobi polynomials are known to reduce to simple trigonometric
(Chebyshev, etc) polynomials
for three specific values of $\alpha$ and $\beta$ (\ref{threeeq}),
in which the zeros
are again ``equally-spaced". We show that these three cases are utilised
for the spin exchange models based on $BC_r$ root system by
Bernard-Pasquier-Serban \cite{ber}.
The  eigenvalues of the $L_K$ (\ref{Ktilde})
and $M$ matrices at the equilibrium are all
``integer-valued".
In particular, The eigenvalue-multiplicity
relation of the  $L_K$ matrix at  equilibrium is
shown to be the same as the height-multiplicity relation of the
chosen set of vectors ${\cal R}$. In this case the `height' is
determined by the `deformed Weyl vector' $\varrho$ (\ref{weylvec})
in contrast to the ordinary Weyl vector $\delta$ (\ref{weyldef})
which determines the height-multiplicity relation for the $M$ matrix
in Calogero system discussed in \S\ref{cdata1}.
The final section is devoted to comments
and discussion. In the Appendix we discuss  a remarkable
  constant matrix $K$ (\ref{QLcomm})  which plays an important
role in many parts of Calogero-Moser theory.
It is a non-negative matrix with integer elements only.
For any root system $\Delta$ and set of vectors ${\cal R}$ its
eigenvalues are all ``integers" with  multiplicities.
The eigenvectors of the $K$ matrix
span representation spaces of the Weyl group whose dimensions are the
multiplicities of the corresponding eigenvalues.

\section{Calogero-Moser Systems}
\label{cal-mo}
\setcounter{equation}{0}

In this section, we briefly summarise the {\em quantum\/}
and {\em classical\/}
Calogero-Moser systems along with as much of the
appropriate notation and background as is necessary
for the main body of the
paper. A   Calogero-Moser model is a
Hamiltonian system associated with a root system \(\Delta\)
of rank \(r\).
This is a set of
vectors in \(\mathbf{R}^{r}\)
invariant under reflections
in the hyperplane perpendicular to each
vector in $\Delta$:
\begin{equation}
   \Delta\ni s_{\alpha}(\beta)=\beta-(\alpha^{\vee}\cdot\beta)\alpha,
    \quad \alpha^{\vee}={2\alpha\over{\alpha^2}},
\quad \alpha, \beta\in\Delta.\label{a1}
\end{equation}
The set of reflections $\{s_{\alpha},\,\alpha\in\Delta\}$ generates a
finite reflection group $G_{\Delta}$, known as a Coxeter (or Weyl) group.
  For
detailed and unified treatment of Calogero-Moser models based various root
systems and various potentials, we refer to \cite{bms,kps}.

The dynamical variables of the Calogero-Moser model are the coordinates
\(\{q_{j}\}\) and their canonically conjugate momenta \(\{p_{j}\}\), with
the canonical commutation (Poisson bracket) relations (throughout
this paper we put $\hbar=1$):
\begin{eqnarray*}
(Q):&&   [q_{j},p_{k}]=i\delta_{j\,k},\qquad [q_{j},q_{k}]=
    [p_{j},p_{k}]=0,\\[-3pt]
&&\hspace{8cm} j,k=1,\ldots,r,\nonumber\\[-3pt]
(C):&&  \{q_{j},p_{k}\}=\delta_{j\,k},\qquad \{q_{j},q_{k}\}=
    \{p_{j},p_{k}\}=0.
\end{eqnarray*}
These will be denoted by vectors in \(\mathbf{R}^{r}\)
\[
    q=(q_{1},\ldots,q_{r}),\qquad p=(p_{1},\ldots,p_{r}).
\]
In quantum theory, the momentum operator \(p_j\) acts as
a differential operator:
\[
    p_j=-i{\partial\over{\partial q_j}}, \quad j=1,\ldots,r.
\]

\subsection{Hamiltonians and Potentials}
\label{ham-pot}

We will concentrate on  those cases in which bound states occur, meaning
those with
discrete spectra. In other words, we deal with  the {\em
rational potential  with harmonic confining force\/} (to be called
Calogero systems
\cite{Cal} for short) and {\em
trigonometric potential\/}
(to be referred to as the Sutherland systems \cite{Sut}):
\begin{eqnarray}
(Q):\quad {\cal H}_Q&=&{1\over 2} p^{2} +V_Q,\quad
V_Q=\left\{\begin{array}{l}
\begin{displaystyle}
{\omega^2\over2}q^2 +{1\over2}\sum_{\rho\in\Delta_+}
    {g_{\rho}(g_{\rho}-1) \rho^{2}\over{(\rho\cdot q)^2}},
\end{displaystyle}
\\[20pt]
\begin{displaystyle}
{1\over2}\sum_{\rho\in\Delta_+}
    {g_{\rho}(g_{\rho}-1) \rho^{2}\over{\sin^2(\rho\cdot q)}},
\end{displaystyle}
\end{array}
\right.\label{QHam}\\[12pt]
(C):\quad {\cal H}_C&=&{1\over 2} p^{2} +V_C,\quad
V_C=\left\{\begin{array}{l}
\begin{displaystyle}
{\omega^2\over2}q^2 +
{1\over2}\sum_{\rho\in\Delta_+}
    {g_{\rho}^2 \rho^{2}\over{(\rho\cdot q)^2}},
\end{displaystyle}
\\[20pt]
\begin{displaystyle}
{1\over2}\sum_{\rho\in\Delta_+}
    {g_{\rho}^2 \rho^{2}\over{\sin^2(\rho\cdot q)}}.
\end{displaystyle}
\end{array}\right.
\label{CHam}
\end{eqnarray}
In these formulae, $\Delta_+$ is the set of positive roots and
  \(g_{\rho}\) are  real {\em positive\/} coupling constants
which are defined on orbits of the corresponding
Coxeter group, {\it i.e.,} they are
identical for roots in the same orbit. For crystallographic root
systems there is one
coupling constant
\(g_{\rho}=g\) for all roots in simply-laced models,
and  there are two {\em independent\/} coupling constants,
\(g_{\rho}=g_L\) for long roots and \(g_{\rho}=g_S\) for
short roots in non-simply laced models.
Throughout this paper we put the scale factor in the trigonometric
functions to
unity for simplicity; instead of the general form
$a^2/\sin^2a(\rho\cdot q)$, we use $1/\sin^2(\rho\cdot q)$.
We also adopt the convention that
long roots have squared length two, $\rho_L^2=2$,
unless otherwise stated.

The Sutherland systems  are integrable,
both at the classical and quantum levels,
for the crystallographic root systems,
  that is
those associated with simple Lie
algebras: \{\(A_{r},\,r\ge 1\}\)%
\footnote{For  $A_r$ models, it is customary to
introduce one more degree of freedom,
$q_{r+1}$ and $p_{r+1}$ and embed
all of the roots in ${\bf R}^{r+1}$.
\label{embedding}}
, \(\{B_{r},\,r\ge 2\}\), \(\{C_{r},\,r\ge
2\}\),
\(\{D_{r},\,r\ge 4\}\), \(E_{6}\), \(E_{7}\), \(E_{8}\), \(F_{4}\) and
\(G_{2}\) and the so-called \(\{BC_{r},\,r\ge 2\}\).
On the other hand, the Calogero systems  are
integrable for any root systems, crystallographic and
non-crystallographic. The latter are \(H_{3}\), \(H_{4}\),
whose Coxeter groups are the
symmetry groups of the icosahedron and four-dimensional 600-cell,
respectively,
and \(\{I_{2}(m),\,m\ge 4\}\) whose Coxeter group is
the dihedral group of order \(2m\).

These potentials, classical and quantum, both rational and
trigonometric, have a hard repulsive singularity
  \(\sim
{1/{(\rho\cdot q)^2}}\) near the reflection hyperplane
\(H_{\rho}=\{q\in\mathbf{R}^{r},\, \rho\cdot q=0\}\).
The strength of the singularity is given by
the coupling constant \(g_{\rho}(g_{\rho}-1)\) (Q),
(\(g_{\rho}^2\)  (C)\,), which is {\em independent\/} of the
choice of the normalisation of the roots.
This repulsive potential is classically (quantum
mechanically, $g_{\rho}>1$) insurmountable. Thus the motion is always
confined within one Weyl chamber both at the classical and quantum
levels. This  feature allows us without loss of generality
to constrain the configuration space,
to the principal Weyl chamber (\(\Pi\) is the set of simple roots):
\begin{equation}
    PW=\{q\in{\bf R}^r|\ \rho\cdot q>0,\quad \rho\in\Pi\}.
    \label{PW}
\end{equation}
In the case of the trigonometric potential, due to the periodicity of
the potential the configuration space is further
limited  to the principal
Weyl alcove
\begin{equation}
    PW_T=\{q\in{\bf R}^r|\ \rho\cdot q>0,\quad \rho\in\Pi,
    \quad \rho_h\cdot q<\pi\},
    \label{PWT}
\end{equation}
where \(\rho_h\) is the highest root.

The potentials of the quantum and classical systems are expressed
neatly in terms of a {\em pre-potential\/} $W$ which is defined
through a ground state wavefunction $\phi_0$ of the quantum
Hamiltonian $H_Q$ (\ref{QHam}).
Since $\phi_0$ can be chosen real  and positive, because it has no
nodes, it can be expressed by a real smooth function $W$, to be called
a {\em pre-potential\/}, in the
principal Weyl chamber ($PW$) (\ref{PW}) or the principal Weyl
alcove ($PW_T$) (\ref{PWT}) by
\begin{eqnarray}
\phi_0&=&e^W,\label{expform}\\
{\cal H}_Q\phi_0&=&{\cal E}_0\phi_0.\label{h0eq}
\end{eqnarray}
The pre-potential $W$ and the ground state energy  ${\cal E}_0$
are expressed
entirely in terms of the coupling constants and roots \cite{bms,kps}:
\begin{eqnarray}
W&=&\left\{
\begin{array}{c}
\begin{displaystyle}
-{\omega\over2}q^2+\sum_{\rho\in\Delta_+}g_{\rho}\log\rho\cdot q,
\end{displaystyle}
\\[16pt]
\begin{displaystyle}
\sum_{\rho\in\Delta_+}g_{\rho}\log\sin(\rho\cdot q),
\end{displaystyle}
\end{array}
\right.
\label{Wform}\\[8pt]
    {\cal E}_0&=& \left\{
    \begin{array}{c}
\begin{displaystyle}
\omega\left({r\over2}+\sum_{\rho\in\Delta_+}g_{\rho}\right),
\end{displaystyle}
\label{minenergy}\\[18pt]
  2\varrho^2.
    \end{array}
    \right.
\end{eqnarray}
The {\em deformed Weyl vector\/} $\varrho$ is defined by
\begin{equation}
    \varrho ={1\over2}\sum_{\rho\in\Delta_+}g_{\rho}\rho,
\label{weylvec}
\end{equation}
which reduces to the {\em Weyl vector\/} $\delta$ when all the coupling
constants are unity:
\begin{equation}
    \delta ={1\over2}\sum_{\rho\in\Delta_+}\rho.
\label{weyldef}
\end{equation}
By plugging  (\ref{expform}) into (\ref{h0eq}) and (\ref{QHam}),
we obtain a simple formula expressing the
quantum potential in terms of the
pre-potential $W$ \cite{bms,kps}:
\begin{eqnarray}
\hspace*{-32mm}(Q):\quad V_Q&=&
{1\over2}\sum_{j=1}^r\left[\left({\partial
W\over{\partial
     q_j}}\right)^2+{\partial^2W\over{\partial q_j^2}}\right]+{\cal
E}_0,\label{qpot}
\end{eqnarray}
and similarly,
\begin{eqnarray}
(C):\quad V_C&=&{1\over2}\sum_{j=1}^r\left({\partial
W\over{\partial
     q_j}}\right)^2+\tilde{\cal E}_0,\quad \tilde{\cal E}_0=\left\{
    \begin{array}{c}
\omega\left(\sum_{\rho\in\Delta_+}g_{\rho}\right),
\\[12pt]
  2\varrho^2.
    \end{array}
    \right.\label{cpot}
\end{eqnarray}
In the context of super-symmetric quantum mechanics \cite{witten,
bms} the quantities $\partial W/\partial q_j$ are called {\em
super-potentials\/}. In this paper we will not discuss
super-symmetry at all and we stick to our notion of $W$ being a
pre-potential.
The difference between the {\em quantum\/} and {\em classical\/}
potential is  $1/2\sum_{j=1}^r\partial^2W/\partial
q_j^2$ plus the zero point energy $\omega r/2$, for the rational
cases. These are both {\em quantum corrections\/}, being of the order
$\hbar$. It should be noted that the quantum
Hamiltonian (\ref{QHam}) with the potential (\ref{qpot}) can be
expressed in a `factorised form'
\begin{equation}
{\cal H}_Q=\sum_{j=1}^r\left(p_j-i{\partial W\over{\partial
     q_j}}\right)\left(p_j+i{\partial W\over{\partial
     q_j}}\right)+{\cal E}_0
=\sum_{j=1}^r\left(p_j+i{\partial W\over{\partial
     q_j}}\right)^\dagger\left(p_j+i{\partial W\over{\partial
     q_j}}\right)+{\cal E}_0,
\end{equation}
which is obviously positive semi-definite apart from the constant
term ${\cal E}_0$. Therefore it is elementary to verify, thanks to
the simple formulae
\begin{equation}
\left(p_j+i{\partial W\over{\partial q_j}}\right)\,e^{W}=0,
\quad j=1,\ldots, r,
\label{grannihi}
\end{equation}
  that
$\phi_0=e^W$ satisfying (\ref{h0eq}) is the lowest energy state.

\subsection{Discrete Spectra}
\label{spectra}
\subsubsection{Rational potentials}
The discrete spectrum of the  Calogero systems is an
{\em integer\/} times $\omega$ plus the ground state energy ${\cal E}_0$.
In other words,  the energy eigenvalue ${\cal
E}$ depends on  the coupling constant
$g_{\rho}$   only via the ground state energy ${\cal E}_0$.
The $integer$ is
specified by  an $r$-tuple of non-negative integers
$\vec{n}=(n_1,\ldots,n_r)$ by \cite{kps}:
\begin{equation}
{\cal E}_{\vec{n}}=\omega N_{\vec{n}}+{\cal E}_0,\quad
N_{\vec{n}}=\sum_{j=1}^rn_j f_j,\quad n_j\in{\bf Z}_+,
\label{ratspectra}
\end{equation}
and the set of integers $\{f_j\}$ are listed in Table 1 for
each root system $\Delta$.
\begin{figure}
     \begin{center}
\begin{tabular}{||c|l||c|l||}
         \hline
          \(\Delta\)& \(f_j=1+e_j\) &\(\Delta\)&\vT \(f_j=1+e_j\)\\
         \hline
         \(A_r\) & \(2,3,4,\ldots,r+1\) & \(E_8\) & \vTm
\(2,8,12,14,18,20,24,30\) \\
         \hline
         \(B_r\) & \(2,4,6,\ldots,2r\) & \(F_4\) & \vTm\(2,6,8,12\) \\
         \hline
         \(C_r\) & \(2,4,6,\ldots,2r\) & \(G_2\) & \vTm\(2,6\) \\
         \hline
       \(D_r\) & \(2,4,\ldots,2r-2;r\) & \(I_2(m)\) &\vTm \(2,m\) \\
       \hline
       \(E_6\) & \(2,5,6,8,9,12\) & \(H_3\) & \vTm\(2,6,10\) \\
       \hline
       \(E_7\) & \(2,6,8,10,12,14,18\) & \(H_4\) &\vTm \(2,12,20,30\) \\
       \hline
     \end{tabular}\\
     \bigskip
Table 1: The degrees \(f_j\) at which independent Coxeter
invariant polynomials exist.
\end{center}
\end{figure}
These are the degrees at which independent Coxeter invariant
polynomials occur. They are related to the {\em exponents\/} $e_j$
of the root system $\Delta$ by
\begin{equation}
    f_j=1+e_j,\quad j=1,\ldots,r.
    \label{coxinvdeg}
\end{equation}
One immediate consequence of the spectra (\ref{ratspectra}) is the
{\em periodicity\/} of motion. Suppose, at time $t=0$,  the system
has the wavefunction $\Psi_0$ then the system returns to the same
physical state after $T=2\pi/\omega$.
Let us introduce a complete set of wavefunctions indexed by the
$r$-tuple of non-negative integers $\vec{n}$
\[
\phi_{\vec{n}},
\]
and express the initial state $\Psi_0$ as the
linear combination
\[
\Psi_0=\sum_{\vec{n}}a_{\vec{n}}\phi_{\vec{n}}.
\]
Then, at time $t$ the wavefunction is given by
\[
\Psi(t)=\sum_{\vec{n}}a_{\vec{n}}\phi_{\vec{n}}
e^{-i{\cal E}_{\vec{n}}t}=
e^{-i{\cal E}_0t}\sum_{\vec{n}}a_{\vec{n}}\phi_{\vec{n}}
e^{-i\omega(\sum_{j=1}^r n_jf_j)t}.
\]
In other words, we have
\[
\Psi(T)=e^{-i{\cal E}_0T}\sum_{\vec{n}}a_{\vec{n}}\phi_{\vec{n}}
e^{-i2\pi(\sum_{j=1}^r n_jf_j)}=e^{-i{\cal
E}_0T}\sum_{\vec{n}}a_{\vec{n}}\phi_{\vec{n}}=e^{-i{\cal
E}_0T}\Psi_0.
\]
For some root systems, the quantum state returns to $\Psi_0$
earlier than  $T=2\pi/\omega$. The corresponding classical theorem,
or rather its generalisation for the entire hierarchy,
  is given as Proposition III.2 in \cite{cfs}.
It is interesting to note that the $1/(\rho\cdot q)^2$
interactions do not disturb the periodicity of the harmonic
potential.

\subsubsection{Trigonometric potentials}
The discrete spectrum of the Sutherland systems is
indexed by a {\em dominant weight\/} $\lambda$ as follows,
\begin{equation}
{\cal E}_{\lambda}=2(\lambda+\varrho)^2,
\end{equation}
in which $\varrho$ is the deformed Weyl vector (\ref{weylvec}).
This spectrum can be interpreted as a ``free" particle energy
\[
{\cal E}={1\over2}p^2,
\]
in which the momentum $p\in{\bf R}^r$ is simply given by
\[
p=2(\lambda+\varrho).
\]
A dominant weight is specified by an $r$-tuple of
non-negative integers $\vec{n}=(n_1,\ldots,n_r)$ by
\begin{equation}
\lambda=\lambda_{\vec{n}}=\sum_{j=1}^rn_j\lambda_j,
\end{equation}
in which $\lambda_j$ is the $j$-th {\em fundamental weight\/}.
We extract explicitly the part of ${\cal E}_\lambda$ which depends
  linearly on
$\vec{n}$,  and write
\begin{equation}
{\cal
E}_{\lambda_{\vec{n}}}=
2(\lambda_{\vec{n}}^2+
\varrho^2+2\sum_{j=1}^rn_j\lambda_j\cdot\varrho).
\label{energyformula}
\end{equation}

\subsection{Classical Lax pairs}
\label{laxpairs}
The classical equations of motion for the Hamiltonian ${\cal H}_C$ are
known to be written in a Lax pair form:
\begin{equation}
    \label{LaxEquation}
  \dot{q}_{j}=  p_{j},\quad  \dot{p}_{j}= -{\partial{\cal
H}_C\over{\partial q_{j}}}
\Longleftrightarrow
    {d\over dt}{L}=[L,M].
\end{equation}
\subsubsection{Universal Lax Pair}

Here we will summarise the
universal formulation applicable to any root system $\Delta$
  for both the rational ($\omega=0$ case) and trigonometric potentials
\cite{bcs2}.  The inclusion of the harmonic confining potential
($\omega\neq0$)
needs a further construction which will be discussed
at the end of this section.
(For the universal quantum Lax pair, which we will not use in this paper,
we refer to \cite{bms,kps}.)
The  universal Lax pair operators read
\begin{eqnarray}
    L(p,q) &=&  p\cdot\hat{H}+X,\qquad
X=i\sum_{\rho\in\Delta_{+}}g_{\rho}
    \,\,(\rho\cdot\hat{H})\,x(\rho\cdot q)\,\hat{s}_{\rho},
    \label{LaxOpDef}\\
    {M}(q) &=&
    {i\over2}\sum_{\rho\in\Delta_{+}}g_{\rho}\rho^2\,y
    (\rho\cdot q)\,(\hat{s}_{\rho}-I),\quad I:\ \mbox{Identity Operator},
    \label{Mdef}
\end{eqnarray}
in which the functions $x(u)$ and $y(u)$ are listed in the Table 2.
\begin{center}
     \begin{tabular}{|l|c|c|c|}
     \hline
      & $V(u)$ & $x(u)$ &\vT $y(u)$ \\
     \hline
     Rational & $1/u^2$ & $1/u$ &\vTm -$1/u^2$ \\
     \hline
     Trigonometric & $1/\sin^2u$ & $\cot u$ &\vTm -$1/\sin^2 u$ \\
     \hline
     \end{tabular}\\
     \bigskip
     Table 2: Functions appearing in the  Lax pair.
\end{center}
The operators $\hat{H}_j$ and $\hat{s}_{\rho}$ obey the following
commutation relations
\begin{eqnarray}
    \label{OpAlgebra1}
    [\hat{H}_{j},\hat{H}_{k}]=0, \\
    \label{OpAlgebra2}
    [\hat{H}_{j},\hat{s}_{\alpha}] = \alpha_{j}
    (\alpha^{\vee}\!\!\cdot\hat{H})\hat{s}_{\alpha},
    \\
    \label{OpAlgebra3}
    \hat{s}_{\alpha}\hat{s}_{\beta}\hat{s}_{\alpha}
    =\hat{s}_{s_{\alpha}(\beta)},\quad
\hat{s}_{\alpha}^2=1,\quad \hat{s}_{-\alpha}=\hat{s}_{\alpha}.
\end{eqnarray}

Let us choose a set of $D$ vectors ${\cal R}$
\begin{equation}
{\cal R}=\{\mu^{(1)},\ldots,\mu^{(D)}|\mu^{(a)}\in {\bf R}^r\},
\end{equation}
which form a single orbit
of  the reflection (Weyl) group $G_\Delta$.
That is, any element of ${\cal R}$ can be obtained from any other  by the
action of the reflection (Weyl) group.
Let us note that all these vectors have the same length,
$(\mu^{(a)})^2=(\mu^{(b)})^2$, $a,b=1,\ldots, D$,
which we denote simply as $\mu^2$.
They form an over-complete basis
\footnote{The $A_r$ case needs a special attention,
since it has one additional degree
of freedom due to the embedding (see footnote on page
\pageref{embedding}).}
\label{compform}
of ${\bf R}^r$:%
\begin{equation}
\sum_{\mu\in{\cal R}}\mu_j\mu_k=\delta_{j\,k}\mu^2D/r,
\qquad j,k=1,\ldots, r.
\end{equation}
In terms of ${\cal R}$, $L$ and $M$ are $D\times D$ matrices whose ingredients
$\hat{H}_{j}$ and $\hat{s}_{\rho}$ are defined by
\begin{equation}
(\hat{H}_{j})_{\mu\nu}=\mu_j\delta_{\mu\nu},\quad
(\hat{s}_{\rho})_{\mu\nu}=\delta_{\mu,s_\rho(\nu)}=
    \delta_{\nu,s_\rho(\mu)}.
\label{sHdef}
\end{equation}
The Lax operators are Coxeter covariant:
\begin{equation}
L(s_{\alpha}(p),s_{\alpha}(q))=\hat{s}_\alpha L(p,q)\hat{s}_\alpha,\quad
M(s_{\alpha}(q))=\hat{s}_\alpha M(q)\hat{s}_\alpha,
\label{lmcoxcov}
\end{equation}
and $L$ ($M$) is (anti-) hermitian:
\begin{equation}
L^\dagger =L,\quad M^\dagger=-M,
\label{hermiticity}
\end{equation}
implying real and pure imaginary eigenvalues of $L$ and $M$,
respectively.
For various examples of the sets of vectors ${\cal R}$ see the
{Appendix}.

\subsubsection{Minimal Type Lax Pair}
A set of weights  $\Lambda=\{\mu\}$ is called {\em minimal\/} if
  the following condition is satisfied:
\begin{equation}
  {2\rho\cdot\mu\over{\rho^2}}=0,\pm 1,
\quad \quad \forall\mu\in\Lambda
  \quad \ \mbox{and}\quad \forall\rho\in\Delta.
  \label{eq:mindef}
\end{equation}
  A representation of Lie algebra $\Delta$ is called
{\em minimal\/} if its weights are minimal.
All the fundamental representations
of the \(A_r\) algebras are minimal.
The vector,
spinor and anti-spinor representations of the \(D_{r}\) algebras are minimal
representations.
There are three minimal representations belonging to the simply-laced
exceptional algebras---the {\bf 27} and \({\bf \overline{27}}\) of \(E_6\)
and the {\bf 56} of \(E_{7}\); $E_8$ has no minimal representations.

When ${\cal R}$ is a set of  minimal weights $\Lambda$,
the representation of the operator $\hat{s}_{\rho}$ simplifies
\begin{equation}
(\hat{s}_{\rho})_{\mu\nu}=\left\{
\begin{array}{lcl}
\delta_{\mu-\nu,\rho},&&\rho^\vee\!\cdot\mu=1,\\
\delta_{\mu-\nu,-\rho},&\quad\mbox{if}\quad&\rho^\vee\!\cdot\mu=-1,\\
\delta_{\mu-\nu,0},&&\rho^\vee\!\cdot\mu=0.\\
\end{array}\right.
\label{srhodef}
\end{equation}
In this case a Lax pair with with a different functional dependence
from the universal Lax pair (\ref{LaxOpDef}) (\ref{Mdef}) is possible for the
trigonometric potential systems,
which we call a {\em minimal type\/} Lax pair
\begin{eqnarray}
  L_m(p,q)  =  p\cdot\hat{H} + X_m ,\qquad
  M_m(q)  =  D+Y_m.
  \label{eq:minLaxform}
\end{eqnarray}
The matrix  \(X_m\)  has the same form as before but
with a different functional dependence on the coordinates $q$,
\begin{equation}
  X_m=i\sum_{\rho\in\Delta_{+}}g_{\rho}
    \,\,(\rho\cdot\hat{H})\,x_m(\rho\cdot q)\,\hat{s}_{\rho},\quad
  x_m(u)=1/\sin u.
  \label{eq:minXYdef}
\end{equation}
The matrix $Y_m$ is an off-diagonal matrix
\begin{equation}
Y_m={i\over2}\sum_{\rho\in\Delta}g_{\rho}\rho^2\,y_m
    (\rho\cdot q)\,\hat{s}_{\rho},\,\quad y_m(u)=x_m'(u)=-\cos u/\sin^2u.
\end{equation}
The diagonal matrix $D$ is defined by
\begin{equation}
  D_{\mu \nu}= \delta_{\mu, \nu}D_{\mu},\quad
  D_{\mu}=-{i\over2}\sum_{\Delta\ni\beta=
\mu-\nu}g_{\beta}\beta^2z(\beta\cdot
q),\quad z(u)=-1/\sin^2u.
  \label{eq:minD}
\end{equation}
This type of Lax pair has been known from the early days of
Calogero-Moser
\cite{OP1}.

\subsubsection{Lax Pair for Calogero Systems}
Lax type representations of the Hamiltonian ${\cal H}_C$
(\ref{CHam}) for the
Calogero systems ($\omega\neq0$)
is obtained from
the rational Lax pair for the $\omega=0$ case discussed above.
The canonical equations of motion are equivalent to the following
Lax equations for \(L^{\pm}\):
\begin{equation}
    {d\over dt}{L^\pm}=\{L^\pm,{\cal H}_C\}=
    [L^{\pm},{M}]\pm i\omega L^{\pm},
    \label{omegaLM}
\end{equation}
in which
\({M}\) is the same as before (\ref{Mdef}), and
  \(L^{\pm}\) and \(Q\) are defined by
\begin{equation}
    L^{\pm}=L\pm i\omega Q, \quad Q=q\cdot\hat{H},
\label{Lpmdef}
\end{equation}
with \(L\), \(\hat{H}\) as earlier (\ref{LaxOpDef}), (\ref{sHdef}).
It is easy to see that the classical commutator $[Q,L]$ is a constant
matrix (see \S4 of \cite{bms} and \S II of \cite{cfs2}):
\begin{equation}
     QL-LQ=iK,
     \quad
     K\equiv
\sum_{\rho\in\Delta_{+}}g_{\rho}(\rho\cdot\hat{H})
     ({\rho}^{\vee}\!\cdot\hat{H})\hat{s}_{\rho}.
     \label{QLcomm}
\end{equation}
We will discuss this interesting matrix $K$ in some detail in the {Appendix}.
  If we
define hermitian operators
\({\cal L}_1\) and
\({\cal L}_2\) by
\begin{equation}
    {\cal L}_1=L^+L^-,\quad {\cal L}_2=L^-L^+,
    \label{defcalL}
\end{equation}
they satisfy Lax-like equations, and classical conserved quantities
are obtained:
    \begin{equation}
    \dot{{\cal L}}_k=[{\cal L}_k,{M}],\quad {d\over{dt}}\mbox{Tr}{\cal
L}_k^n=0,\quad k=1,2.
\label{Lpmeq}
\end{equation}

This  completes the brief summary of Calogero-Moser systems,
the quantum and classical Hamiltonians, the discrete spectra and
their classical
Lax representations.

\section{Classical Equilibrium and Spin Exchange Models}
\label{cequili}
\setcounter{equation}{0}
Here we discuss the properties of the classical potential
$V_C$, the pre-potential $W$, and Lax matrices $L$, $M$, ${\cal
L}_{1,2}$ near the classical equilibrium point:
\begin{equation}
p=0,\quad q=\bar{q}.
\label{clasequil}
\end{equation}


For the classical potential the point $\bar{q}$ is characterised as its
{\em minimum} point:
\begin{equation}
\left.{\partial V_C\over{\partial
q_j}}\right|_{\bar{q}}=0,\quad j=1,\ldots, r,
\label{Vmin}
\end{equation}
whereas it is a {\em maximal\/} point of the pre-potential $W$ and of
the ground state wavefunction
$\phi_0=e^W$:
\begin{equation}
\left.{\partial W\over{\partial
q_j}}\right|_{\bar{q}}=0,\quad j=1,\ldots, r.
\label{Wmax}
\end{equation}
In this connection, it should be noted that the condition (\ref{grannihi})
$(p+i\partial W/\partial q_j)e^W=0$
is also satisfied classically at this point.
In the Lax representation it is a point at which two Lax matrices commute:
\begin{equation}
0=[\bar{L},\bar{M}],\quad 0=[\bar{L}_m,\bar{M}_m],
\quad 0=[\bar{\cal L}_{(1,2)},\bar{M}],
\label{commlax}
\end{equation}
in which $\bar{L}=L(0,\bar{q})$, $\bar{M}=M(\bar{q})$
etc and $d\bar{L}/dt=0$, etc at
the equilibrium point. The value of a quantity $A$
at the equilibrium is expressed
by $\bar{A}$.

By differentiating (\ref{cpot}), we obtain
\begin{equation}
{\partial V_C\over{\partial
q_j}}=\sum_{l=1}^r{\partial^2 W\over{\partial
q_j\partial q_l}}{\partial W\over{\partial q_l}}.
\label{Vder}
\end{equation}
Since $\partial^2W/\partial q_j\partial q_k$ is {\em negative definite}
everywhere,
\begin{equation}
{\partial^2W
\over{\partial q_j\partial q_k}}=\left\{
\begin{array}{c}
\begin{displaystyle}
-\omega\delta_{j\,k}
-\sum_{\rho\in\Delta_{+}}g_{\rho}{\rho_j\rho_k\over{(\rho\cdot\!q)^2}},
\end{displaystyle}
\\[18pt]
\begin{displaystyle}
-\sum_{\rho\in\Delta_{+}}
g_{\rho}{\rho_j\rho_k\over{\sin^2(\rho\cdot\!q)}},
\end{displaystyle}
\end{array}
\right.
\label{W2der}
\end{equation}
we find the equilibrium point of $W$ is a maximum and that
the two conditions (\ref{Vmin}) and (\ref{Wmax}) are equivalent:
\begin{equation}
\left.{\partial V_C\over{\partial
q_j}}\right|_{\bar{q}}=0,\quad j=1,\ldots, r,
\Longleftrightarrow
\left.{\partial W\over{\partial
q_j}}\right|_{\bar{q}}=0,\quad j=1,\ldots, r.
\label{VWequiv}
\end{equation}
By differentiating (\ref{Vder}) again, we obtain
\[
{\partial^2 V_C\over{\partial q_j\partial q_k}}
=\sum_{l=1}^r{\partial^2 W\over{\partial q_j\partial q_l}}
{\partial^2 W\over{\partial q_l\partial q_k}}+
\sum_{l=1}^r{\partial^3 W\over{\partial q_j\partial q_k\partial q_l}}
{\partial W\over{\partial q_l}}.
\]
Thus at the equilibrium point of the classical potential
$V_C$, the following
relation holds:
\begin{equation}
\left.{\partial^2 V_C\over{\partial q_j\partial q_k}}\right|_{\bar{q}}
=\sum_{l=1}^r\left.{\partial^2 W\over
{\partial q_j\partial q_l}}\right|_{\bar{q}}
\left.{\partial^2 W\over{\partial q_l\partial q_k}}\right|_{\bar{q}}.
\end{equation}
If we define the following two symmetric $r\times r$ matrices
$\tilde{V}$ and
$\tilde{W}$,
\begin{equation}
\tilde{V}=\mbox{Matrix}\left[\left.\
{\partial^2 V_C\over{\partial q_j\partial
q_k}}\right|_{\bar{q}}\right],
\quad
\widetilde{W}=\mbox{Matrix}\left[\left.\
{\partial^2 W\over{\partial q_j\partial
q_k}}\right|_{\bar{q}}\right],
\label{VWmat}
\end{equation}
we have
\begin{equation}
\tilde{V}=\widetilde{W}^2,
\label{VWrel}
\end{equation}
and
\begin{eqnarray}
\begin{array}{rclll}
\mbox{Eigenvalues}(\tilde{V})&=&\{w_1^2,\ldots,w_r^2\},&&\\[6pt]
\mbox{Eigenvalues}(\widetilde{W})&=&\{-w_1,\ldots,-w_r\},
&\quad w_j>0,& j=1,\ldots,r.
\end{array}
\label{Vposdef}
\end{eqnarray}
That is $\tilde{V}$ is {\em positive definite} and
the point $\bar{q}$ is actually a minimal point of
$V_C$.

As mentioned above, the classical potential
$V_C$ tends to plus infinity at all the
boundaries (including the infinite point in $PW$) of $PW$ ($PW_T$). Since
it is positive definite (see (\ref{CHam})),
$V_C$ has at least one equilibrium
(minimal) point in  $PW$ ($PW_T$).
Next we show that it is unique in $PW$ ($PW_T$).
Suppose there are two classical equilibrium points
$\bar{q}^{(1)}$ and $\bar{q}^{(2)}$
\[
\left.{\partial W\over{\partial
q_j}}\right|_{\bar{q}^{(1)}}=\left.{\partial W\over{\partial
q_j}}\right|_{\bar{q}^{(2)}}=0,\quad j=1,\ldots, r,
\]
then (see (\ref{cpot}))
\[
V_C(\bar{q}^{(1)})=V_C(\bar{q}^{(2)})=\tilde{\cal E}_0.
\]
Let us consider a space $P$ of paths of finite length $q(t)$,
($0\le t\le 1$),
connecting these two
equilibrium points, $q(0)=\bar{q}^{(1)}$ and
$q(1)=\bar{q}^{(2)}$. For each path $q(t)$ there is maximum
\[
m[q(t)]=
  \max_{0<t<1}
V_C(q(t)).
\]
Since $m[q(t)]>\tilde{\cal E}_0$, there is a minimum of $m[q(t)]$
in the space of
paths $P$:
\[
\mbox{Min}=\min_{q(t)\in P}m[q(t)].
\]
Let us denote the extremal path achieving Min by
$q_C(t)$ and $q_C(t_M)=\bar{q}_C$ be
its maximal point.
By definition of $\bar{q}_C$, it is an extremal point of $V_C$ with  one
negative eigenvalue of $\partial^2V_C/\partial q_j\partial q_k$
in the direction of
$q_C(t)$. However, from (\ref{Vposdef}) we know it is impossible.
Thus the assumption
of two extremal points $\bar{q}^{(1)}$ and $\bar{q}^{(2)}$ is false.

A few remarks are in order. Most of the discussion in this section,
except for those
depending on the explicit form of $W$ (\ref{W2der}), are valid in any
classical potentials of multiparticle quantum mechanical systems.
The dynamics of the pre-potentials $W$
  (\ref{Wform}), or rather that of $-W$,
for the rational and trigonometric and hyperbolic potentials has been
discussed by Dyson
\cite{dyson} from a different point of view.
It was also introduced by Calogero and collaborators \cite{calmat, calpere}
  in the
context of determining the equilibrium but without the connection
with the quantum
ground state wavefunction.

\bigskip
At the end of this section let us briefly
summarise the basic ingredients of the spin
exchange models associated with the Calogero
(Sutherland) system based on the root system
$\Delta$ and with the set of vectors ${\cal R}$, \cite{is1}.
They are defined at the equilibrium points
(\ref{clasequil}) of the corresponding
classical systems.
Here we call each element
$\mu$ of ${\cal R}$ a {\em site\/} to which a dynamical
degree of freedom called {\em
spin\/} is attached. The spin takes a finite set of
discrete values.
In the simplest, and typical case, they are an up
($\uparrow$) and a down ($\downarrow$).
The dynamical  state of the spin exchange model
is represented by a vector $\psi_{Spin}$  which takes values in the
tensor product of $D$
copies of a vector space ${\cal V}$ whose basis consists of
  an up
($\uparrow$) and a down ($\downarrow$):
\begin{equation}
\psi_{Spin}\in {\buildrel D\over\otimes}{\cal V}_{\mu}.
\end{equation}
The Hamiltonian of the spin exchange model ${\cal H}_{Spin}$  is
\begin{equation}
{\cal H}_{Spin}=\left\{
\begin{array}{l}
\begin{displaystyle}
{1\over2}\sum_{\rho\in\Delta_{+}}g_{\rho}\rho^2\,
{1\over{(\rho\cdot\bar{q})^2}}
(1-\hat{\cal P}_\rho)
\end{displaystyle},
\\[6pt]
\begin{displaystyle}
{1\over2}\sum_{\rho\in\Delta_{+}}g_{\rho}\rho^2\,
{1\over{\sin^2(\rho\cdot\bar{q})}}
(1-\hat{\cal P}_\rho)
\end{displaystyle},
\end{array}
\right.
\label{spinHam}
\end{equation}
in which $\{\hat{\cal P}_{\rho}\}$, $\rho\in\Delta_+$ are
the dynamical variables
called spin exchange operators. The operator $\hat{\cal P}_{\rho}$
exchanges the spins of sites $\mu$ and
$s_{\rho}(\mu)$, $\forall\mu\in{\cal R}$.
In terms of the operator-valued Lax pairs
\begin{equation}
L_{Spin}=\left\{
\begin{array}{l}
\begin{displaystyle}
i\sum_{\rho\in\Delta_{+}}g_{\rho}
    \,\,(\rho\cdot\hat{H})\,{1\over{\rho\cdot \bar{q}}}\,\hat{\cal
P}_{\rho}\hat{s}_{\rho}
\end{displaystyle},
\\[6pt]
\begin{displaystyle}
i\sum_{\rho\in\Delta_{+}}g_{\rho}
    \,\,(\rho\cdot\hat{H})\,\cot(\rho\cdot \bar{q})\,\hat{\cal
P}_{\rho}\hat{s}_{\rho}
\end{displaystyle},
\end{array}
\right.
\end{equation}
\begin{equation}
M_{Spin}=\left\{
\begin{array}{l}
\begin{displaystyle}
-{i\over2}\sum_{\rho\in\Delta_{+}}g_{\rho}\rho^2\,{1\over{(\rho\cdot
\bar{q})^2}}\,\hat{\cal P}_{\rho}\left(\hat{s}_{\rho}-I\right),
\end{displaystyle}
\\[6pt]
\begin{displaystyle}
-{i\over2}\sum_{\rho\in\Delta_{+}}g_{\rho}\rho^2\,
{1\over{\sin^2(\rho\cdot
\bar{q})}}\,\hat{\cal P}_{\rho}\left(\hat{s}_{\rho}-I\right),
\end{displaystyle}
\end{array}
\right.
\end{equation}
the Heisenberg equations of motion for the trigonometric spin exchange
model can be written in a
matrix form
\begin{equation}
i[{\cal H}_{Spin},L_{Spin}]=[L_{Spin},M_{Spin}].
\end{equation}
Since the $M_{Spin}$ matrix satisfies a {\em sum up to zero} condition,
\begin{equation}
    \sum_{\mu\in{\cal R}}({M_{Spin}})_{\mu\nu}=
    \sum_{\nu\in{\cal R}}({M_{Spin}})_{\mu\nu}=0,
    \label{sumMEzero}
\end{equation}
one obtains conserved quantities  via the
{\em total sum} of $L_{Spin}^k$:
\begin{equation}
[{\cal H}_{Spin}, \mbox{Ts}(L_{Spin}^k)]=0,\quad
\mbox{Ts}(L_{Spin}^k)\equiv\sum_{\mu,\nu\in{\cal
R}}(L_{Spin}^k)_{\mu\nu},
\quad k=3,\ldots
\label{totsum2}
\end{equation}
These are necessary ingredients for complete integrability.

The rational spin exchange model needs some modification similar to
those for the Calogero systems. We define
\begin{equation}
    L^{\pm}_{Spin}=L_{Spin} \pm i\omega \bar{Q},
    \quad \bar{Q}=\bar{q}\cdot\hat{H},
\label{Lpmspindef}
\end{equation}
then the Heisenberg equations of motion   in a
matrix form read
\begin{equation}
i[{\cal H}_{Spin},L_{Spin}^+L_{Spin}^-]=[L_{Spin}^+L_{Spin}^-,M_{Spin}]
\end{equation}
and conserved quantities are given by
\[
\mbox{Ts}\left((L_{Spin}^+L_{Spin}^-)^k\right)\equiv\sum_{\mu,\nu\in{\cal
R}}\left(L_{Spin}^+L_{Spin}^-\right)_{\mu\nu}^k,
\quad k=3,\ldots,.
\]

Let us emphasise that the current definition of completely
integrable spin exchange
models is universal, in the sense that it applies to any
root system $\Delta$ and to
an arbitrary choice of the set of vectors ${\cal R}$. It contains all
the known examples of spin exchange models as subcases.
For the $A_r$ root system and for the
set of vector weights,
${\cal R}={\bf V}$, (\ref{arvecwei}),
the trigonometric spin exchange model reduces to the
well-known Haldane-Shastry model
\cite{halsha}, the rational spin exchange model reduces
to the so-called Polychronakos
model \cite{fmp}.
For the $BC_r$ root systems with trigonometric interactions,
a spin model has been proposed with
${\cal R}$ chosen to be the set of short roots $\Delta_S$,
or rather, to be more precise, its $r$-dimensional degeneration.
In this case, complete integrability is known
only for three different values of
the coupling ratio $g_S/g_L$ \cite{ber}.
For $BC_r$ root systems with rational
interactions, a spin model with $r$ sites has been proposed \cite{yam}.

As is clear from the formulation, the dynamics of spin exchange models
depends on the details of the classical potential $V_C$ or $W$ at the
{\em equilibrium point\/} and on ${\cal R}$.
It is quite natural to expect that the highly
organised spectra of the known spin exchange models \cite{halsha}-\cite{is2}
are correlated with the remarkable properties of
the Lax matrices at the equilibrium point---for example,
the integer eigenvalues and their high degeneracies.
These will be explored in the following two sections.

A determination of the energy spectrum of specific spin exchange models
is not pursued in the present paper.

\section{Classical Data I: Rational Potential}
\label{cdata1}
\setcounter{equation}{0}
Next, we will obtain various data of the classical Calogero systems
extracted from the potentials, pre-potentials, Lax matrices etc., near the
equilibrium point. First, we will derive universal properties which are
valid in any root system. Those results depending on specific root
systems will be discussed afterwards.

\subsection{Minimum Energy}
\label{ratminener}
Let us consider the equation (\ref{Vmin})
and (\ref{Wmax}) for determining the classical equilibrium, which for the
rational case reads:
\begin{eqnarray}
\left.{\partial V_C\over{\partial
q_j}}\right|_{\bar{q}}=0&\Rightarrow&
\sum_{\rho\in\Delta_+}g_{\rho}^2{\rho^2\rho_j\over
{(\rho\cdot\!\bar{q})^3}}=\omega^2\bar{q}_j,
\label{calVeq}
\\[-3pt]
&&\hspace*{6cm} j=1,\ldots, r.\nonumber\\[-3pt]
\left.{\partial W\over{\partial
q_j}}\right|_{\bar{q}}=0&\Rightarrow&
\sum_{\rho\in\Delta_+}g_{\rho}{\rho_j\over{(\rho\cdot\!\bar{q})}}
=\omega\bar{q}_j,
\label{calWeq}
\end{eqnarray}
By multiplying $\bar{q}_j$ to both equations,
we obtain the virial theorem for
the  classical potential $V_C$
\begin{equation}
\sum_{\rho\in\Delta_+}g_{\rho}^2{\rho^2\over{(\rho\cdot\!\bar{q})^2}}
=\omega^2
\bar{q}^2,
\end{equation}
and a relationship
\begin{equation}
\omega
\bar{q}^2=\sum_{\rho\in\Delta_+}g_{\rho}{\rho\cdot\!\bar{q}
\over{(\rho\cdot\!\bar{q})}}
=\sum_{\rho\in\Delta_+}g_{\rho}.
\end{equation}
By combining these we arrive at the minimal value of the classical potential
(\ref{cpot}):
\begin{equation}
V_C(\bar{q})=\omega^2
\bar{q}^2=\omega(\sum_{\rho\in\Delta_+}g_{\rho})=\tilde{\cal E}_0.
\label{genminval}
\end{equation}
As stated before, this is the ground state energy ${\cal E}_0$
  minus the zero
point energy $\omega r/2$.
Although it is a classical quantity, it has the general structure of
a {\em coupling
constant\/}(s) times an {\em integer\/}:
\begin{equation}
\tilde{\cal E}_0=\left\{
\begin{array}{lr}
\omega g\times \#\Delta/2, &\qquad\mbox{simply laced,}\\[3pt]
\omega\left(g_L\times\#\Delta_L+g_S\times\#\Delta_S\right)/2,&
\qquad\mbox{non-simply laced.}
\end{array}\right.
\end{equation}
Here, $\#\Delta$ is the total number of  roots,
$\#\Delta_L$ ($\#\Delta_S$) is the number of  long (short) roots, and
$\#\Delta=\#\Delta_L+\#\Delta_S$.

\subsection{Determination of  the Equilibrium Point and Eigenvalues  of
$\widetilde{W}$}
\label{ratdetequi}
Once the  equilibrium position
$\bar{q}=(\bar{q}_1,\ldots,\bar{q}_r)$ of the pre-potential $W$ is
determined, one can define a Coxeter invariant  polynomial of
one variable, say $x$, to encode the data. For $A_r$ it is
$\prod_{j=1}^r(x-\bar{q}_j)$ and for $B_r$ ($D_r$) it is
$\prod_{j=1}^r(x-\bar{q}_j^2)$, since the set of
$\{\bar{q}_1,\ldots,\bar{q}_r\}$ and $\{\bar{q}_1^2,\ldots,\bar{q}_r^2\}$
(or rather $\{\pm\bar{q}_1,\ldots,\pm\bar{q}_r\}$)
are invariant under the Weyl group of $A_r$ and $B_r$, respectively.
As shown below, these are classical orthogonal polynomials for the
classical root systems (after suitably scaling $x$): the Hermite
polynomials for $A_r$
\cite{calmat,OP1}, and the associated Laguerre polynomials for $B_r$ ($C_r$
and $D_r$) \cite{OP1}.
For an arbitrary root system $\Delta$, such polynomials can be defined
through a Lax matrix for a proper choice of
${\cal R}$ by det$(yI-\bar{Q})$,
in which $\bar{Q}=\bar{q}\cdot\hat{H}$ is
the diagonal matrix $Q$ (\ref{Lpmdef}) at
equilibrium.
In fact, for $A_r$ and the choice of vector weights, this is the Hermite
polynomial $H_{r+1}(x)$ (with $x=y$),
and for $B_r$ ($D_r$) and the set of short
roots (vector weights),
it is the Laguerre polynomial $L_{r}^{(\alpha)}(x)$
(with $x=y^2$).
For the exceptional and non-crystallographic root systems,
the polynomials have not to the best of our
knowledge been identified or named.
We strongly believe and have several pieces of numerical
evidence that the polynomials
for non-classical root systems have ``integer coefficients" like the Hermite
and Laguerre polynomials. We have not been able
to determine these polynomials
exactly, except in the case $I_2(m)$ with special coupling
ratios $g_e=g_o=g$,  (\ref{i2poly1})-(\ref{i2poly3}).

\bigskip
After determining $\bar{q}$ we will evaluate the eigenvalues of
$\widetilde{W}\equiv W^{\prime\prime}|_{\bar{q}}$ and $\tilde{V}\equiv
V^{\prime\prime}|_{\bar{q}}$ in this subsection  and various Lax pair
matrices
$\bar{L}$,
$\bar{M}$, ${\cal L}_k$ etc in
\S\ref{rateiglax}.

To the  integer spaced quantum spectrum (\ref{ratspectra}), apart from
the ${\cal E}_0$ term, one could simply associate the following
{\em effective quadratic potential\/}
\begin{equation}
V_{eff}={1\over2}\sum_{j=1}^r\left(\omega f_j\right)^2\tilde{q}_j^2,
\end{equation}
in certain normal coordinates $\tilde{q}_j$.
We will show later that the classical potential $V_C$ has
the same behaviour as
above when expanded at the equilibrium point,  in other words:
\begin{equation}
\mbox{Spec}(\tilde{V})=\omega^2\{f_1^2,\ldots,f_r^2\}.
\end{equation}
  Considering the relation
$\tilde{V}=\widetilde{W}^2$ (\ref{VWrel}), this is equivalent to showing
\begin{equation}
\mbox{Spec}(\widetilde{W})=-\omega\{f_1,\ldots,f_r\}=
-\omega\{1+e_1,\ldots,1+e_r\}.
\end{equation}
Since the exponents $\{e_j\}$ satisfy the relation
\begin{equation}
\sum_{j=1}^r e_j=\#\Delta/2=hr/2,\quad
\end{equation}
where $h$ is the Coxeter number, we have a simple sum rule
(see footnote on page \pageref{compform})%
\begin{equation}
\mbox{Tr}(\widetilde{W})=-\omega(r+\#\Delta/2)=-\omega r(1+h/2).
\label{trWformula}
\end{equation}

\subsubsection{$A_r$}
\label{arratWspec}
Calogero and collaborators discussed this problem
about quarter of a century ago
\cite{calmat,OP1}.
In these cases the root vectors embedded in ${\bf R}^{r+1}$ are
given by:
\begin{equation}
A_r=\{{\bf e}_j-{\bf e}_k,\
j,k=1,\ldots,r+1|{\bf e}_j\in{\bf R}^{r+1}, {\bf e}_j\cdot
{\bf e}_k=\delta_{jk}\}.
\end{equation}
The equations (\ref{calWeq}) read
\begin{equation}
\sum_{k\ne j}^{r+1}{1\over{\bar{q}_j-\bar{q}_k}}=
{\omega\over{g}}\bar{q}_j,
\quad j=1,\ldots,r+1.
\label{herroots}
\end{equation}
These determine $\{\bar{x}_j=\sqrt{\omega\over g}\bar{q}_j\}$,
$j=1,\ldots,r+1$ to be the zeros of the Hermite  polynomial $H_{r+1}(x)$
\cite{szego}.

The matrix $\widetilde{W}$ is given by
\begin{equation}
\widetilde{W}_{jk}=-\left(\omega+g\sum_{l\ne
j}{1\over{(\bar{q}_j-\bar{q}_l)^2}}\right)\delta_{jk}
+g{1\over{(\bar{q}_j-\bar{q}_k)^2}},
\end{equation}
and is equal to $(-\omega I+i\bar{M})_{jk}$
for the representation of the Lax
matrix $\bar{M}$ (\ref{Mdef}) in terms of the $A_r$ vector weights
(\ref{arvecwei}). From the general result in \S\ref{rateiglax}
(\ref{specME}), we obtain
\begin{equation}
A_r:\quad \mbox{Spec}(\widetilde{W})=-\omega\left\{ 1, 2,
\ldots,r+1\right\}.
\label{specarW}
\end{equation}
The general result on the relationship between $\bar{q}$
and $\tilde{\cal E}_0$ (\ref{genminval})  translates, in this case, to the
classical result
\begin{equation}
\sum_{j=1}^{r+1}\bar{x}_j^2={r(r+1)\over2},
\label{herrootsum}
\end{equation}
in which $\{\bar{x}_j\}$, $j=1,\ldots,r+1$ are the zeros of $H_{r+1}(x)$.
These are some of the earliest results concerning
integer eigenvalues associated
with the Calogero-Moser classical equilibrium points.  The original results
\cite{calmat, ahmcal} depended heavily on specific properties  of Hermite
polynomials. Here we have emphasised the universal structure rather than
particular properties of specific systems.

\subsubsection{$B_r$ ($D_r$)}
\label{brratWspec}
In this case, the root vectors are expressed neatly in terms of an
orthonormal basis of  ${\bf
R}^{r}$ by:
\begin{equation}
B_r=\{\pm{\bf e}_j \pm{\bf e}_k,\quad \pm {\bf e}_j,\quad
j,k=1,\ldots,r|{\bf e}_j\in{\bf R}^{r}, {\bf e}_j\cdot
{\bf e}_k=\delta_{jk}\}.
\label{brroots}
\end{equation}
Let us note that the rational $C_r$ and $BC_r$ systems are identical with
the $B_r$ system.  Assuming $\bar{q}_j\ne0$, the equations (\ref{calWeq})
read
\begin{equation}
\sum_{k\ne
j}^{r}{1\over{\bar{q}_j^2-\bar{q}_k^2}}
+{g_S/2g_L\over{\bar{q}_j^2}}={\omega\over{2g_L}},
\quad j=1,\ldots,r,
\label{lagroots}
\end{equation}
and determine $\{\bar{q}_j^2\}$, $j=1,\ldots,r$, as the
zeros of the associated Laguerre  polynomial $L_{r}^{(\alpha)}(cx)$,
with $\alpha=g_S/g_L-1$, $c=\omega/g_L$, \cite{szego,OP1}.
The general result on the relationship between $\bar{q}$
and $\tilde{\cal E}_0$ (\ref{genminval})  translates in this case to the
classical result
\begin{equation}
\sum_{j=1}^{r}\bar{x}_j=r(r+\alpha),
\label{lagrootsum}
\end{equation}
in which $\{\bar{x}_j\}$, $j=1,\ldots,r$, are the zeros of
$L_{r}^{(\alpha)}(x)$.
  The subcase with $g_S=0$, that is
$D_r$, $\{\bar{q}_j^2\}$, $j=1,\ldots,r$, are the zeros of the
associated Laguerre  polynomial \cite{szego,OP1},
\begin{equation}
r\,L_{r}^{(-1)}(cx)=-cxL_{r-1}^{(1)}(cx)
\label{lagiden}
\end{equation}
  for which one of the $\bar{q}_j$ is zero.
(This also means that the $\{\bar{q}_j\}$ of $B_r$ for $g_S/g_L=2$ or
$\alpha=1$ are the same as the non-vanishing $\{\bar{q}_j\}$ of $D_{r+1}$.
This can be understood easily from the Dynkin diagram folding
$D_{r+1}\to B_r$. Let us note that another Weyl invariant of $D_r$,
$\bar{q}_1\cdots\bar{q}_r$, is trivial (zero) in the present case.)
By summing (\ref{lagroots}) over $j$, we obtain another sum rule
for the inverse square of the zeros
\begin{equation}
{g_S}\sum_{j=1}^r{1\over{\bar{q}_j^{2}}}={r\omega},
\quad \mbox{or} \quad \sum_{j=1}^r{1\over{\bar{x}_j}}={r\over{\alpha+1}}.
\label{invsum}
\end{equation}
This formula implies that the $D_r$ limit, or $g_S\to0$ ($\alpha\to-1$)
limit is singular. In other words, in this limit one of the $\bar{q}_j$ must
vanish, otherwise the left hand side of (\ref{invsum}) goes to zero, whereas the
right hand side is a constant. This singularity explains the difference of the
spectrum of
$\widetilde{W}$ for $B_r$ and $D_r$ in units of $-\omega$:
\begin{equation}
\begin{tabular}{|c|l|}
\hline
$\Delta$&\vT \hspace*{10mm}Spec($\widetilde{
W}$)\\
\hline
$B_r$&\vTm 2, \, 4, \, 6,\, \ldots, \, $2r-2$, \, $2r$\\
\hline
$D_r$&\vTm 2, \, 4, \, 6,\, \ldots, \, $2r-2$, \, $r$\\
\hline
\end{tabular}
\end{equation}
verified by direct computation. This is to be compared with
Table 1.
It is easy to understand via a   scaling the coupling constant independence of
the spectrum of $\widetilde{W}$   for  $A_r$ and $D_r$ root sytems.
However, for the non-simply laced root systems, $B_r$, $F_4$ and
$I_2(even)$, with two independent coupling constants,
$g_L$ and $g_S$, the  coupling  independence of
the spectrum of $\widetilde{W}$ is rather non-trivial.

\bigskip
In fact, Szeg\"o derived the equations (\ref{herroots}) and (\ref{lagroots})
while tackling the problem of maximising $\phi_0=e^{W}$ (Theorem 6.7.2
\cite{szego}) in a slightly different notation and setting---without $V_C$,
or quantum mechanics, or the Lax pairs. However, he did not
mention (\ref{invsum}).

\subsubsection{Exceptional root systems ($F_4$ and $E_r$, $r=6,7,8$)}
\label{excepratWspec}
In each of these cases we have calculated the equilibrium position numerically,
and evaluated the spectrum of $\widetilde{W}$.
For $F_4$,  various ratios of $g_S/g_L$
have been tried  and we have verified that the
spectrum of $\widetilde{W}$ is independent of the coupling ratio.
The results are tabulated  in  units of $-\omega$:
\begin{equation}
\begin{tabular}{|c|c|l|}
\hline
$\Delta$&${\ r}$&\vT \hspace*{14mm}Spec($\widetilde{
W}$)\\
\hline
$F_4$&4&\vTm 2, \,6, \,8, \,12\\
\hline
$E_6$&6&\vTm 2, \,5, \,6, \,8, \,9, \,12\\
\hline
$E_7$&7&\vTm 2, \,6, \,8, \,10, \,12, \,14, \,18\\
\hline
$E_8$&8&\vTm 2, \,8, \,12, \,14, \,18, \,20, \,24, \,30\\
\hline
\end{tabular}
\end{equation}

\subsubsection{$I_2(m)$}
\label{i2Wspec}
In this case the root vectors are given by:
\begin{equation}
    I_2(m)=\left\{\vT \sqrt{2}\left(\vTm
    \cos((j-1)\pi/m),\sin((j-1)\pi/m)\right)\in{\bf
R}^2,\quad j=1,\ldots,2m\right\}.
    \label{dihedroots}
\end{equation}
The complete set of quantum eigenfunctions are obtained by
separation of variables in terms of  two dimensional polar coordinates
\cite{OP1,kps}:
\begin{equation}
(q_1,q_2)=r(\sin\varphi,\cos\varphi),
\label{polar}
\end{equation}
which of course separate the pre-potential, the ground state wavefunction:
\begin{equation}
W=\left\{\begin{array}{ll}
g(m\log r+\log\sin m\varphi)-{\omega\over2}r^2,&m\ \mbox{odd,}\\[6pt]
\!g_e\!\left(\!\vTm m/2\log r+\log\sin(m\varphi/2)\right)+
g_o\!\left(\!\vTm m/2\log r+\log\cos(m\varphi/2)\right)-{\omega\over2}r^2,&m\
\mbox{even.}\\
\end{array}
\right.
\end{equation}
The equilibrium points are easily obtained:
\begin{equation}
\bar{r}^2={mg\over{\omega}},\quad
\bar{\varphi}={\pi\over{2m}},\quad m\ \mbox{odd,}
\qquad
\bar{r}^2={m(g_e+g_o)\over{2\omega}},\quad
\tan{m\bar{\varphi}\over2}=\sqrt{g_e\over{g_o}},
\quad m\ \mbox{even,}
\end{equation}
together with the values of the second derivatives:
\begin{eqnarray}
\left.{\partial^2 W\over{\partial r^2}}\right|_{\bar{q}}&=&-2\omega,
\quad
\left.{\partial^2 W\over{\partial\varphi^2}}\right|_{\bar{q}}=-m^2g,
\hspace{21.5mm} m:\ \mbox{odd,}
\\[3pt]
\left.{\partial^2 W\over{\partial r^2}}\right|_{\bar{q}}&=&-2\omega,
\quad
\left.{\partial^2
W\over{\partial\varphi^2}}\right|_{\bar{q}}=-m^2(g_e+g_o)/2,
\quad m\ \mbox{even.}
\end{eqnarray}
These translate into the spectrum of $\widetilde{W}$ in  cartesian
coordinates:
\begin{equation}
I_2(m):\quad \mbox{Spec}(\widetilde{W})=-\omega\left\{2,m\right\},\quad
m\ \mbox{odd or even.}
\label{specaI2W}
\end{equation}
The case
$G_2$  is the $m=6$   dihedral root system treated above.

It is relatively easy to derive the explicit form the Coxeter invariant
polynomials for $I_2(m)$ when
\begin{equation}
g_e=g_o=g.
\end{equation}
In these cases the pre-potential $W$ and the equations for
the equilibrium position look the same for even or odd $m$.
If we choose for ${\cal R}$ the set of vertices $R_m$ of
the regular $m$-gon associated with $I_2(m)$, (\ref{rmvert}),
we obtain a degree $m$ polynomial in $y$, det$(yI-\bar{Q})$.
By scaling
\(
y=\sqrt{2mg\over{\omega}}x
\)
we obtain
\begin{equation}
\mbox{det}(yI-\bar{Q})
\propto \left\{
\begin{array}{ll}
\prod_{k=1}^m
\left(x-\sin\left[{2k\pi\over m}+{\pi\over{2m}}\right]\right)&
m\,\mbox{even}
\\[8pt]
\prod_{k=1}^m
\left(x-\sin\left[{2k\pi\over m}+{\pi\over{m}}\right]\right)&
m\,\mbox{odd}
\end{array}
\right\}
\propto T_m(x),
\label{i2poly1}
\end{equation}
in which $T_m$ is the Chebyshev polynomial of the first
kind (\ref{2chebfirst}),
\begin{equation}
T_m(x)=\cos m\varphi,\qquad x=\cos\varphi.
\end{equation}
They satisfy the orthogonality
\begin{equation}
\int_{-1}^{1}{T_m(x)T_l(x)\over{\sqrt{1-x^2}}}dx
\propto \delta_{m\,l}.
\label{i2poly3}
\end{equation}
The general theory of Coxeter invariant polynomials for
the arbitrary coupling case $g_e\ne g_o$ will be published elsewhere.

\subsubsection{$H_r$}
\label{hrratWspec}
In this case, we have evaluated the equilibrium points numerically
and verified the following
\begin{eqnarray}
H_3\quad \mbox{Spec}(\widetilde{W})&=&-\omega\{2,6,10\},\\
H_4\quad \mbox{Spec}(\widetilde{W})&=&-\omega\{2,12,20,30\}.
\end{eqnarray}
In all the cases, from  \S\ref{arratWspec} to
\S\ref{hrratWspec} the 1 part in
the spectrum of $\widetilde{W}$, {\em i.e.,}
$f_j=1+e_j$, is always due to the confining  harmonic potential
$-\omega q^2/2$ in the pre-potential $W$.

\subsection{Eigenvalues of Lax Matrices}
\label{rateiglax}
The Lax pair operators $L$, $M$, $L^\pm$ etc (\ref{LaxOpDef}), (\ref{Mdef}),
(\ref{Lpmdef}) are $D\times D$
matrices  if a set of vectors ${\cal R}$ forming  a single orbit of the
reflection (Coxeter) group with $D$ elements is chosen.

\subsubsection{Universal Spectrum of $M$}

Let us denote by ${\bf
v}_0$ a special vector in ${\bf R}^D$ with each element unity:
\begin{equation}
{\bf v}_0=(1,1,\ldots,1)^T\in{\bf R}^D,\quad D=\#{\cal R},
\qquad \mbox{or}\quad
{\bf v}_{0\mu}=1,\quad \forall \mu\in{\cal R}.
\label{v0def}
\end{equation}
Let us note that the condition
for classical equilibrium (\ref{calWeq}) can be written simply in terms
of
$L^{-}$ as
\[
\sum_{\nu\in{\cal R}}(\bar{L}^{-})_{\mu\nu}=0,
\qquad \bar{L}^-\equiv L^-(0,\bar{q}),
\]
since $\sum_{\nu\in{\cal R}}(\hat{s}_\rho)_{\mu\nu}=1$.
Similarly from (\ref{Mdef}) we obtain {\em sum up to zero\/} conditions
\[
\sum_{\nu\in{\cal R}}M_{\mu\nu}=0,\quad \sum_{\mu\in{\cal R}}M_{\mu\nu}=0.
\]
It should be stressed that the above two conditions are essential for
deriving the {\em quantum conserved quantities\/} \cite{bms,kps}.
These can be expressed neatly in matrix-vector notation as
\begin{equation}
\bar{L}^-{\bf v}_0=0,\quad {\bf v}_0^T\bar{L}^+=0,
\quad \bar{M}{\bf v}_0=0,\quad
{\bf v}_0^T\bar{M}=0,
\end{equation}
inspiring  the idea that ${\bf v}_0$ is the classical
  (Coxeter
invariant) {\em ground\/} state of a matrix counterpart of
the Hamiltonian ($\bar{M}$) and that
$\bar{L}^-$ is an {\em annihilation\/} operator.
The analogy goes further when we evaluate the Lax equation for $L^{\pm}$
(\ref{omegaLM}) at the classical equilibrium to obtain
\begin{equation}
[\bar{M},\bar{L}^{\pm}]=\pm i\omega \bar{L}^{\pm}.
\label{mlpmrel}
\end{equation}
However, the commutator of $L^+$ and $L^-$ does not produce $\bar{M}$
but the constant matrix $K$ (see (\ref{QLcomm}) and the Appendix):
\begin{equation}
[\bar{L}^+,\bar{L}^-]=[\bar{L}+i\omega \bar{Q},
\bar{L}-i\omega \bar{Q}]=-2\omega K,
\label{lpmk}
\end{equation}
together with the relation
\begin{equation}
[\bar{M},[\bar{L}^+,\bar{L}^-]]=0,
\end{equation}
since $K$ and $\bar{M}$ commute (\ref{kmcomm}).
The  relation (\ref{mlpmrel})
simply means that the eigenvalues of $\bar{M}$ are integer spaced in
units of $i\omega$. We obtain
\begin{equation}
\bar{M}{\bf v}_0=0,\quad \bar{M}\bar{L}^+
{\bf v}_0=i\omega \bar{L}^+{\bf v}_0,\quad
\ldots ,
\quad
\bar{M}(\bar{L}^+)^n{\bf v}_0=in\omega (\bar{L}^+)^n{\bf v}_0,
\end{equation}
implying  $\bar{L}^+$ is a corresponding {\em creation\/} operator.
This also means there is a {\em universal\/} formula:
\begin{equation}
\mbox{Spec}(\bar{M})=i\omega\left\{0, 1, 2,
\ldots,\right\},
\label{specME}
\end{equation}
with possible degeneracies. The following sum rules (trace formulae)
(\ref{trME}) and (\ref{caltraceform})
are useful.
Let us note the simple formula
\begin{equation}
(\hat{s}_{\rho}-I)_{\mu\,\mu}=\left\{
\begin{array}{cl}
-1, & \qquad \rho\cdot\!\mu\ne 0,\\[6pt]
0,&\qquad\rho\cdot\!\mu= 0
\end{array}
\right.
\label{FRdef}
\end{equation}
and the fact that, for a fixed $\rho$, the number of $\mu$'s
in ${\cal R}$ such that
$\rho\cdot\!\mu\ne 0$ is almost independent of $\rho$,
depending only on its orbit
($|\rho|$) and ${\cal
R}$. Let us denote this number by $F_{\rho}^{\cal R}$.
On taking the trace of $\bar{M}$,
we obtain
\begin{equation}
\mbox{Tr}(\bar{M})={i\over2}\sum_{\rho\in\Delta_+}
g_{\rho}F_{\rho}^{\cal R}{\rho^2
\over{(\rho\cdot\!\bar{q})^2}}.
\label{trME}
\end{equation}
This formula simplifies for the simply laced root systems.
In those cases, we arrive at a simple relation between
Tr($\widetilde{W}$), which is independent of ${\cal R}$,
and Tr($\bar{M}$), which depends on the choice of ${\cal R}$,
by comparing with (\ref{W2der}):
\begin{equation}
\mbox{Tr}(\bar{M})=-F^{\cal
R}{i\over2}\left(\omega r+\mbox{Tr}(\widetilde{W})\right)
={i\over4}\omega rhF^{\cal R},
\qquad \Delta:\ \mbox{simply laced},
\label{caltraceform}
\end{equation}
in which (\ref{trWformula}) is used.
This formula provides a non-trivial check for the
numerical evaluation of the eigenvalues of $\bar{M}$, since the right
hand side (except for the factor $i\omega/4$) is an integer determined
by
$\Delta$ and ${\cal R}$.

For $A_r$ ($B_r$) with vector
weights (or with short roots) $\bar{M}$ has no degeneracy but high
multiplicities occur for the $D_r$ vector or spinor weights.
Here is the
summary of the spectrum of $\bar{M}$ (in  units of $i\omega$) with
[multiplicity] for the classical root systems:
\begin{equation}
\begin{tabular}{|c|c|c|l|}
\hline
$\Delta$&${\cal R}$&$D$&\vT \hspace*{25mm}Spec($\bar{M}$)\\
\hline
$A_r$&{\bf V}&$r+1$&\vTm 0, \,1, \ldots, \,$r-1$, \,$r$\\
\hline
$B_r$&$\Delta_S$&$2r$&\vTm 0, \,1, \,2, \ldots, \,$2r-1$ \\
\hline
$D_r$&{\bf V}&$2r$&\vTm 0, \,1, \,2, \ldots, $r-1$[2], \ldots, \,$2r-2$\\
\hline
$D_4$&{\bf S}&8&\vTm 0,  \,1,  \,2,  \,3[2], \,4,  \,5,  \,6\\
\hline
$D_5$&{\bf S}&16&\vTm 0,  \,1,  \,2,  \,3[2], \,4[2], \,5[2], \,6[2],
\,7[2], \,8,
      \,9,  \,10\\
\hline
$D_6$&{\bf S}&32&\vTm 0,  \,1,  \,2,  \,3[2], \,4[2],
\,5[3], \,6[3], \,7[3], \,8[3],
\,9[3],  \\
&&&\vTm \ 10[3],  \,11[2], \,12[2], \,13,  \,14, \,15\\
\hline
\end{tabular}
\label{Madmin}
\end{equation}
For the minimal weights of the exceptional root systems $E_r$, we obtain:
\begin{equation}
\begin{tabular}{|c|c|c|l|}
\hline
$\Delta$&${\cal R}$&$D$&\vT \hspace*{33mm}Spec($\bar{M}$)\\
\hline
$E_6$&{\bf 27}&27&\vTm
0, \,1,  \,2,  \,3,  \,4[2], \,5[2], \,6[2], \,7[2],
  \,8[3], \,9[2], \,10[2],   \\
&&& \vTm\ \,11[2], \,12[2], \,13,  \,14,  \,15,
    \,16 \\
\hline
$E_7$&{\bf 56}&56&\vTm
0,  \,1, \,2,  \,3,  \,4,  \,5[2], \,6[2], \,7[2], \,8[2], \,9[3],
  \,10[3], \,11[3], \\
&&&\vTm \ 12[3], \, 13[3], \,14[3], \,15[3], \,16[3], \,17[3],
  \,18[3], \,19[2], \\
&&&\vTm \  20[2], \, 21[2], \,22[2], \,23,  \,24,  \,25,  \,26,
\,27 \\
\hline
\end{tabular}
\label{e67minM}
\end{equation}
These are consistent with  the trace formula for
$\bar{M}$ (\ref{caltraceform}), since
\begin{equation}
F^{\bf V}(A_r)=2,\quad F^{\bf V}(D_r)=4,\quad
F^{\bf S}(D_r)=2^{r-2},\quad
F^{\bf 27}(E_6)=12,\quad
F^{\bf 56}(E_7)=24.
\label{minF}
\end{equation}
For example, $F^{\bf V}(A_r)=2$
can be seen easily. For the typical choice
$\rho={\bf e}_1-{\bf e}_2$, $\mu={\bf e}_1$ and $\mu={\bf e}_2$ are
the only two vectors having non-vanishing scalar product with $\rho$.

Let us define the {\em height\/} of a vector $\mu\in{\bf
R}^r$ by  its scalar product with the {\em Weyl vector\/}, {\em i.e.,\/}
(\ref{weyldef})
\begin{equation}
\delta\cdot\mu\in{\bf R}.
\end{equation}
The set of heights of all vectors in  ${\cal R}$, denoted by
$\delta\cdot{\cal R}$,
\begin{equation}
\delta\cdot{\cal R}=\{\delta\cdot\mu |\ \mu\in{\cal R}\},
\end{equation}
together with its  maximum $h_{max}\equiv\max(\delta\cdot{\cal R})$, are
independent of the choice of the positive roots.
The above results on the  spectrum of the Lax matrix $\bar{M}$ at
equilibrium (defined by ${\cal R}$) (\ref{Madmin})-(\ref{e67minM}) are
summarised neatly by the set of heights of the vectors in ${\cal R}$
shifted by $h_{max}$:
\begin{equation}
\mbox{Spec}(\bar{M})=i\omega\{\delta\cdot\mu+h_{max} |\mu\in{\cal
R}\}.
\label{Mspecform}
\end{equation}

\bigskip
The eigenvalues and multiplicities of $\bar{M}$
in the root type Lax pairs
of simply laced crystallographic root systems
can also be understood as the {\em height\/} and multiplicities
of $\Delta$.
These are mirror symmetric with respect to the midpoint,
including multiplicity. The maximum multiplicity
is always the rank $r$ occurring twice in the
middle. The highest eigenvalue (in  units of $i\omega$) is $2h-3$,
where $h$ is the Coxeter number.
If the lower half is shifted by $-(h-1)$, ($h-1$ is the maximal height),
and the higher half by $-(h-2)$, then
the eigenvalues range from
$-(h-1)$ to $h-1$, precisely the range of the {\em height\/} of the roots.
The multiplicities of $\bar{M}$ are just the numbers of roots with that
height. The following tables summarize  the spectrum
of $\bar{M}$ in the root type
Lax pairs for the simply laced classical root
systems obtained by direct computation:
\begin{equation}
\begin{tabular}{|c|c|c|c|l|}
\hline
$\Delta$&$h$&${\cal R}$&$D$&\vT \hspace*{38mm}Spec($\bar{M}$)\\
\hline
$A_r$&$r+1$&$\Delta$&$r(r+1)$&\vTm 0, \,1[2],
\ldots, $r-1$[$r$], \,$r$[$r$],
$r+1$[$r-1$], \ldots, $2r-2$[2], \,$2r-1$\\
\hline
$D_4$&$6$&$\Delta$&24&\vTm 0,  \,1,  \,2[3], \,3[3], \,4[4], \,5[4],
  \,6[3], \,7[3], \,8,
      \,9, \\
\hline
$D_5$&8&$\Delta$&40&\vTm 0,  \,1,  \,2[2], \,3[3], \,4[4], \,5[4], \,6[5],
\,7[5], \,8[4],
\,9[4], \,10[3], \,11[2],\\
&&&&\vTm \ 12, \,13, \\
\hline
$D_6$&10&$\Delta$&60&\vTm
0,  \,1,  \,2[2], \,3[2], \,4[4], \,5[4], \,6[5], \,7[5],
\,8[6], \,9[6], \,10[5], \,11[5],\\
&&&& \ 12[4], \,13[4], \,14[2],
\,15[2], \,16,  \,17\\
\hline
$D_7$&12&$\Delta$&84&\vTm
0,  \,1,  \,2[2], \,3[2], \,4[3], \,5[4], \,6[5], \,7[5],
  \,8[6], \,9[6], \,10[7], \,11[7],\\
&&&& \ 12[6], \,13[6], \,14[5],
  \,15[5], \,16[4], \,17[3], \,18[2], \,19[2], \,20,  \,21, \\
\hline
\end{tabular}
\label{adrrootM}
\end{equation}
The spectrum of $\bar{M}$ with the long roots of the
$B_r$ is very interesting.
The highest eigenvalue is $2h-5$$=2h^\vee-3$,
in which $h^\vee$ is the {\em dual Coxeter
number\/} and the highest multiplicity is
$r-1$, the number of the long simple roots.
The spectrum is mirror symmetric with
respect to the midpoint.
If the lower half is shifted by $-(h^\vee-1)$ and the higher
half by $-(h^\vee-2)$,
the eigenvalues range from
$-(h^\vee-1)$ to $h^\vee-1$, which is the range of the
{\em height\/} of the roots.
Thus we conclude that the spectrum of $\bar{M}$  with the long roots is again
the same as the distribution of the
$B_r$ long roots   with respect to the {\em height\/}:
\begin{equation}
\begin{tabular}{|c|c|c|c|c|l|}
\hline
$\Delta$&$h$&$h^\vee$&${\cal R}$&$D$&\vT \hspace*{33mm}Spec($\bar{M}$)\\
\hline
$B_4$&$8$&$7$&$\Delta_L$&24&\vTm 0,  \,1,  \,2[2], \,3[2],
\,4[3], \,5[3], \,6[3], \,7[3],
\,8[2], \,9[2], \,10,  \,11\\
\hline
$B_5$&10&9&$\Delta_L$&40&\vTm 0,  \,1,  \,2[2], \,3[2],
\,4[3], \,5[3], \,6[4],
  \,7[4], \,8[4], \,9[4], \,10[3],\\
&&&&&\vTm  \ 11[3], \,12[2], \,13[2],
\,14,  \,15   \\
\hline
$B_6$&12&11&$\Delta_L$&60&\vTm 0, \,1,  \,2[2], \,3[2],
\,4[3], \,5[3], \,6[4], \,7[4],
\,8[5], \,9[5], \,10[5],\\
&&&&&\vTm \ 11[5], \,12[4], \,13[4], \,14[3],
\,15[3], \,16[2], \,17[2], \,18,  \,19\\
\hline
$B_7$&14&13&$\Delta_L$&84&\vTm 0, \,1,  \,2[2], \,3[2],
\,4[3], \,5[3], \,6[4], \,7[4],
\,8[5], \,9[5], \,10[6], \\
&&&&&\vTm \ 11[6], \,12[6], \,13[6], \,14[5],
  \,15[5], \,16[4], \,17[4], \,18[3],\\
&&&&&\vTm  \ 19[3], \,20[2], \,21[2], \,22,  \,23\\
\hline
\end{tabular}
\label{brrootM}
\end{equation}
The spectrum of root type $\bar{M}$ (\ref{adrrootM}),(\ref{brrootM}),
(\ref{f4rootM}), (\ref{excrootM}) can
be expressed succinctly  in terms of the Weyl vector $\delta$:
\begin{equation}
\mbox{Spec}(\bar{M})=\left\{
\begin{array}{lc}
\delta\cdot \mu+h_{max}, & \mbox{for } \delta\cdot\mu<0\\
\delta\cdot \mu+h_{max}-1, & \mbox{for } \delta\cdot\mu>0\\
\end{array}
\right|\left.\vTb \ \mu\in\Delta\ (\Delta_L)\right\},
\label{Mspecform2}
\end{equation}
in which as before $h_{max}\equiv\max(\delta\cdot{\Delta})$ or
$\max(\delta\cdot{\Delta_L})$.
The spectra for $F_4$ in terms of $\Delta_L$ and
$\Delta_S$ are the same, reflecting the
self-duality of the $F_4$ root system.
The situation is about the same as in the $B_r$ cases.
The highest multiplicity is 2, which is the number of long (short)
simple roots.
\begin{equation}
\begin{tabular}{|c|c|c|c|c|l|}
\hline
$\Delta$&$h$&$h^\vee$&${\cal R}$&$D$&\vT
\hspace*{38mm}Spec($\bar{M}$)\\
\hline
$F_4$&12&9&$\Delta_L$&24&\vTm
0,  \,1,  \,2,  \,3,  \,4[2],
  \,5[2], \,6[2], \,7[2], \,8[2], \,9[2], \,10[2], \,11[2], \\
&&&&&\vTm \ 12,  \,13,  \,14,  \,15, \\
\hline
$F_4$&12&9&$\Delta_S$&24&\vTm
0,  \,1,  \,2,  \,3,  \,4[2],
  \,5[2], \,6[2], \,7[2], \,8[2], \,9[2], \,10[2], \,11[2], \\
&&&&&\vTm \ 12,  \,13,  \,14,  \,15, \\
\hline
\end{tabular}
\label{f4rootM}
\end{equation}
\begin{equation}
\begin{tabular}{|c|c|c|c|l|}
\hline
$\Delta$&$h$&${\cal R}$&$D$&\vT \hspace*{38mm}Spec($\bar{M}$)\\
\hline
$E_6$&12&$\Delta$&72&\vTm
0,  \,1, \,2,  \,3[2], \,4[3], \,5[3], \,6[4], \,7[5], \,8[5],
  \,9[5], \,10[6], \,11[6], \\
&&&& \vTm\  12[5], \,13[5], \,14[5],
\,15[4], \,16[3], \,17[3], \,18[2], \,19,  \,20,  \,21   \\
\hline
$E_7$&18&$\Delta$&126&\vTm
0,  \,1,  \,2,  \,3,  \,4[2], \,5[2], \,6[3], \,7[3],
\,8[4], \,9[4], \,10[5], \,11[5],\\
&&&&\vTm \ 12[6], \,13[6], \,14[6], \,15[6],
\,16[7], \,17[7], \,18[6], \,19[6], \,20[6],\\
&&&&\vTm  \ 21[6], \,22[5],
\,23[5], \,24[4], \,25[4], \,26[3], \,27[3], \,28[2], \,29[2],
\\
&&&&\vTm
\ 30,  \,31,  \,32,  \,33\\
\hline
$E_8$&30&$\Delta$&240&\vTm
0, \,1, \,2, \,3, \,4, \,5, \,6[2], \,7[2], \,8[2],
\,9[2], \,10[3], \,11[3], \,12[4],\\
&&&&\vTm \ 13[4], \,14[4], \,15[4],
\ 16[5], \,17[5], \,18[6], \,19[6], \,20[6], \,21[6],\\
&&&&\vTm  \ 22[7],
\,23[7], \,24[7], \,25[7], \,26[7], \,27[7], \,28[8], \,29[8],
\,30[7],\\
&&&&\vTm  \ 31[7], \,32[7], \,33[7], \,34[7], \,35[7], \,36[6],
\,37[6], \,38[6], \,39[6],\\
&&&&\vTm \ 40[5], \,41[5], \,42[4], \,43[4],
\,44[4], \,45[4], \,46[3], \,47[3],\\
&&&&\vTm \ 48[2], \,49[2], \,50[2],
\,51[2], \,52, \,53, \,54, \,55, \,56, \,57 \\
\hline
\end{tabular}
\label{excrootM}
\end{equation}
The eigenvalue (the {\em height\/} of the root) where the multiplicity
changes corresponds to the exponent. When the multiplicity changes by two
units, which occurs only in $D_{even}$, there are two equal exponents.
We do not have analytic proofs of these facts.

\bigskip
The situation for the non-crystallographic root systems
is  different since the ``integral heights" are not defined for the roots.
The highest eigenvalue is not
$2h-3$.   The places where the multiplicity changes, counted from the
center of the spectrum, are not the exponents but 3, 5 and 7
($3+7=10=5+5=h$ for $H_3$) and 7, 13, 17 and 23 ($7+23=13+17=30=h$ for
$H_4$). It is known that $H_3$ ($H_4$) is obtained from $D_6$ ($E_8$) by
``folding".  The
above integers are the exponents of
$D_6$ and $E_8$. The rest of the exponents of $D_6$ ($E_8$) are inherited
by $H_3$ ($H_4$). The pair
$D_6$ and
$H_3$ ($E_8$ and
$H_4$) share the same Coxeter number $h$. For other aspects of the
$\bar{M}$ spectra of root type Lax pairs of $H_r$,
  we do not have an explanation to offer. Here is the summary of the
spectrum of
$\bar{M}$ for the root type Lax pairs of $H_r$:
\begin{equation}
\begin{tabular}{|c|c|c|c|l|}
\hline
$\Delta$&$h$&${\cal R}$&$D$&\vT \hspace*{33mm}Spec($\bar{M}$)\\
\hline
$H_3$&10&$\Delta$&30&\vTm 0, \,1, \,2[2], \,3[2],
\,4[3], \,5[3], \,6[3], \,7[3],
\,8[3], \,9[3], \\
&&&&\vTm \ 10[2], \,11[2], \,12,\, 13\\
\hline
$H_4$&30&$\Delta$&120&\vTm
0, \,1, \,2, \,3,  \,4,  \,5,
\,6[2],
\,7[2],
\,8[2], \,9[2], \,10[3], \,11[3],\\
&&&&\vTm  \ 12[3],
  \,13[3], \,14[3], \,15[3],
\,16[4], \,17[4], \,18[4], \,19[4],\\
&&&&\vTm \ 20[4], \,21[4],
  \,22[4],
\,23[4], \,24[4], \,25[4], \,26[4], \,27[4],\\
&&&& \vTm  \ 28[4], \,29[4],
\,30[3], \,31[3], \,32[3], \,33[3], \,34[3], \,35[3],\\
&&&&\vTm \ 36[2],
\,37[2], \,38[2], \,39[2], \,40, \,41,  \,42,  \,43,
     \,44,  \,45 \\
\hline
\end{tabular}
\label{e67rootM}
\end{equation}
In all these root type cases the highest multiplicity is equal to the
rank
$r$.
The spectra of $\bar{M}$ for the simply laced root systems are
consistent with  the trace formula for
$\bar{M}$ (\ref{caltraceform}), since
\begin{eqnarray}
F^{\Delta}(A_r)&=&2(2r-1),\quad
F^{\Delta}(D_r)=8r-14,\qquad
F^{\Delta}(E_6)=42,\nonumber\\
F^{\Delta}(E_7)&=&66,\quad
F^{\Delta}(E_8)=114,\quad
F^{\Delta}(H_3)=26,\quad
F^{\Delta}(H_4)=90.
\label{fDelta}
\end{eqnarray}
For  crystallographic root systems, {\em i.e.\/}, $A_r$, $D_r$ and $E_r$,
$F^{\Delta}=4h-6$ and $F^{\Delta}$ is twice the maximal eigenvalue of
$\bar{M}$ for all the cases listed above.

Finally, for $I_2(m)$  in the  $m$ dimensional
representation (\ref{rmvert}):
\begin{equation}
\begin{tabular}{|c|c|c|c|l|}
\hline
$\Delta$&$h$&${\cal R}$&$D$&\vT \hspace*{10mm}Spec($\bar{M}$)\\
\hline
$I_2(2n+1)$&$2n+1$&$R_{2n+1}$&$2n+1$&\vT
0, \,1,  \ldots, \,$2n-1$, \,$2n$\\
\hline
$I_2(2n)$&$2n$&$R_{2n}$&$2n$&\vTm 0, \,1,  \ldots, \,$2n-2$, \,$2n-1$\\
\hline
\end{tabular}
\label{I2mM}
\end{equation}

\subsubsection{Spectrum of $\bar{\cal L}_{1}$ and $\bar{\cal L}_{2}$}
Next let us consider the spectra of
$\bar{\cal L}_{1}=\bar{L}^+\bar{L}^-$ and
$\bar{\cal L}_{2}=\bar{L}^-\bar{L}^+$,
the generators of the conserved quantities
(\ref{Lpmeq}).
  Note first that
a classical analogue of the creation-annihilation
operator commutation relation
of a harmonic oscillator reads $[L^+,L^-]=-2\omega K$, (see (\ref{lpmk}).
By using the information on $K$ in the Appendix, we can derive the
spectrum of
$\bar{\cal L}_{1}=\bar{L}^+\bar{L}^-$ and
$\bar{\cal L}_{2}=\bar{L}^-\bar{L}^+$ for specific choices of ${\cal R}$.

Let us explain the method using the simplest examples.
First, $A_r$ with vector weights embedded in ${\bf R}^{r+1}$
(\ref{arvecwei}).   The $K$
matrix has the following form,
\[
K=g({\bf v}_0{\bf v}_0^T-I),
\]
with the highest eigenvalue at ${\bf
v}_0$  (\ref{ArvecK}), (\ref{pFvec}):
\[
K{\bf v}_0=gr{\bf v}_0.
\]
Since $\bar{\cal L}_{1,\,2}$ are simultaneously diagonalisable with $\bar{M}$
(\ref{commlax}), it is natural to assume that
$\{(\bar{L}^+)^m{\bf v}_0\}$ form
the eigenvectors for $\bar{\cal L}_{1,2}$. In fact, we have:
\begin{eqnarray}
\bar{\cal L}_{1}{\bf v}_0&=&0,\quad
\bar{\cal L}_{1}\bar{L}^+{\bf
v}_0=\bar{L}^+\left([\bar{L}^-,\bar{L}^+]+\bar{L}^+
\bar{L}^-\right){\bf v}_0=2\omega
\bar{L}^+K{\bf v}_0=2\omega gr\bar{L}^+{\bf v}_0,
\nonumber\\
\bar{\cal L}_{2}{\bf v}_0&=&\left([\bar{L}^-,\bar{L}^+]+
\bar{L}^+\bar{L}^-\right){\bf
v}_0=2\omega K{\bf v}_0=2\omega gr{\bf v}_0,
\end{eqnarray}
and we arrive at
\begin{eqnarray}
A_r\ ({\bf V}):\quad \mbox{Spec}(\bar{\cal
L}_{1})&=&2g\omega\{0,r,r-1,\ldots,1\},\\
A_r\ ({\bf V}):\quad \mbox{Spec}(\bar{\cal
L}_{2})&=&2g\omega\{r,r-1,\ldots,1,0\}.
\end{eqnarray}
In this case, it is easy to see
$\bar{L}^2+\omega^2\bar{Q}^2$ also has integer eigenvalues.

Next, let us consider $D_r$ with vector weights (\ref{drvecs}), or $B_r$
with the  short roots (\ref{brshorts}). In these cases we have
(\ref{drk}) and (\ref{bcrk}):
\[
  D_r\ ({\bf V}):\ K=g\left({\bf v}_0{\bf v}_0^T-I-{}_SI\right), \qquad
B_r\ (\Delta_S):\ K=g_L\left({\bf v}_0{\bf v}_0^T-I-{}_SI\right)
+2g_S\,{}_SI,
\]
in which ${}_SI$ is the second identity matrix. It is 1 for the
elements
$({\bf e}_j,-{\bf e}_j)$, $(-{\bf e}_j,{\bf e}_j)$, $j=1,\ldots,r$
and zero otherwise.
The $L^\pm$ satisfy simple commutation relation with ${}_SI$:
and
\begin{equation}
{}_SI(L^\pm)^m=(-1)^m(L^\pm)^m{}_SI.
\end{equation}
We have (in units of $2g\omega$ for the simply laced root systems):
\begin{equation}
\begin{tabular}{|c|c|c|l|}
\hline
$\Delta$&${\cal R}$&$D$&\vT \hspace*{28mm}Spec($\bar{\cal
L}_{1}$)\\
\hline
$A_r$&{\bf V}&$r+1$&\vTm 0, \, $r$, \, $r-1$, \ldots, \,2, \,1\\
\hline
$D_r$&{\bf V}&$2r$&\vTm 0[2], \, $2(r-1)$[2],  \ldots, \, 2[2]\\
\hline
$B_r$&$\Delta_S$&$2r$&\vTm 0,
$2(r-1)g_L+2g_S$, \, $2(r-1)g_L$, \, $2(r-2)g_L+2g_S$,  \\ &&&\vTm \,
$2(r-2)g_L$, \,\ldots, \,$2g_L+2g_S$,
\,$2g_L$, \,
$2g_S$\\
\hline
\end{tabular}
\end{equation}
In these cases, the spectrum of $\bar{L}^2+\omega^2\bar{Q}^2$ also consists of
integer eigenvalues.

\bigskip
It is interesting to note for other cases the spectrum of $\bar{\cal
L}_{1}$ does not always consist of  integers. For
example, the spinor weights of
$D_r$, the set of roots for
$A_r$,
$D_r$, etc and for
the exceptional $E_r$ and non-crystallographic root systems
$H_r$.
Here we list only the integer eigenvalues of $\bar{\cal L}_{1}$
in units of
$2g\omega$ (the total number of integer eigenvalues including multiplicity
is denoted by
$\# I$):

\begin{equation}
\begin{tabular}{|c|c|c|c|l|}
\hline
$\Delta$&${\cal R}$&$D$&$\# I$&\vT \hspace*{24mm}Spec($\bar{\cal
L}_{1}$)\\
\hline
$A_3$&$\Delta$&12&10&\vTm 0[3], \,2,  \,4[2], \,6[3], \,8 \\
\hline
$A_4$&$\Delta$&20&14&\vTm 0[4], \,4[3], \,5, \,6[3],  \,9,  \,10[2] \\
\hline
$A_5$&$\Delta$&30&20&\vTm 0[5],  \,4[3], \,6[6],  \,8[2], \,10[2],
  \,12[2] \\
\hline
$A_6$&$\Delta$&42&30&\vTm 0[6],  \,3, \,4[5],  \,6[5], \,7,  \,8[3],
\,10[4],  \,14[3], \,15[2]\\
\hline
\end{tabular}
\end{equation}
\begin{equation}
\begin{tabular}{|c|c|c|c|l|}
\hline
$\Delta$&${\cal R}$&$D$&$\# I$&\vT \hspace*{28mm}Spec($\bar{\cal
L}_{1}$)\\
\hline
$D_4$&{\bf S}&8&8&\vTm 0[2], \,2[2], \,4[2], \,6[2]\\
\hline
$D_5$&{\bf S}&16&8&\vTm 0[2], \,2,  \,4,   \,6,  \,10[2], \,12\\
\hline
$D_6$&{\bf S}&32&16&\vTm 0[3], \,1[2],  \,3,   \,4[2],  \,5,   \,7,
\,8,  \,11,  \,12,   \,15,   \,19,  \,20\\
\hline
$D_7$&{\bf S}&64&14&\vTm 0[5],  \,2,  \,3,   \,5,  \,6,
       \,9,   \,15,  \,21,   \,30,  \,31\\
\hline
$D_4$&$\Delta$&24&24&\vTm
0[4], \,2[3], \,4[4], \,6[4], \,8[3], \,10, \,12[4], \,16\\
\hline
$D_5$&$\Delta$&40&26&\vTm0[5], \,2[2],  \,4[3],
  \,6[6], \,8[2], \,10[3],  \,12,   \,14,  \,16[2],  \,18\\
\hline
$D_6$&$\Delta$&60&38&\vTm 0[6],  \,2,   \,4[4],  \,6[7], \,8[2], \,10[7],
\,12[3], \,14[2],
  \,16[2],\\
&&&&\vTm \ 20[2], \,24[2]\\
\hline
$D_7$&$\Delta$&84&49&\vTm 0[7],  \,2, \,4[5], \,6[8], \,8[3], \,10[7],
\,12[3],  \,14[4],  \,16[2], \\
&&&&\vTm \ 18[2], \,20[3],  \,24[3],  \,30\\
\hline
\end{tabular}
\end{equation}
The results for the exceptional root systems are in  units of
  $2\omega$ for $F_4$:
\begin{equation}
\begin{tabular}{|c|c|c|c|l|}
\hline
$\Delta$&${\cal R}$&$D$&$\# I$&\vT \hspace*{22mm}Spec($\bar{\cal
L}_{1}$)\\
\hline
$F_4$&$\Delta_L$&24&12&\vTm 0[2],  \,$6g_L$,
  \,$2(g_L+2g_S)$[2],  \,$4(2g_L+g_S)$,
\,$4(g_L+2g_S)$,\\
&&&&\vTm \ $2(5g_L+4g_S)$, \,$8(2g_L+g_S)$, \,$12(g_L+g_S)$[3]\\
\hline
$F_4$&$\Delta_S$&24&12&\vTm 0[2],   \,$2g_L+g_S$[2], \,$3g_S$,
  \,$2(2g_L+g_S)$, \,$2(g_L+2g_S)$,\\
&&&&\vTm \  $2g_L+5g_S$, \,$6(g_L+g_S)$[3], \,$4(g_L+2g_S)$\\
\hline
\end{tabular}
\end{equation}
For the simply laced $E_r$ in  units of $2g\omega$:
\begin{equation}
\begin{tabular}{|c|c|c|c|l|}
\hline
$\Delta$&${\cal R}$&$D$&$\# I$&\vT \hspace*{22mm}Spec($\bar{\cal
L}_{1}$)\\
\hline
$E_6$&{\bf 27}&27&15&\vTm 0[3],  \,2[3],  \,4,   \,6,  \,8,
\,10,  \,16[3], \,18,  \,20\\
\hline
$E_7$&{\bf 56}&56&23&\vTm 0[3], \,1[2], \,3,   \,4[2],  \,5,
       \,7,   \,8[2],  \,9,   \,11,  \,12, \\
&&&&\vTm    \ 15,   \,16,   \,18,
  \,20,   \,27,   \,32,  \,35,  \,36 \\
\hline
$E_6$&$\Delta$&72&29&\vTm 0[6],  \,6[9],  \,12[8],
  \,18[2], \,24,  \,30[2],  \,36\\
\hline
$E_7$&$\Delta$&126&31&\vTm 0[7],  \,6[8],  \,8,  \,10,
  \,12[3],  \,14[2],  \,16[2],\\
&&&&\vTm   \,18[2],   \,24,
      \,36,  \,48,   \,50,  \,56\\
\hline
$E_8$&$\Delta$&240&55&\vTm 0[8], \,6[11], \,12[6], \,18[3],
\,24[5], \,30[9],\\
&&&&\vTm \ 36[4],  \,54,   \,60[2], \,84[3], \,90[2], \,96\\
\hline
\end{tabular}
\end{equation}
The results for the non-crystallographic  root systems are:
\begin{equation}
\begin{tabular}{|c|c|c|c|l|}
\hline
$\Delta$&${\cal R}$&$D$&$\# I$&\vT \hspace*{24mm}Spec($\bar{\cal
L}_{1}$)\\
\hline
$H_3$&$\Delta$&30&15&\vTm 0[3],  \,2,  \,3,   \,5[6],  \,8[2], \,10[2]\\
\hline
$H_4$&$\Delta$&120&48&\vTm 0[4], \,5[4], \,10[7],
\,15[18], \,20[3], \,25[2], \,30[10]\\
\hline
\end{tabular}
\end{equation}
All the eigenvalues are ``integers" for $I_2(m)$  in the  $m$ dimensional
representation (\ref{rmvert}):
\begin{equation}
\begin{tabular}{|c|c|c|c|l|}
\hline
$\Delta$&${\cal R}$&$D$&$\# I$&\vT \hspace*{25mm}Spec($\bar{\cal
L}_{1}$)\\
\hline
$I_2(2n+1)$&$R_{2n+1}$&$2n+1$&$2n+1$&\vT 0, \,$4n+2$[$2n-1$], \,$8n+4$\\
\hline
$I_2(2n)$&$R_{2n}$&$2n$&$2n$&\vTm 0, \,$8g_en$, \,$8g_on$,
\,$4(g_e+g_0)n$[$2n-4$],
\,$8(g_e+g_0)n$\\
\hline
\end{tabular}
\end{equation}

\section{Classical Data II: Trigonometric Potential}
\label{cdata2}
\setcounter{equation}{0}

\subsection{Minimum Energy}
\label{trigminener}
Let us start this subsection by recalling that the  classical minimum energy
$2\varrho^2$ (\ref{minenergy})  is, in fact, ``quantised".
In this section we discuss only the crystallographic root system
$\Delta$ to which a Lie algebra ${\mathfrak g}_\Delta$ is associated.
If all the coupling constants
are unity $g_{\rho}=1$, $\varrho=\delta$, and  the
Freudenthal-de Vries (``strange") formula
leads to
\begin{equation}
2\varrho^2={\mbox{dim}({\mathfrak g}_\Delta)\rho^2_h\,h^\vee\over12},
\end{equation}
in which dim$({\mathfrak g}_\Delta)$ is the dimension of the Lie
algebra ${\mathfrak g}_\Delta$, $\rho_h$ is the highest root and
$h^\vee$ is the dual Coxeter number.
This gives the classical minimum energy formula for the
simply laced root systems
(in the unit of $g^2$ and with $\alpha^2=2$):
\begin{equation}
\begin{tabular}{||c|c||c|c||c|c||c|c||c|c||}
\hline
$\Delta$&\vT ${\cal E}_0$&$\Delta$& ${\cal E}_0$&$\Delta$& ${\cal E}_0$
&$\Delta$& ${\cal E}_0$&$\Delta$& ${\cal E}_0$\\
\hline
\vTm $A_r$&$r(r+1)(r+2)/6$&$D_r$&$r(r-1)(2r-1)/3$&$E_6$&156&$E_7$&399
&$E_8$&1240\\
\hline
\end{tabular}
\label{suthenergSimp}
\end{equation}
For the non-simply laced root systems,
the classical minimum energy formula is given by:
\begin{equation}
\begin{tabular}{|c|c|}
\hline
$\Delta$&\vT ${\cal E}_0$\\
\hline
\vTm $B_r$&$r\left(\vTm 2g_L^2 + 4r^2g_L^2 - 6g_Lg_S + 3g_S^2 +
     r (-6\ g_L^2 + 6g_Lg_S)\right)/6$\\
\hline
\vTm $C_r$&$ r(g_S^2 - 6g_Sg_L + 6g_L^2 - 3g_S^2r +
         6g_Sg_Lr + 2g_S^2r^2)/3$\\
\hline
\vTm $F_4$& $28g_L^2 + 36g_Lg_S + 14g_S^2$\\
\hline
\vTm $G_2$& $4g_L^2 + 4g_Lg_S + 4g_S^2/3$\\
\hline
\end{tabular}
\end{equation}
in which long roots have $\rho_L^2=2$, except for the $C_r$ case where
a different normalisation $\rho_L^2=4$ is chosen.

By taking the trace of $\widetilde{W}$ (\ref{W2der}), we obtain
\begin{equation}
\mbox{Tr}(\widetilde{W})=-\sum_{\rho\in\Delta_+}
    {g_{\rho} \rho^2\over{\sin^2(\rho\!\cdot\! \bar{q})}}.
\end{equation}
For the simply laced root systems, this is related to $V_C(\bar{q})$
(\ref{CHam}) and thus
to
${\cal E}_0$ (\ref{cpot}):
\begin{equation}
\mbox{Tr}(\widetilde{W})=-2V_C(\bar{q})/g=-2{\cal E}_0/g=-4{\varrho^2}/g,
\qquad \Delta:\ \mbox{simply laced}.
\label{trigtrW}
\end{equation}
As in the Calogero systems (\ref{caltraceform}), Tr($\bar{M}$) is related to
Tr($\widetilde{W}$). By taking the trace of $\bar{M}$, we obtain
\begin{equation}
\mbox{Tr}(\bar{M})={i\over2}\sum_{\rho\in\Delta_+}
g_{\rho}F_{\rho}^{\cal R}{\rho^2
\over{\sin^2(\rho\!\cdot\!\bar{q})^2}},
\end{equation}
on recalling the earlier definition of $F_{\rho}^{\cal R}$ (\ref{FRdef}).
This formula simplifies for the simply
laced root systems to:
\begin{equation}
\mbox{Tr}(\bar{M})=-{i\over2}F^{\cal R}\mbox{Tr}(\widetilde{W})
=2iF^{\cal R}{\varrho^2}/g,
\qquad \Delta:\ \mbox{simply laced}.
\label{trigMWrel}
\end{equation}
As in the calogero case, this formula provides a non-trivial check
for the numerical evaluation of the eigenvalues of $\bar{M}$.
Since the Lax matrix $\bar{L}$ is off-diagonal,
$(\bar{L})_{\mu\,\mu}=0$ and we have a trivial
trace formula:
\begin{equation}
\mbox{Tr}(\bar{L})=0.
\end{equation}

\subsection{Determination of the Equilibrium Point and Eigenvalues  of
$\widetilde{W}$}
\label{trigdetequi}
Since the quantum energy levels of the Sutherland systems
are not integers (time a
constant)  spaced but  (\ref{energyformula})
\[
{\cal
E}_{\lambda_{\vec{n}}}=
2(\lambda_{\vec{n}}^2+
\varrho^2+2\sum_{j=1}^rn_j\lambda_j\!\cdot\!\varrho),
\]
it is not obvious what to expect for the eigenfrequencies of
the small oscillations near
the equilibrium point.
In other words, what are the corresponding spectra of
$\tilde{V}$ or equivalently of
$\widetilde{W}$? An educated guess would be that,
just as in the rational
potential situation,
we assume  the parts of the spectra which are {\em linear\/}  in the integer
labels $\vec{n}$ correspond to the eigenfrequencies of
the small oscillations near the
equilibrium point. That is, we expect
\begin{equation}
\mbox{Spec}(\tilde{V})=\{(4\lambda_1\!\cdot\!\varrho)^2,
\ldots,(4\lambda_r\!\cdot\!\varrho)^2\}
\end{equation}
and
\begin{equation}
\mbox{Spec}(\tilde{W})=-\{4\lambda_1\!\cdot\!\varrho,
\ldots,4\lambda_r\!\cdot\!\varrho\},
\label{simweight}
\end{equation}
which we will show presently. For the simply laced root systems,
we have a simple relation
\begin{equation}
\varrho={g\over2}\sum_{\rho\in{\Delta_+}}\rho=g\sum_{j=1}^r\lambda_j,
\end{equation}
which implies a simple sum rule
\begin{equation}
\mbox{Tr}(\tilde{W})=-{4\varrho^2\over g}=-{2{\cal E}_0\over g},
\qquad \Delta:\
\mbox{simply laced},
\label{trWsumrule}
\end{equation}
which has been derived before (\ref{trigtrW}) via a different route.
The equations
determining the equilibrium position (\ref{Wmax}) read
\[
\sum_{\rho\in\Delta_+}g_{\rho}\cot(\rho\!\cdot\!\bar{q})\,
\rho_j=0,\quad j=1,\ldots,r,
\]
and can be expressed in terms of the $L$ matrix at
equilibrium:
\begin{equation}
\bar{L}{\bf v}_0=0={\bf v}_0^T\bar{L}.
\label{lzerovec}
\end{equation}
The ``ground state" ${\bf v}_0$ (\ref{v0def})
is also annihilated by $\bar{M}$;
\[
\bar{M}{\bf v}_0=0={\bf v}_0^T\bar{M}.
\]
These relations are valid for any ${\cal R}$.
As in the Calogero  case,
the equilibrium positions
$\bar{q}=(\bar{q}_1,\ldots,\bar{q}_r)$ can be easily
identified for the classical
root systems. For the exceptional root systems
the equilibrium positions are determined
numerically. We
shall discuss each case in turn.

\subsubsection{$A_r$}
\label{artrigWspec}

In this case, the equilibrium position and the
eigenvalues of the Lax matrices can be
obtained explicitly. This is the reason
why the Haldane-Shastry model
is better understood than other spin exchange models.
The equations determining the equilibrium position (\ref{Vmin})
and (\ref{Wmax}) read:
\[
\sum_{k\neq j}^{r+1}{\cos\,[\bar{q}_j-\bar{q}_k]
\over{\sin^3[\bar{q}_j-\bar{q}_k]}}=0,
\qquad \sum_{k\neq j}^{r+1}\cot[\bar{q}_j-\bar{q}_k]=0,
\quad j=1,\ldots,r+1,
\]
and
the equilibrium position is
``{\em equally-spaced\/}"
\begin{equation}
\bar{q}=\pi(0,1,\ldots,r-1,r)/(r+1)+\xi {\bf v}_0,\qquad \xi\in{\bf R}:
\mbox{arbitrary},
\label{eqspaced}
\end{equation}
due to the well-known trigonometric identities:
\[
  \sum_{k\neq
j}^{r+1}{\cos\,[\pi(j-k)/(r+1)]\over{\sin^3[\pi(j-k)/(r+1)]}}=0,\qquad
\sum_{k\neq j}^{r+1}\cot[\pi(j-k)/(r+1)]=0,
\quad j=1,\ldots,r+1.
\]
This enables us to calculate most quantities exactly. For example, we have
\begin{equation}
\widetilde{W}_{j\,k}=g{(1-\delta_{j\,k})\over{\sin^2[(j-k)\pi/(r+1)]}}-
g\delta_{j\,k}\sum_{l\ne j}{1\over{\sin^2[(j-l)\pi/(r+1)]}},
\quad j,k=1,\ldots,r+1
\end{equation}
and
\begin{equation}
A_r:\quad \mbox{Spec}(\widetilde{W})=-2g\left\{ r, (r-1)2,
\ldots,(r+1-j)j, \ldots, 2(r-1), r\right\},
\label{specartrigW}
\end{equation}
in which the trivial eigenvalue 0, coming from the
translational invariance, is removed.
This agrees with the general formula (\ref{simweight})
of the $\widetilde{W}$
spectrum ({\em i.e.\/} the $j$-th entry is $4\lambda_j\cdot\varrho$,
and
obviously satisfies the above sum rule (\ref{suthenergSimp}),
(\ref{trWsumrule})).
The spectrum (\ref{specartrigW}) is symmetric with respect
to the middle point,
$\lambda_j\leftrightarrow \lambda_{r+1-j}$,
reflecting the symmetry of the $A_r$ Dynkin diagram.
It is easy to see that
$\widetilde{W}$ is essentially the same as the Lax matrix
$\bar{M}$ with the vector weights
(${\cal R}={\bf V}$, see (\ref{arvecwei})):
\begin{equation}
\bar{M}=-i\widetilde{W}.
\label{WeqMar}
\end{equation}
(This is consistent with (\ref{trigMWrel}), since $F^{\bf V}=2$, see
(\ref{minF})).

\paragraph{$A_r$ Universal Lax pair (${\bf V}$)}

The other Lax matrix with the vector weights reads ($j,k=1,\ldots,r+1$):
\begin{equation}
(\bar{L})_{j\,k}=ig(1-\delta_{j\,k})\cot[\pi(j-k)/(r+1)],
\end{equation}
\begin{equation}
A_r\ ({\bf V}):\quad \mbox{Spec}(\bar{L})=g\left\{
\begin{array}{ll}
0[2], \,\pm2, \,\pm4, \ldots, \,\pm(r-1)& r:\ \mbox{odd}\\
0, \quad \, \pm1, \,\pm3, \ldots, \,\pm(r-1)& r:\ \mbox{even}
\end{array}
\right\},
\label{specartrigL}
\end{equation}
with the common  eigenvectors ($h=r+1$):
\begin{equation}
u^{(a)}, \quad (u^{(a)})_j=e^{2iaj\pi/h},
\quad a=0,1,\ldots,r,\quad u^{(0)}\equiv {\bf v}_0,
\end{equation}
satisfying
\begin{eqnarray}
\bar{L}u^{(a)}&=& g \lambda_a u^{(a)}, \qquad \lambda_a=\left\{
\begin{array}{ll}
0,&a=0,\\
r+1-2a,&a\ne 0,\\
\end{array}
\right.\\
\bar{M}u^{(a)}&=& ig \mu_a u^{(a)}, \quad \ \ \mu_a=2a(r+1-a).
\end{eqnarray}
These are well-known results \cite{OP1,calpere}.

\paragraph{$A_r$ Minimal type Lax pair (${\bf V}$)}

The minimal Lax pair matrices in the vector weights read
($j,k=1,\ldots,r+1$):
\begin{equation}
(\bar{L}_m)_{j\,k}=ig(1-\delta_{j\,k})/\sin[\pi(j-k)/(r+1)],
\end{equation}
\begin{equation}
(\bar{M}_m)_{j\,k}=ig{(1-\delta_{j\,k})\over{\sin^2[(j-k)\pi/(r+1)]}}-
ig\delta_{j\,k}\sum_{l\ne j}{\cos[(j-l)\pi/(r+1)]\over
{\sin^2[(j-l)\pi/(r+1)]}}.
\end{equation}
They  have common eigenvectors with integer eigenvalues ($h=r+1$):
\begin{equation}
v^{(a,\,\pm)}, \quad (v^{(a,\pm)})_j=e^{\pm iaj\pi/h},
\quad a=1, 3, 5, \ldots,\le h,
\end{equation}
\begin{eqnarray}
\bar{L}_mv^{(a,\,\pm)}&=&\pm g(h-a) v^{(a,\,\pm)},
\label{arLmesoec}\\
\bar{M}_mv^{(a,\,\pm)}&=& ig\left(ah- (a^2+1)/2\right) v^{(a,\,\pm)}.
\end{eqnarray}
The above spectrum of $\bar{M}_m$ can be derived easily from
the following relation between $\bar{L}_m$ and $\bar{M}_m$
  (see eq.(5.8) of \cite{cfs})
\begin{equation}
R^{1/2}\bar{M}_mR^{-1/2}-R^{-1/2}\bar{M}_mR^{1/2}=-i\left(
R^{1/2}\bar{L}_mR^{-1/2}+R^{-1/2}\bar{L}_mR^{1/2}\right),
\end{equation}
in which $\quad R\equiv e^{2i\bar{Q}}$.
We note $R^{\pm a/2}{\bf v}_0=v^{(a,\,\pm)}$ and use the spectrum of
$\bar{L}_m$.
The above relationship is a special case of the general formulae which are
valid in any root systems having minimal weights:
\begin{eqnarray}
R^{-1/2}\bar{L}_mR^{1/2}&=&\bar{L}
+K,\quad R^{1/2}\bar{L}_mR^{-1/2}=\bar{L}-K,
\label{lm-lequiv}\\
R^{1/2}\bar{M}_mR^{-1/2}&=&\bar{M}-iR^{1/2}\bar{L}_mR^{-1/2},\\
R^{-1/2}\bar{M}_mR^{1/2}&=&\bar{M}+iR^{-1/2}\bar{L}_mR^{1/2},
\end{eqnarray}
in which the constant matrix $K$ is defined in (\ref{QLcomm}).
These  mean, for example, that the spectrum of
$\bar{L}_m$ and $\bar{L}\pm K$ are the same
and those of $\bar{M}$ and $\bar{M}_m\pm i\bar{L}_m$ are the same.
We will see many examples later.

\paragraph{$A_r$ Root type Lax pair}
The $\bar{L}$-matrices of the $A_r$ root type Lax pair
do not have integer eigenvalues, although the quantities $\bar{L}^2$ do.
Let us tentatively say that $\bar{L}$ has  $\sqrt{\mbox{integer}}$
eigenvalues. (Recall that Tr($L^2$) is proportional to the
Hamiltonian.) However, a new type of
$L$-matrix having all integer eigenvalues can be defined by
\begin{equation}
\bar{L}_K=\bar{L}+ \tilde{K},
\qquad
\tilde{K}=
\sum_{\rho\in\Delta_{+}}g_{\rho}|\rho\cdot\hat{H}|
     \hat{s}_{\rho},\qquad [\tilde{K},\bar{M}]=0,
     \label{Ktilde}
\end{equation}
in which $\tilde{K}$ is a non-negative matrix closely related to the $K$-matrix
defined by  (\ref{QLcomm}).
The absolute value in the definition of $\tilde{K}$ means
$\tilde{K}_{\mu\,\nu}=
\sum_{\rho\in\Delta_{+}}g_{\rho}|\rho\cdot\mu|
     (\hat{s}_{\rho})_{\mu\,\nu}$, $\mu,\,\nu\in{\cal R}$.
This type of Lax matrix has been obtained
(see \S8.3 eq.(8.22) in \cite{kps}) by
incorporating a spectral parameter ($\xi$) into the
Lax pair and taking a limit (say,
$\xi\to-i\infty$). For ${\cal R}=\Lambda$ \{set of minimal weights\}, we have
$\tilde{K}\equiv K$
and $\bar{L}_K$ has the same spectrum as the minimal type
$\bar{L}_m$ due to the relation
(\ref{lm-lequiv}).
The spectra of $\bar{L}_K$ are very simple, whereas those of
$\bar{M}$ of the root type are
sums of those of $\widetilde{W}$, {\em i.e.\/}
  $4\lambda_j\!\cdot\!\varrho$, with varied
multiplicities:
\begin{equation}
A_r\ (\Delta):\quad \mbox{Spec}(\bar{L}_K)=g\left\{
  \,\pm\,2[r], \,\pm\,4[r-1], \ldots, \,\pm\,2(r-1)[2], \,\pm\, r
\right\},
\label{specartrigLK}
\end{equation}
\begin{equation}
\begin{tabular}{|c|c|c|l|}
\hline
$\Delta$&${\cal R}$&$D$&\vT \hspace*{15mm}Spec($\bar{M}$)\\
\hline
$A_3$&$\Delta$&12&\vTm 0, \,6[4], \,8[3], \,12[2],
\,14[2]\\
\hline
$A_4$&$\Delta$&20&\vTm 0, \,8[4], \,12[6], \,16[2],
\,20[6], \,24\\
\hline
$A_5$&$\Delta$& 30&\vTm 0, \,10[4], \,16[6],
\,18[3], \,20[2], \,26[6], \,28[4], \,32[2], \,34[2]\\
\hline
$A_6$& $\Delta$&42&\vTm 0,  \,12[4], \,20[6], \,24[8], \,32[6],
\,36[8], \,40[2], \,44[6],
\,48 \\
\hline
$A_7$&$\Delta$&56&\vTm 0,  \,14[4], \,24[6], \,28[2], \,30[6],
\,32[3], \,38[6],
\,44[8], \,46[4],\\
&&&\vTm  \,\quad \,48[2], \,54[6], \,56[4], \,60[2], \,62[2]\\
\hline
$A_8$&$\Delta$&72&\vTm 0, \,16[4], \,28[6], \,32[2], \,36[6],
  \,40[6], \,44[6],
\,52[8], \,56[10], \\
&&&\vTm  \,\quad \,64[6], \,68[8], \,72[2], \,76[6], \,80 \\
\hline
\end{tabular}
\label{arrootMei}
\end{equation}
The eigenvalues of
$\bar{M}$ are of the form $i\sum_{j=1}^r a_j(4\varrho\cdot\lambda_j)$, in which
$a_j=0,1$.
The relation between Tr($\widetilde{W}$) and Tr($\bar{M}$),
(\ref{trigMWrel}) is satisfied, since
\(
F^{\Delta}(A_r)=2(2r-1)
\)---see (\ref{fDelta}).
\subsubsection{$BC_r$ and $D_r$}
\label{bcrtrigWspec}
The analytical treatment of the classical
equilibrium position of the $BC_r$ and $D_r$
Sutherland system has not been reported, to the best of
our knowledge,
except for the aforementioned three cases when the coupling ratio
$g_S/g_L$ takes  special values \cite{is1,ber}.
We will show in this subsection, that the equilibrium position
is given in terms of
the zeros of Jacobi polynomials.
The Jacobi polynomials $P_r^{(\alpha,\beta)}$ are
known to reduce to elementary
trigonometric polynomials,  Chebyshev polynomials, etc. for three cases:
\begin{equation}
(i)\ \alpha=\beta=-1/2,\qquad (ii)\ \alpha=\beta=1/2,
\qquad (iii)\ \alpha=1/2, \quad
\beta=-1/2,
\label{threeeq}
\end{equation}
  which will be identified later with the three cases discussed in
  \cite{is1,ber}.

Let us start from the pre-potential of the $BC_r$ Sutherland system
\begin{eqnarray}
W&=&g_M\sum_{j<k}^r\log \left[\vTm \sin(q_j-q_k)\,\sin(q_j+q_k)\right]
+\sum_{j=1}^r\left\{g_S\log\sin q_j+g_L\log\sin 2q_j\right\}\\
&=&g_M\sum_{j<k}^r\log\left[\vTm (-1/2)(\cos 2q_j-\cos2q_k)\right]
+\sum_{j=1}^r\left\{g_S\log\sin q_j+g_L\log\sin 2q_j\right\}
\end{eqnarray}
which depends on
three independent coupling constants, $g_L$, $g_M$
and $g_S$, for the long, middle
and short roots, respectively. Here we have adopted the following
representation of the $BC_r$ roots
in terms of an orthonormal basis of ${\bf R}^r$:
\begin{equation}
BC_r=\{\pm{\bf e}_j \pm{\bf e}_k,\quad \pm {\bf e}_j,\quad
\pm 2{\bf e}_j,\quad
j,k=1,\ldots,r|{\bf e}_j\in{\bf R}^{r}, {\bf e}_j\cdot
{\bf e}_k=\delta_{jk}\}.
\end{equation}
We look for the solution $\{\bar{q}_j\}$ of (\ref{Wmax})
\[
{\partial W\over{\partial q_j}}=0,\quad j=1,\ldots,r,
\]
which read
\begin{equation}
-2g_M\sum_{k\ne j}^r{\sin2\bar{q}_j\over
{\cos2\bar{q}_j-\cos2\bar{q}_k}}+g_S{\cos
\bar{q}_j\over{\sin \bar{q}_j}}
+2g_L{\cos2\bar{q}_j\over{\sin2\bar{q}_j}}=0, \quad
j=1,\ldots,r.
\end{equation}
For non-vanishing $g_S$ and $g_L$, $\sin2\bar{q}_j=0$
cannot satisfy the above equation.
Thus by dividing by $\sin2\bar{q}_j$ we obtain
\begin{equation}
\sum_{k\ne j}^r{1\over{\bar{x}_j-\bar{x}_k}}+
{g_S+g_L\over{2g_M}}{1\over{\bar{x}_j-1}}
+{g_L\over2g_M}{1\over{\bar{x}_j+1}}=0, \quad j=1,\ldots,r,
\end{equation}
in which
\begin{equation}
\bar{x}_j\equiv \cos2\bar{q}_j.
\label{defxbar}
\end{equation}
These are the equations satisfied by the zeros $\{\bar{x}_j\}$ of Jacobi
polynomial
$P_r^{(\alpha,\beta)}(x)$  \cite{szego} with
\begin{equation}
\alpha=(g_L+g_S)/g_M-1, \quad \beta=g_L/g_M-1.
\end{equation}
The  solution (the equilibrium position)
  is shown to be unique.

Next let us consider the $D_r$ case, the pre-potential is simply
\begin{equation}
W=g\sum_{j<k}^r\log\left[\vTm (-1/2)(\cos 2q_j-\cos2q_k)\right]
\end{equation}
and the equations for its equilibrium point read:
\begin{equation}
\sin2\bar{q}_j\sum_{k\ne j}^r{1\over{\cos2\bar{q}_j-\cos2\bar{q}_k}}=0,
\quad j=1,\ldots,r.
\end{equation}
These can be decomposed into two parts:
\begin{equation}
\sin2\bar{q}_1=0=\sin2\bar{q}_r \ \Longleftrightarrow \
\cos2\bar{q}_1=1,\ \cos2\bar{q}_r=-1
\end{equation}
and
\begin{equation}
\sum_{k=2, \ne
j}^{r-1}{1\over{\bar{x}_j-\bar{x}_k}}+{1\over{\bar{x}_j-1}}
+{1\over{\bar{x}_j+1}}=0, \quad j=2,\ldots,r-1,
\label{drroots}
\end{equation}
in which $\{\bar{x}_j\}$, $j=2,\ldots,r-1$ are defined as before
(\ref{defxbar}).
The latter part (\ref{drroots}) are  the equations that
the zeros $\{\bar{x}_j\}$ of the Jacobi
polynomial
$P_{r-2}^{(1,1)}(x)$ or the Gegenbauer polynomial
$C_{r-2}^{3/2}(x)$ satisfy.

Note, the problem of finding the maximal point of
the $D_r$ pre-potential $W$
is the same as the classical problem of maximising the
van der Monde determinant
\begin{equation}
VdM(x_1,\ldots, x_r)=\prod_{j<k}^r(x_j-x_k),
\end{equation}
under the boundary conditions
\begin{equation}
1=x_1>x_2>\cdots>x_{r-1}>x_r=-1.
\end{equation}

Now let us show that the three special cases (\ref{threeeq})
are also characterised by
{\em equally-spaced\/} $\bar{q}_j$'s, that is
$\bar{q}_j-\bar{q}_{j+1}$ is independent of $j$.
\begin{enumerate}
\item[$(i)$]
For $\alpha=\beta=-1/2$ $\Leftrightarrow$ $g_L/g_M=1/2$,
$g_S=0$, which is
a special case of $C_r$  obtained from the Dynkin diagram folding
$A_{2r-1}\to C_r$.   Jacobi polynomial $P_r^{(-1/2,-1/2)}(x)$ is known to
be proportional to  Chebyshev polynomial of the first kind $T_r(x)$, which can
be expressed as
\begin{equation}
T_r(x)=\cos r\varphi,\qquad x=\cos\varphi.
\label{2chebfirst}
\end{equation}
The zeros are {\em equally-spaced\/} in $\varphi$:
\begin{equation}
\bar{\varphi}_j={(2j-1)\pi\over{2r}} \Leftrightarrow
\cos2\bar{q}_j=\cos{(2j-1)\pi\over{2r}}
\Leftrightarrow \bar{q}_j={(2j-1)\pi\over{4r}},\quad j=1,\ldots,r.
\label{crequi}
\end{equation}
The Dynkin diagram folding $A_{2r-1}\to C_r$ explains this situation neatly.
By imposing the following restrictions on the dynamical variables
\begin{equation}
q_j=-q_{2r+1-j},\quad j=1,\ldots, r,
\label{arcrred}
\end{equation}
in the pre-potential of $A_{2r-1}$ Sutherland system,
\[
W_{A_{2r-1}}=g\sum_{j<k}^{2r}\log\sin(q_j-q_k),
\]
it reduces to to that of $C_r$ with the coupling relation $g_L/g_M=1/2$:
\begin{equation}
W_{A_{2r-1}\to C_r}=2g\sum_{j<k}^r\log \left[\vTm
\sin(q_j-q_k)\,\sin(q_j+q_k)\right]
+g\sum_{j=1}^r\log\sin 2q_j.
\end{equation}
The equilibrium point of the above $A_{2r-1}$
pre-potential is given in general by
\begin{equation}
\bar{q}_j={j\pi\over{2r}}+\xi,\quad j=1,\ldots, 2r,
\label{xieq}
\end{equation}
in which $\xi$ is an arbitrary real constant,
reflecting the ``translational invariance"
of the pre-potential. By imposing the  restrictions
(\ref{arcrred}) on the above
equilibrium point,
we find $\xi=-(2r+1)\pi/4r$, which turns the general
$A_{2r-1}$ equilibrium point
(\ref{xieq}) to that of $C_r$ (\ref{crequi}).
It is interesting to note that the
above equilibrium point (\ref{crequi}) is given by the deformed Weyl vector
(\ref{weylvec})  with
$g_M=\pi/2r$, $g_L=\pi/4r$ and a choice of positive roots
$\{{\bf e}_j-{\bf e}_k\}$ for $j<k$.
\item[$(ii)$]
For $\alpha=\beta=1/2$ $\Leftrightarrow$
$g_L/g_M=3/2$, $g_S=0$, which is
also a special case of $C_r$. Jacobi polynomial
$P_r^{(1/2,1/2)}(x)$ is known
to be proportional to  Chebyshev polynomial  of the second kind
$U_r(x)$ which has a simple
expression  as a trigonometric polynomial:
\begin{equation}
U_r(x)={\sin(r+1)\varphi\over{\sin\varphi}},
\qquad x=\cos\varphi.
\end{equation}
The zeros are {\em equally-spaced\/} in $\varphi$:
\begin{equation}
\bar{\varphi}_j={j\pi\over{r+1}} \Leftrightarrow
\cos2\bar{q}_j=\cos{j\pi\over{r+1}}
\Leftrightarrow \bar{q}_j={j\pi\over{2(r+1)}},\quad j=1,\ldots,r.
\end{equation}
\item[$(iii)$]
For $\alpha=1/2, \beta=-1/2$ $\Leftrightarrow$
$g_L/g_M=1/2$, $g_S/g_M=1$.
In this case we have
\[
P_r^{(1/2,-1/2)}(x)\propto
{\sin[(2r+1)\varphi/2]\over{\sin[\varphi/2]}},
\qquad x=\cos\varphi.
\]
The zeros are {\em equally-spaced\/} in $\varphi$:
\begin{equation}
\bar{\varphi}_j={2j\pi\over{2r+1}} \Leftrightarrow
\cos2\bar{q}_j=\cos{2j\pi\over{2r+1}}
\Leftrightarrow \bar{q}_j={j\pi\over{2r+1}},\quad j=1,\ldots,r.
\end{equation}
This equilibrium point is also obtained as a deformed
Weyl vector (\ref{weylvec}) for
$g_M=\pi/(2r+1)$.
For $\alpha=-1/2$, $\beta=1/2$, Jacobi polynomial
$P_r^{(\alpha,\beta)}$ is proportional to
  another trigonometric polynomial. But this case is not compatible
with {\em positive\/} coupling constants $g_{\rho}$
and will not be discussed here.
\end{enumerate}

In this connection, let us remark on the dynamical
implications of another well-known
Dynkin diagram folding, $D_{r+1}\to B_r$. By restricting one
of the dynamical
variables  of $D_{r+1}$ Sutherland system to its equilibrium position
\begin{equation}
q_{r+1}=0,
\label{drbrred}
\end{equation}
its pre-potential,
\[
W_{D_{r+1}}=g\sum_{j<k}^{r+1}\log \left[\vTm \sin(q_j-q_k)\,
\sin(q_j+q_k)\right],
\]
reduces to  that of $B_r$ with the coupling relation $g_S/g_M=2$:
\begin{equation}
W_{D_{r+1}\to B_r}=g\sum_{j<k}^r\log \left[\vTm \sin(q_j-q_k)\,
\sin(q_j+q_k)\right]
+2g\sum_{j=1}^r\log\sin q_j.
\label{drbrred2}
\end{equation}
This means that the equilibrium position of the reduced $B_r$ system
(\ref{drbrred}),
the zeros of $P_r^{(1,-1)}(x)$, is
given by that of the original $D_{r+1}$ system, {\em i.e.\/} the zeros of
$P_{r-1}^{(1,1)}(x)$ plus $x=-1$. In other words,
the following identities hold:
\begin{equation}
(r+1)(x+1)\,P_{r-1}^{(1,1)}(x)=2r\,P_r^{(1,-1)}(x),\quad r=1, 2, \ldots,
\end{equation}
which are trigonometric counterparts of (\ref{lagiden}).

\bigskip
In the following, we summarise  the spectra of $\widetilde{W}$,
$\bar{L}$, $\bar{M}$,
$\bar{L}_m$, $\bar{M}_m$ of the $D_r$ Sutherland system
which are evaluated numerically, with
the vector weights ${\bf V}$ and the  roots
$\Delta$.
The spectra are all ``integer-valued", except for $\bar{L}$. The
combinations (\ref{Ktilde})
$\bar{L}_{K}=\pm \tilde{K}+\bar{L}$ are integer valued having the same
spectra as
$\bar{L}_m$ for ${\cal R}={\bf V}$,  see (\ref{lm-lequiv}). It is interesting
to note that  $\bar{L}^2$
is integer-valued for ${\cal R}={\bf V}$,
but the eigenvalues are not all integers
for ${\cal R}=\Delta$.
The spectrum of $\widetilde{W}$ is:
\begin{eqnarray}
D_r:\quad \mbox{Spec}(\widetilde{W})&=&-g\left\{\vTm 4(r-1), \,4(2r-3),
\ldots, 2j(2r-1-j), \ldots, \right.\nonumber\\
&&\left. \qquad\qquad  \vTm 2(r-2)(r+1), \,r(r-1)[2]\right\},
\label{specdrtrigW}
\end{eqnarray}
which agrees with the general formula (\ref{simweight})
of the $\widetilde{W}$
spectrum, {\em i.e.,\/} the $j$-th entry is $4\lambda_j\cdot\varrho$, and
obviously satisfies the  sum rule (\ref{suthenergSimp}), (\ref{trWsumrule}).
The two-fold degeneracy reflects the Dynkin diagram symmetry
corresponding to
the spinor and anti-spinor fundamental weights,
$\lambda_{S}\leftrightarrow \lambda_{\bar{S}}$.

\paragraph{$D_r$ Universal Lax pair (${\bf V}$)}

The spectrum of $\bar{M}$ is
\begin{eqnarray}
D_r\ ({\bf V}):\quad \mbox{Spec}(\bar{M})&=&ig\left\{\vTm 0,
\,4(r-1)[2], \,4(2r-3)[2],
\ldots, 2j(2r-1-j)[2], \ldots, \right.\nonumber\\
&&\left. \qquad\qquad  \vTm 2(r-2)(r+1)[2], \,r(r-1)[2], \,2r(r-1)\right\},
\end{eqnarray}
which is essentially the duplication of  that of $\widetilde{W}$,
except for the
lowest, {\em i.e.,\/} 0, and the highest eigenvalues, $2r(r-1)$.
The latter is exactly twice the eigenvalue of those belonging to $\lambda_S$
($\lambda_{\bar{S}}$).
Let us note that the identity between the traces of $\widetilde{W}$
and $\bar{M}$ (\ref{trigMWrel}) is also satisfied,
since $F^{\bf V}(D_r)=4$, see (\ref{minF}). As in the $A_r$
vector weight case (\ref{WeqMar}),
these can be understood by the close relationship
between
$\widetilde{W}$ and
$\bar{M}$:
\begin{equation}
\bar{M}=i\pmatrix{A&B\cr B&A\cr},\quad \widetilde{W}=-A+B.
\end{equation}
The $r\times r$ matrices $A$ and $B$ are
\begin{eqnarray}
A_{j\,j}&=&g\sum_{k\ne j}^r\left(
{1\over{\sin^2(\bar{q}_j-\bar{q}_k)}}+
{1\over{\sin^2(\bar{q}_j+\bar{q}_k)}}\right)
=-\widetilde{W}_{j\,j},
\quad
B_{j\,j}=0,\\
  A_{j\,k}&=&-g{1\over{\sin^2(\bar{q}_j-\bar{q}_k)}},
\qquad\qquad
B_{j\,k}=-g{1\over{\sin^2(\bar{q}_j+\bar{q}_k)}}.
\end{eqnarray}
Thus to each eigenvector $v$ of $\widetilde{W}$ with eigenvalue -$\lambda$,
$\widetilde{W}v=(-A+B)v=-\lambda v$, corresponds an eigenvector
V with eigenvalue
$i\lambda$:
\begin{equation}
V=\pmatrix{v\cr -v\cr},\quad
\bar{M}V=i\pmatrix{A&B\cr B&A\cr}\pmatrix{v\cr -v\cr}=i\lambda V.
\end{equation}
The $\bar{L}$ matrix
with the vector weights has the following decomposition:
\begin{equation}
\bar{L}=\pmatrix{C&D\cr -D&-C\cr},\quad {}_SI\bar{L}=-\bar{L}{}_SI,
\quad {}_SI=\pmatrix{0&I_r\cr I_r&0\cr},
\end{equation}
in which $I_r$ is the $r\times r$ identity matrix and
\begin{equation}
C_{j\,k}=ig(1-\delta_{j\,k})\cot(\bar{q}_j-\bar{q}_k),\quad
D_{j\,k}=ig(1-\delta_{j\,k})\cot(\bar{q}_j+\bar{q}_k), \quad j,k=1,\ldots,r.
\end{equation}
Since $\bar{L}$ commutes with $\bar{M}$, $\bar{L}V$
provides another independent eigenvector
with the same eigenvalue $\bar{M}(\bar{L}V)=i\lambda(\bar{L}V)$,
\[
\bar{L}V=\pmatrix{C&D\cr -D&-C\cr}\pmatrix{v\cr -v}=(C-D)v\pmatrix{1\cr 1},
\]
except for the duplicated
eigenvalue
$r(r-1)$ and the lowest and the highest.
The zero mode (the eigenvector corresponding to the lowest eigenvalue) is ${\bf
v}_0$ which is annihilated by $\bar{L}$.
The eigenvectors
of $\widetilde{W}$ belonging to the duplicated eigenvalue $-r(r-1)$ are
\[
v_s=(1,0,\ldots,0)^T,\quad
v_{\bar{s}}=(0,0,\ldots,1)^T,
\]
corresponding to the conditions $\cos2q_1=1$ and $\cos2q_r=-1$.
The corresponding eigenvectors of $\bar{M}$
are both annihilated by $\bar{L}$,
\[
\bar{L}\pmatrix{v_s\cr -v_s}=0,\quad
\bar{L}\pmatrix{v_{\bar{s}}\cr -v_{\bar{s}}}=0.
\]

The spectrum of $\bar{L}$ is $\sqrt{\mbox{integer}}$\,:
\begin{eqnarray}
D_r\ ({\bf V}):\quad \mbox{Spec}(\bar{L})&=&
g\left\{\vTm 0[4], \,\pm \,2\sqrt{2},
\,\pm \,2\sqrt{6}, \,\pm \,4\sqrt{3},
\ldots, \pm \,2\sqrt{(j-1)(j-2)},\right.\nonumber\\
&&\left. \qquad\qquad\qquad  \vTm  \ldots,
  \,\pm \,2\sqrt{(r-1)(r-2)}\right\}.
\label{drlsspec}
\end{eqnarray}

\paragraph{$D_r$ Minimal type Lax pair (${\bf V}$)}

The minimal Lax pairs have integer spectrum:
\begin{equation}
D_r:\quad \mbox{Spec}(\bar{L}_m)=g\left\{\vTm 0[2],
\,\pm \,2, \,\pm \,4,
\ldots, \,\pm \,2(r-2), \,\pm \,2(r-1)\right\}
\label{drlmspec}
\end{equation}
and
\begin{equation}
\begin{tabular}{|c|c|c|c|l|}
\hline
$\Delta$&$h$&${\cal R}$&$D$&\vT \hspace*{25mm}Spec($\bar{M}_m$)\\
\hline
$D_4$&6&${\bf V}$&8&\vTm 6[2], \,12[2], \,16[2], \,22[2]\\
\hline
$D_5$&8&${\bf V}$&10&\vTm 8[2], \,20[2], \,22[2], \,32[2], \,38[2]\\
\hline
$D_6$&10&${\bf V}$&12&\vTm 10[2], \,28[2], \,30[2],
\,42[2], \,52[2], \,58[2]\\
\hline
$D_7$&12&${\bf V}$&14&\vTm 12[2], \,34[2], \,42[2],
\,52[2], \,66[2], \,76[2], \,82[2]\\
\hline
$D_8$&14&${\bf V}$&16&\vTm 14[2], \,40[2], \,56[2],
\,62[2], \,80[2], \,94[2], \,104[2],
\,110[2]\\
\hline
\end{tabular}
\end{equation}
The lowest eigenvalue is $h$, the Coxeter number,
and the highest eigenvalue is $rh-2$.
The two-fold degenerate eigenvalues of $\widetilde{W}$,
$r(r-1)$ are always contained.

\paragraph{$D_r$ Root type Lax pair}
The $\bar{L}_K$  matrices have simple spectra.
They are mirror symmetric with respect to zero.
The highest multiplicity is the rank $r$ and the highest (lowest) eigenvalue
is
$2(h-1)$, with interval 2. Thus the multiplicity distribution of
the eigenvalues of $\bar{L}_K$ of the root type Lax matrix is
the number of roots having the specified (2 times the) height.
We have encountered the same distributions (shifted parallelly) in the
eigenvalues of $\bar{M}$ in Calogero systems.

\begin{equation}
\begin{tabular}{|c|c|c|c|l|}
\hline
$\Delta$&$h$&${\cal R}$&$D$&\vT \hspace*{35mm}Spec($\bar{L}_K$)\\
\hline
$D_4$&6&$\Delta$&24&\vTm  \,$\pm$\,2[4], \,$\pm$\,4[3], \,$\pm$\,6[3],
\,$\pm$\,8, \,$\pm$\,10\\
\hline
$D_5$&8&$\Delta$& 40&\vTm  \,$\pm$\,2[5], \,$\pm$\,4[4], \,$\pm$\,6[4],
\,$\pm$\,8[3], \,$\pm$\,10[2], \,$\pm$\,12, \,$\pm$\,14\\
\hline
$D_6$&10& $\Delta$&60&\vTm  \,$\pm$\,2[6], \,$\pm$\,4[5], \,$\pm$\,6[5],
\,$\pm$\,8[4], \,$\pm$\,10[4], \,$\pm$\,12[2],
\,$\pm$\,14[2],  \,$\pm$\,16, \,$\pm$\,18\\
\hline
$D_7$&12&$\Delta$&84&\vTm  \,$\pm$\,2[7], \,$\pm$\,4[6], \,$\pm$\,6[6],
\,$\pm$\,8[5], \,$\pm$\,10[5], \,$\pm$\,12[4],
\,$\pm$\,14[3],  \,$\pm$\,16[2],\\
&&&&\vTm
\quad \,$\pm$\,18[2], \,$\pm$\,20, \,$\pm$\,22\\
\hline
$D_8$&14&$\Delta$&112&\vTm \,$\pm$\,2[8], \,$\pm$\,4[7], \,$\pm$\,6[7],
\,$\pm$\,8[6], \,$\pm$\,10[6], \,$\pm$\,12[5],
\,$\pm$\,14[5],  \,$\pm$\,16[3], \\
&&&&\vTm  \quad \,$\pm$\,18[3], \,$\pm$\,20[2],
\,$\pm$\,22[2],  \,$\pm$\,24, \,$\pm$\,26\\
\hline
\end{tabular}
\label{drrootLKei}
\end{equation}
Here is the summary of the spectra of the $\bar{L}_K$ ($L_m$) matrices
(\ref{arLmesoec}), (\ref{specartrigLK}), (\ref{drlmspec}),
(\ref{drrootLKei}).  The eigenvalues are
2 times the `height' which is determined by the deformed Weyl vector
$\varrho$:
\begin{equation}
\mbox{Spec}(\bar{L}_K)=\{2\varrho\cdot\mu |\mu\in{\cal R}\}.
\label{lkform}
\end{equation}
This formula applies to all the other $\bar{L}_K$ ($L_m$) matrices,
(\ref{specbrslk}), (\ref{brrootLKei}),
(\ref{speccrvlk}), (\ref{crrootLKei}), (\ref{ErminLei}),
(\ref{ErrootLei}), (\ref{F4uniLKei}), (\ref{G2uniLKei}). This is to be
compared with the formulae for the spectra of
$\bar{M}$ matrices of Calogero system (\ref{Mspecform}),
(\ref{Mspecform2}), in which the `height' is determined by the Weyl
vector $\delta$.  The difference is visible in the non-simply laced root
systems (\ref{specbrslk}), (\ref{brrootLKei}),
(\ref{speccrvlk}), (\ref{crrootLKei}), (\ref{F4uniLKei}),
(\ref{G2uniLKei}).
\begin{equation}
\begin{tabular}{|c|c|c|l|}
\hline
$\Delta$&${\cal R}$&$D$&\vT \hspace*{35mm}Spec($\bar{M}$)\\
\hline
$D_4$&$\Delta$&24&\vTm 0, \,12[6], \,20[6], \,24[6], \,32[3], \,36[2]\\
\hline
$D_5$&$\Delta$& 40&\vTm 0, \,16[2], \,20[4], \,28[5],
\,36[10], \,40[2], \,44,
\,48[4], \,52[4], \,56[4], \,64, \,68[2]\\
\hline
$D_6$& $\Delta$&60&\vTm 0, \,20[2], \,30[4], \,36[5],
\,48[5], \,50[4], \,56[7],
\,60[2], \,66[4], \,68[4], \,76[4],\\
&&&\vTm  \quad  \,78[4], \,80[2], \,84,
  \,86[2], \,92[4], \,96[2], \,104,  \,108[2]\\
\hline
$D_7$&$\Delta$&84&\vTm 0, \,24[2], \,42[4], \,44[5],
\,60[5], \,66[4], \,68,
\,72[5], \,80[6], \,84[6], \,86[4],\\
&&&\vTm \quad \,96[4], \,102[4], \,104[5],
\,108[2], \,114[4], \,116[4], \,122[2], \,124[4],\\
&&&\vTm \quad \,128[2],
\,132, \,140[4], \,144[2], \,152, \,156[2]\\
\hline
$D_8$&$\Delta$&112&\vTm 0,  \,28[2], \,52[5], \,56[4],
\,72[5], \,80,  \,84[4],
\,88[5], \,100[9], \,108[10], \,112[2], \\
&&&\vTm \quad \,116[4], \,124,
\,128[8], \,136[4], \,140[6], \,144[4], \,152[4], \,156[4],
\,160[5], \\
&&&\vTm \quad \,164[4], \,172[4], \,180[4], \,184[2], \,188,
  \,196[4], \,200[2], \,208,  \,212[2]\\
\hline
\end{tabular}
\label{drrootMei}
\end{equation}
The relation between Tr($\widetilde{W}$) and Tr($\bar{M}$),
(\ref{trigMWrel}) are satisfied, since
\(
F^{\Delta}(D_r)=8r-14,
\)
see (\ref{fDelta}).

\bigskip
Next let us summarise  the spectra of $\widetilde{W}$,
$\bar{M}$ and $\bar{L}$
  of the $B_r$ Sutherland system
which are evaluated numerically. The set of short roots
$\Delta_S$ is chosen for the Lax
pairs. They are all ``integer valued", except for $\bar{L}$.
As in the $D_r$ case
$\bar{L}^2$ is integer valued.
The spectrum of $\widetilde{W}$ is:
\begin{eqnarray}
B_r:\quad \mbox{Spec}(\widetilde{W})&=&-\left\{\vTm 4(r-1)g_L+2g_S,
\,4((2r-3)g_L+g_S),
\ldots, \right.\nonumber\\
&& \vTm \qquad 2j((2r-1-j)g_L+g_S), \ldots,
   2(r-1)(rg_L+g_S),\nonumber\\
&&\left.\vTm \qquad\qquad\qquad\qquad \,r((r-1)g_L+g_S)\right\},
\label{specbrtrigW}
\end{eqnarray}
which agrees with the general formula (\ref{simweight})
  of the $\widetilde{W}$
spectrum, {\em i.e.\/} the $j$-th entry is $4\lambda_j\cdot\varrho$, and
obviously satisfies the  sum rule (\ref{suthenergSimp}),
(\ref{trWsumrule}).
The last piece corresponds to
the spinor  fundamental weight.

\paragraph{$B_r$ Root type Lax pair ($\Delta_S$)}
The spectrum of $\bar{M}$ is
\begin{eqnarray}
B_r\ (\Delta_S):\quad \mbox{Spec}(\bar{M})&=&
i\left\{\vTm 0, 4(r-1)g_L+2g_S[2],
\,4((2r-3)g_L+g_S)[2],
\ldots, \right.\nonumber\\
&& 2j((2r-1-j)g_L+g_S)[2], \ldots,
   2(r-1)(rg_L+g_S)[2],\nonumber\\
&&\left.\vTm \qquad\qquad\qquad\qquad \,r((r-1)g_L+g_S)\right\}\\
&=&i4\varrho\cdot\left\{0, \ \lambda_1[2], \,\lambda_2[2],
\ldots, \,\lambda_{r-1}[2],
\,\lambda_r\right\},
\end{eqnarray}
which is essentially the duplication of  that of $\widetilde{W}$,
except for the
lowest, {\em i.e.,\/} 0, and the  eigenvalue belonging to the spinor weight.
Let us note that the identity between the traces of $\widetilde{W}$
and $\bar{M}$
(\ref{trigMWrel}) is not satisfied, since $B_r$ is not simply laced.
It is simply the lack of the contribution from the ``anti-spinor weight"
which is removed by the Dynkin diagram folding (\ref{drbrred}).
The eigenvectors
of $\widetilde{W}$ belonging to the degenerate eigenvalue
$-r((r-1)g_L+g_S)$ are
\[
v_s=(1,0,\ldots,0)^T,
\]
corresponding to the condition $\cos2q_1=1$.
The explanation of the duplication of the $\bar{M}$ spectrum is
essentially the same as
in the $D_r$ case. The spectrum of $\bar{L}$ is $\sqrt{\mbox{integer}}$:
\begin{eqnarray}
B_r\ (\Delta_S):\quad \mbox{Spec}(\bar{L})&=&
\sqrt{g_L}\left\{\vTm 0[2], \,\pm
\,2\sqrt{g_S}, \,\pm
\,2\sqrt{2(g_L+g_S)},
\ldots,\right.\\
&&\left.\vTm \pm 2\sqrt{j((j-1)g_L+g_S)}, \ldots, \,\pm
\,2\sqrt{(r-1)((r-2)g_L+g_S)}\right\}.\nonumber
\end{eqnarray}
To be more precise, $\sqrt{\mbox{integer}}$ means that the spectrum of
$\bar{L}^2$ is a quadratic polynomial in $g_L$ and $g_S$ with integer
coefficients. For $g_L=g_S=g$ it reduces to that of minimal Lax matrix
$\bar{L}_m$ of $D_r$
(\ref{drlmspec}). For $g_S=0$ it reduces to that of the Lax matrix
  $\bar{L}$ of
$D_r$ (\ref{drlsspec}). The modified Lax matrix $\bar{L}_K$
  (see (\ref{Ktilde}))  has
simple integer spectrum, see formula (\ref{lkform}):
\begin{eqnarray}
B_r\ (\Delta_S):\quad \mbox{Spec}(\bar{L}_K)&=&\left\{\vTm \pm
\,{g_S}, \,\pm
\,(2g_L+g_S), \,\pm\,(4g_L+g_S),
\ldots, \right.\nonumber\\
&& \left.\vTm \qquad\qquad \pm\, (2(r-1)g_L+g_S)\right\}.
\label{specbrslk}
\end{eqnarray}

\paragraph{$B_r$ Root type Lax pair ($\Delta_L$)}
The $\bar{L}_K$ matrices have simple spectrum
\begin{equation}
\begin{tabular}{|c|c|c|l|}
\hline
$\Delta$&${\cal R}$&$D$&\vT \hspace*{35mm}Spec($\bar{L}_K$)\\
\hline
$B_4$&$\Delta_L$&24&\vTm  \,$\pm$\,2$g_L$[3], \,$\pm$\,4$g_L$[2],
\,$\pm$\,6$g_L$,
\,$\pm$\,2($g_L+g_S)$, \,$\pm$\,2($2g_L+g_S)$, \\
&&&\vTm \ $\pm$\,2(3$g_L+g_S)$[2], \,$\pm$\,2(4$g_L+g_S)$,
\,$\pm$\,2(5$g_L+g_S)$,\\
\hline
$B_5$&$\Delta_L$& 40&\vTm  \,$\pm$\,2$g_L$[4], \,$\pm$\,4$g_L$[3],
\ldots, \,$\pm$\,8$g_L$,
\,$\pm$\,2($g_L+g_S)$, \,$\pm$\,2($2g_L+g_S)$,\\
&&&\vTm \ $\pm$\,2(3$g_L+g_S)$[2], \,$\pm$\,2(4$g_L+g_S)$[2],
\,$\pm$\,2(5$g_L+g_S)$[2],
\,$\pm$\,2(6$g_L+g_S)$\\
&&&\vTm \ $\pm$\,2(7$g_L+g_S)$\\
\hline
$B_6$& $\Delta_L$&60&\vTm  \,$\pm$\,2$g_L$[5], \,$\pm$\,4$g_L$[4],
\ldots, \,$\pm$\,10$g_L$,
\,$\pm$\,2($g_L+g_S)$, \,$\pm$\,2($2g_L+g_S)$,\\
&&&\vTm \ $\pm$\,2(3$g_L+g_S)$[2], \,$\pm$\,2(4$g_L+g_S)$[2],
  \,$\pm$\,2(5$g_L+g_S)$[3],
\,$\pm$\,2(6$g_L+g_S)$[2]\\
&&&\vTm \ $\pm$\,2(7$g_L+g_S)$[2], \,$\pm$\,2(8$g_L+g_S)$,
\,$\pm$\,2(9$g_L+g_S)$\\
\hline
\end{tabular}
\label{brrootLKei}
\end{equation}
The spectra of $\bar{M}$ can be expressed succinctly
in terms of the fundamental weights
$\{\lambda_j\}$, whose  expression in terms of
the coupling constants $g_L$ and $g_S$
can be found in (\ref{specbrtrigW}). The
entry
$\lambda_j$ means that the corresponding eigenvalue is
$i4\varrho\cdot\lambda_j$, etc:
\begin{equation}
\begin{tabular}{|c|c|c|l|}
\hline
$\Delta$&${\cal R}$&$D$&\vT \hspace*{35mm}Spec($\bar{M}$)\\
\hline
$B_4$&$\Delta_L$&24&\vTm 0, \,$\lambda_1$[2], \,$\lambda_4$[2],
\,$\lambda_2$[5],
\,$\lambda_3$[6], \,($\lambda_1+\lambda_2)$, \,($\lambda_2+\lambda_4)[2]$,
\,($\lambda_1+\lambda_3)$[4],\\
&&& \,($\lambda_2+\lambda_3)$\\
\hline
$B_5$&$\Delta_L$& 40&\vTm  0, \,$\lambda_1$[2], \,$\lambda_5$[2],
\,$\lambda_2$[5],
\,$\lambda_3$[5], \,($\lambda_1+\lambda_5)$[2],
\,$\lambda_4$[4], \,($\lambda_1+\lambda_2)$,\\
&&&\vTm
\,($\lambda_2+\lambda_5)$[2], \,($\lambda_1+\lambda_3)$[4],
\,($\lambda_1+\lambda_4)$[4], \,($\lambda_3+\lambda_5)$[2],
\,($\lambda_2+\lambda_3)$,
\\
  &&&\vTm \,($\lambda_2+\lambda_4)$[4], \,($\lambda_3+\lambda_4)$\\
\hline
$B_6$& $\Delta_L$&60&\vTm  0, \,$\lambda_1$[2], \,$\lambda_6$[2],
\,$\lambda_2$[5],
\,$\lambda_3$[5], \,($\lambda_1+\lambda_6)$[2], \,($\lambda_1+\lambda_2)$,
\,$\lambda_4$[5], \,$\lambda_5$[4],\\
&&&\vTm
\,($\lambda_2+\lambda_6)$[2], \,($\lambda_1+\lambda_3)$[4],
\,($\lambda_1+\lambda_4)$[4], \,($\lambda_1+\lambda_5)$[4],
\,($\lambda_2+\lambda_3)$,
\\
  &&&\vTm  \,($\lambda_3+\lambda_6)$[2],
\,($\lambda_4+\lambda_6)$[2], \,($\lambda_2+\lambda_4)$[4],
\,($\lambda_2+\lambda_5)$[4],  \,($\lambda_3+\lambda_4)$,\\
  &&&\vTm \,($\lambda_3+\lambda_5)$[4],
\,($\lambda_4+\lambda_5)$\\
\hline
$B_7$&$\Delta_L$&84&\vTm   0, \,$\lambda_1$[2],
\,$\lambda_2$[5], \,$\lambda_7$[2],
\,$\lambda_3$[5], \,($\lambda_1+\lambda_2)$, \,($\lambda_1+\lambda_7)$[2],
\,$\lambda_4$[5], \,$\lambda_5$[5],\\
&&&\vTm
  \,($\lambda_1+\lambda_3)$[4],  \,$\lambda_6$[4],
   \,($\lambda_2+\lambda_7)$[2],
\,($\lambda_1+\lambda_4)$[4], \,($\lambda_2+\lambda_3)$,
\,($\lambda_3+\lambda_7)$[2],
\\
  &&&\vTm  \,($\lambda_1+\lambda_5)$[4], \,($\lambda_1+\lambda_6)$[4],
  \,($\lambda_2+\lambda_4)$[4], \,($\lambda_4+\lambda_7)$[2],
\,($\lambda_2+\lambda_5)$[4], \\
  &&&\vTm \,($\lambda_5+\lambda_7)$[2], \,($\lambda_2+\lambda_6)$[4],
  \,($\lambda_3+\lambda_4)$,  \,($\lambda_3+\lambda_5)$[4],
   \,($\lambda_3+\lambda_6)$[4],
\\
&&&\vTm
\,($\lambda_4+\lambda_5)$,
\,($\lambda_4+\lambda_6)$[4], \,($\lambda_5+\lambda_6)$\\
\hline
\end{tabular}
\label{brrootMei}
\end{equation}

\paragraph{$C_r$  Lax pair (${\bf V}$)}

\bigskip
Here let us summarise  the spectra of $\widetilde{W}$,
  $\bar{M}$ and $\bar{L}$
  of the $C_r$ Sutherland system
which are evaluated numerically.
The set of vector weights ${\bf V}$ is chosen for the Lax
pairs. They are all ``integer valued", except for $\bar{L}$.
As in the $B_r$ and $D_r$
case
$\bar{L}^2$ is integer valued.
The spectrum of $\widetilde{W}$ is:
\begin{eqnarray}
C_r:\quad \mbox{Spec}(\widetilde{W})&=&-\left\{\vTm 4((r-1)g_M+g_L),
\,4((2r-3)g_M+2g_L),
\ldots, \right.\nonumber\\
&& \vTm \qquad 2j((2r-1-j)g_M+2g_S), \ldots,
   2(r-1)(rg_M+2g_L),\nonumber\\
&&\left.\vTm \qquad\qquad\qquad\qquad \,2r((r-1)g_M+2g_L)\right\},
\label{speccrtrigW}
\end{eqnarray}
which agrees with the general formula (\ref{simweight})
  of the $\widetilde{W}$
spectrum, {\em i.e.\/} the $j$-th entry is $4\lambda_j\cdot\varrho$, and
obviously satisfies the  sum rule (\ref{suthenergSimp}),
(\ref{trWsumrule}).
  The spectrum of $\bar{M}$ is
\begin{eqnarray}
C_r\ ({\bf V}):\quad \mbox{Spec}(\bar{M})&=&i\left\{\vTm 0,
4((r-1)g_M+g_L)[2],
\,4((2r-3)g_M+2g_L)[2],
\ldots, \right.\nonumber\\
&& \vTm \quad 2j((2r-1-j)g_M+2g_S)[2], \ldots,
   2(r-1)(rg_M+2g_L)[2],\nonumber\\
&&\left.\vTm \qquad\qquad\qquad\qquad \,2r((r-1)g_M+2g_L)\right\}\\
&=&i4\varrho\cdot\left\{0, \ \lambda_1[2], \,\lambda_2[2],
  \ldots, \,\lambda_{r-1}[2],
\,\lambda_r\right\},
\end{eqnarray}
which is essentially the  duplication of  that of $\widetilde{W}$,
except for the
lowest, {\em i.e.,\/} 0, and the  highest eigenvalue corresponding to the
fundamental weight of the long simple root.
This degeneracy pattern reflects the Dynkin diagram
folding $A_{2r-1}\to C_r$.
Let us note that the identity between the traces of
$\widetilde{W}$ and $\bar{M}$
(\ref{trigMWrel}) is not satisfied, since $C_r$ is not simply laced.
The spectrum of $\bar{L}$ is $\sqrt{\mbox{integer}}$\,:
\begin{eqnarray}
C_r\ ({\bf V}):\quad \mbox{Spec}(\bar{L})&=&
\sqrt{g_M}\left\{\vTm 0[2], \,\pm
\,2\sqrt{2g_L},
\,\pm
\,2\sqrt{2(g_M+2g_L)},
\ldots, \right.\nonumber\\
&&\left.\vTm \hspace{-24mm}\,\pm 2\sqrt{j((j-1)g_M+2g_L)},\ldots, \,\pm
\,2\sqrt{(r-1)((r-2)g_M+2g_L)}\right\}.
\end{eqnarray}
The modified Lax matrices $\bar{L}_K$ (see (\ref{Ktilde}))  have
simple integer spectra:
\begin{eqnarray}
C_r\ ({\bf V}):\quad \mbox{Spec}(\bar{L}_K)&=&\left\{\vTm \pm
\,{2g_L}, \,\pm
\,(2g_M+2g_L), \,\pm\,(4g_M+2g_L),
\ldots, \right.\nonumber\\
&& \left.\vTm \qquad\qquad \pm\, (2(r-1)g_M+2g_L)\right\}.
\label{speccrvlk}
\end{eqnarray}

\paragraph{$C_r$ Root type Lax pair ($\Delta_L$)}
The $\bar{L}_K$ matrices have simple spectrum
\begin{equation}
\begin{tabular}{|c|c|c|l|}
\hline
$\Delta$&${\cal R}$&$D$&\vT \hspace*{35mm}Spec($\bar{L}_K$)\\
\hline
$C_4$&$\Delta_M$&24&\vTm  \,$\pm$\,2$g_M$[3], \,$\pm$\,4$g_M$[2],
\,$\pm$\,6$g_M$,
\,$\pm$\,2($g_M+2g_L)$, \,$\pm$\,2($2g_M+2g_L)$, \\
&&&\vTm \ $\pm$\,2(3$g_M+2g_L)$[2], \,$\pm$\,2(4$g_M+2g_L)$,
\,$\pm$\,2(5$g_M+2g_L)$,\\
\hline
$C_5$&$\Delta_M$& 40&\vTm  \,$\pm$\,2$g_M$[4], \,$\pm$\,4$g_M$[3], \ldots,
\,$\pm$\,8$g_M$,
\,$\pm$\,2($g_M+2g_L)$, \,$\pm$\,2($2g_M+2g_L)$,\\
&&&\vTm \ $\pm$\,2(3$g_M+2g_L)$[2], \,$\pm$\,2(4$g_M+2g_L)$[2],
\,$\pm$\,2(5$g_M+2g_L)$[2],
\\
&&&\vTm \ \,$\pm$\,2(6$g_M+2g_L)$, \,$\pm$\,2(7$g_M+2g_L)$,\\
\hline
$C_6$& $\Delta_M$&60&\vTm  \,$\pm$\,2$g_M$[5], \,$\pm$\,4$g_M$[4], \ldots,
\,$\pm$\,10$g_M$,
\,$\pm$\,2($g_M+2g_L)$, \,$\pm$\,2($2g_M+2g_L)$,\\
&&&\vTm \ $\pm$\,2(3$g_M+2g_L)$[2], \,$\pm$\,4(2$g_M+g_L)$[2],
\,$\pm$\,2(5$g_M+2g_L)$[3],
\\
&&&\vTm \ $\pm$\,4(3$g_M+g_L)$[2], $\pm$\,2(7$g_M+2g_L)$[2],
\,$\pm$\,2(8$g_M+2g_L)$,
\,$\pm$\,2(9$g_M+2g_L)$\\
\hline
$C_7$&$\Delta_M$&84&\vTm  \,$\pm$\,2$g_M$[6], \,$\pm$\,4$g_M$[5],
\ldots, \,$\pm$\,12$g_M$,
\,$\pm$\,2($g_M+2g_L)$, \,$\pm$\,2($2g_M+2g_L)$,\\
&&&\vTm
\ $\pm$\,2(3$g_M+2g_L)$[2], \,$\pm$\,2(4$g_M+2g_L)$[2],
\,$\pm$\,2(5$g_M+2g_L)$[3],
\\
&&&\vTm \,$\pm$\,2(6$g_M+2g_L)$[3],
\ $\pm$\,2(7$g_M+2g_L)$[3], \,$\pm$\,2(8$g_M+2g_L)$[2],
\\
&&&\vTm \ \,$\pm$\,2(9$g_M+2g_L)$[2],
\,$\pm$\,2(10$g_M+2g_L)$, \,$\pm$\,2(11$g_M+2g_L)$,\\
\hline
\end{tabular}
\label{crrootLKei}
\end{equation}
The interpretation of the root type $\bar{L}_K$ eigenvalues in terms
of the ``height" of the roots is also valid for $C_r$, too.
The spectra of $\bar{M}$ can be expressed succinctly in terms of
the fundamental weights
$\{\lambda_j\}$, whose  expression in terms of the coupling
constants $g_L$ and $g_M$
can be found in (\ref{speccrtrigW}). The
entry
$\lambda_j$ means that the corresponding eigenvalue is
$i4\varrho\cdot\lambda_j$, etc:
\begin{equation}
\begin{tabular}{|c|c|c|l|}
\hline
$\Delta$&${\cal R}$&$D$&\vT \hspace*{35mm}Spec($\bar{M}$)\\
\hline
$C_4$&$\Delta_M$&24&\vTm 0, \,$\lambda_1$[2],  \,$\lambda_2$[5],
\,$\lambda_3$[4], \,$\lambda_4$[2],
\,($\lambda_1+\lambda_2)$, \,($\lambda_1+\lambda_3)$[4],
\,($\lambda_1+\lambda_4)$[2],\\
&&&  \,($\lambda_2+\lambda_3)$, \,($\lambda_2+\lambda_4)[2]$\\
\hline
$C_5$&$\Delta_M$& 40&\vTm  0, \,$\lambda_1$[2],  \,$\lambda_2$[5],
\,$\lambda_3$[5], \,$\lambda_4$[4],
\,($\lambda_1+\lambda_2)$, \,$\lambda_5$[2],
\,($\lambda_1+\lambda_3)$[4],
\\
&&&\vTm \,($\lambda_1+\lambda_4)$[4],
\,($\lambda_1+\lambda_5)$[2], \,($\lambda_2+\lambda_3)$,
\,($\lambda_2+\lambda_4)$[4], \,($\lambda_2+\lambda_5)$[2],
  \\
  &&&\vTm \,($\lambda_3+\lambda_4)$, \,($\lambda_3+\lambda_5)$[2]\\
\hline
$C_6$& $\Delta_M$&60&\vTm  0, \,$\lambda_1$[2],  \,$\lambda_2$[5],
\,$\lambda_3$[5], \,($\lambda_1+\lambda_2)$, \,$\lambda_4$[5],
  \,$\lambda_5$[4], \,($\lambda_1+\lambda_3)$[4],  \,$\lambda_6$[2],
\\
&&&\vTm \,($\lambda_1+\lambda_4)$[4],  \,($\lambda_1+\lambda_5)$[4],
\,($\lambda_2+\lambda_3)$, \,($\lambda_1+\lambda_6)$[2],
\,($\lambda_2+\lambda_4)$[4],
\\
  &&&\vTm   \,($\lambda_2+\lambda_5)$[4], \,($\lambda_2+\lambda_6)$[2],
  \,($\lambda_3+\lambda_4)$,
  \,($\lambda_3+\lambda_5)$[4],
\,($\lambda_2+\lambda_6)$[2],
   \\
  &&&\vTm \,($\lambda_4+\lambda_5)$,
\,($\lambda_4+\lambda_6)$[2]\\
\hline
$C_7$&$\Delta_M$&84&\vTm   0, \,$\lambda_1$[2],  \,$\lambda_2$[5],
\,$\lambda_3$[5], \,($\lambda_1+\lambda_2)$, \,$\lambda_4$[5],
  \,$\lambda_5$[5], \,($\lambda_1+\lambda_3)$[4],  \,$\lambda_6$[4],
\\
&&&\vTm  \,$\lambda_7$[2],
\,($\lambda_1+\lambda_4)$[4], \,($\lambda_2+\lambda_3)$,
\,($\lambda_1+\lambda_5)$[4],  \,($\lambda_1+\lambda_6)$[4],
\\
  &&&\vTm \,($\lambda_1+\lambda_7)$[2], \,($\lambda_2+\lambda_4)$[4],
   \,($\lambda_2+\lambda_5)$[4],  \,($\lambda_2+\lambda_6)$[4],
  \,($\lambda_3+\lambda_4)$,
  \\
  &&&\vTm  \,($\lambda_2+\lambda_7)$[2],
  \,($\lambda_3+\lambda_5)$[4],  \,($\lambda_3+\lambda_6)$[4],
  \,($\lambda_3+\lambda_7)$[2],  \,($\lambda_4+\lambda_5)$,
\\
&&&\vTm \,($\lambda_4+\lambda_6)$[4],  \,($\lambda_4+\lambda_7)$[2],
  \,($\lambda_5+\lambda_6)$,  \,($\lambda_5+\lambda_7)$[2]\\
\hline
\end{tabular}
\label{crrootMei}
\end{equation}

\bigskip
In the rest of this section, we list the results
on the exceptional root systems using tables since most of the  methods and 
concepts
have now been explained.
\subsubsection{$E_r$}
\label{errtrigWspec}
First we list the eigenvalues of $\widetilde{W}$ (the coupling constant and
minus sign removed):
\begin{equation}
\begin{tabular}{|c|c|l|}
\hline
$\Delta$&$r$&\vT \hspace*{13mm}Spec($\widetilde{W}$)\\
\hline
$E_6$&6&\vTm 32[2], \,44, \,60[2], \,84\\
\hline
$E_7$&7&\vTm 54, \,68, \,98, \,104, \,132, \,150, \,192\\
\hline
$E_8$&8&\vTm 116, \,184, \,228, \,272, \,336, \,364, \,440, \,540\\
\hline
\end{tabular}
\label{ErWei}
\end{equation}
Of course they are equal to $\{4\varrho\cdot\lambda_j\}$.
The degeneracies
in $E_6$ spectrum reflect the symmetry in Dynkin diagram.
First let us show the eigenvalues of the Lax matrices for the set of minimal
weights ${\bf 27}$ of $E_6$ and ${\bf 56}$ of $E_7$. In these cases the
spectrum of minimal type $L$ matrix and that of modified $L_K$ are the same:
\begin{equation}
\begin{tabular}{|c|c|c|l|}
\hline
$\Delta$&${\cal R}$&$D$&\vT \hspace*{13mm}Spec($\bar{L}_m$)\\
\hline
$E_6$&${\bf 27}$&27&\vTm 0[3], \,$\pm$\,2[2], \,$\pm$\,4[2], \,$\pm$\,6[2],
\,$\pm$\,8[2], \,$\pm$\,10,
\,$\pm$\,12, \,$\pm$\,14,
      \,$\pm$\,16, \\
\hline
$E_7$&${\bf 56}$&56&\vTm \,$\pm$\,1[3], \,$\pm$\,3[3], \,$\pm$\,5[3],
  \,$\pm$\,7[3],
      \,$\pm$\,9[3], \,$\pm$\,11[2], \,$\pm$\,13[2], \\
&&&\vTm     \,$\pm$\,15[2], \,$\pm$\,17[2], \,$\pm$\,19,  \,$\pm$\,21,
       \,$\pm$\,23,  \,$\pm$\,25,  \,$\pm$\,27,\\
\hline
\end{tabular}
\label{ErminLei}
\end{equation}
The eigenvalues of minimal $\bar{M}_m$  and those of $\bar{M}$ are
slightly different. The latter are $4\varrho\!\cdot\!\lambda_j$ and
their sums but those of the former are different:
\begin{equation}
\begin{tabular}{|c|c|c|l|}
\hline
$\Delta$&${\cal R}$&$D$&\vT \hspace*{13mm}Spec($\bar{M}_m$)\\
\hline
$E_6$&${\bf 27}$&27&\vTm 0[3], 16[2],  \,32,   \,38[2],  \,46[2],  \,52[2],
\,60, \,62[2],
\,68[2], \,72[2], \\
&&&\vTm \,80[2], \,88[2], \,92, \,94[2], \,110[2], \,118[2] \\
\hline
$E_7$&${\bf 56}$&56&\vTm 27[2], \,61[2], \,7[2], \,83[2],
\,101[2], \,113[2],
\,115[2], \,127[4], \,141[2],  \,149[2], \\
&&&\vTm  \,151[2], \,161[2],  \,171[2], \,179[4], \,189[2],
  \,197[2], \,203[2], \,205[2],
\,211[2],  \\
&&&\vTm \,227[2], \,241[2], \,245[2], \,247[2], \,271[2],
\,289[2], \,299[2]\\
\hline
\end{tabular}
\label{ErminMei}
\end{equation}
\begin{equation}
\begin{tabular}{|c|c|c|l|}
\hline
$\Delta$&${\cal R}$&$D$&\vT \hspace*{13mm}Spec($\bar{M}$)\\
\hline
$E_6$&${\bf 27}$&27&\vTm 0,  \,32[4], \,44[2], \,60[6], \,64,
  \,76[2], \,84[4], \\
&&&\vTm  \,92[2], \,104[2], \,116[2], \,120, \\
\hline
$E_7$&${\bf 56}$&56&\vTm 0,  \,54[3], \,68[2], \,98[3],
\,104[4], \,122[2], \,132[4],
\,150[5],
  \\
&&&\vTm  \,152,  \,166[2], \,172[2], \,186[2], \,192[6],
\,202[2], \,204,  \,218[2], \\
&&&\vTm  \,230[2], \,236[2], \,246[2],
\,248, \,260[2], \,282[2], \,296[2], \,302
\\
\hline
\end{tabular}
\label{EruniMei}
\end{equation}
The relation between Tr($\widetilde{W}$) and Tr($\bar{M}$),
(\ref{trigMWrel}) are satisfied, since
$F^{\bf 27}=12$ in $E_6$ and
$F^{\bf 56}=24$ in $E_{7}$, see (\ref{minF}).

\paragraph{Root type Lax pair}
\begin{equation}
\begin{tabular}{|c|c|c|c|l|}
\hline
$\Delta$&$h$&${\cal R}$&$D$&\vT \hspace*{33mm}Spec($\bar{L}_K$)\\
\hline
$E_6$&12&$\Delta$&72&\vTm  \,$\pm$\,2[6], \,$\pm$\,4[5], \,$\pm$\,6[5],
\,$\pm$\,8[5], \,$\pm$\,10[4],
\,$\pm$\,12[3], \,$\pm$\,14[3],
\,$\pm$\,16[2]  \\
&&&&\vTm  \,$\pm$\,18,  \,$\pm$\,20, \,$\pm$\,22  \\
\hline
$E_7$&18&$\Delta$&126&\vTm  \,$\pm$\,2[7], \,$\pm$\,4[6],
\,$\pm$\,6[6], \,$\pm$\,8[6], \,$\pm$\,10[6],
\,$\pm$\,12[5], \,$\pm$\,14[5],
\,$\pm$\,16[4],
  \\
&&&&\vTm  \,$\pm$\,18[4], \,$\pm$\,20[3], \,$\pm$\,2[3],
\,$\pm$\,24[2], \,$\pm$\,26[2], \,$\pm$\,28,  \,$\pm$\,30,
     \,$\pm$\,32,  \,$\pm$\,34 \\
\hline
$E_8$&30&$\Delta$&240&\vTm $\pm$\,2[8], \,$\pm$\,4[7],
  \,$\pm$\,6[7], \,$\pm$\,8[7],
     \,$\pm$\,10[7], \,$\pm$\,12[7], \,$\pm$\,14[7],
\,$\pm$\,16[6],
\\
&&&&\vTm  \,$\pm$\,18[6], \,$\pm$\,20[6], \,$\pm$\,22[6],
\,$\pm$\,24[5], \,$\pm$\,26[5], \,$\pm$\,28[4],
\,$\pm$\,30[4],
\\
&&&&\vTm \,$\pm$\,32[4], \,$\pm$\,34[4], \,$\pm$\,36[3],
\,$\pm$\,38[3], \,$\pm$\,40[2],
\,$\pm$\,42[2],
\,$\pm$\,44[2],
\\
&&&&\vTm \,$\pm$\,46[2], \,$\pm$\,48, \,$\pm$\,50,
\,$\pm$\,52,  \,$\pm$\,54,
  \,$\pm$\,56,  \,$\pm$\,58
\\
\hline
\end{tabular}
\label{ErrootLei}
\end{equation}
In all cases, the highest multiplicity of $\bar{L}_K$
is the rank $r$ and the highest
eigenvalue is $2(h-1)$ with interval 2. Thus the multiplicity distribution of
the eigenvalues of $\bar{L}_K$ of the root type Lax matrix is
the number of roots having the specified (2 times the) height.
Like in all the other cases, the eigenvalues of
$\bar{M}$ are of the form $i\sum_{j=1}^r a_j(4\varrho\cdot\lambda_j)$,
in which
$a_j=0,1$.
\begin{equation}
\begin{tabular}{|c|c|c|c|l|}
\hline
$\Delta$&$h$&${\cal R}$&$D$&\vT \hspace*{33mm}Spec($\bar{M}$)\\
\hline
$E_6$&12&$\Delta$&72&\vTm  0, \,32[4], \,44[4], \,60[10], \,64[2],
\,76[4], \,84[8],
\,92[10], \,104[4],
   \\
&&&&\vTm  \,108[2],  \,116[8],
\,120[4], \,128,
\,136[4], \,144[2], \,148[2], \,164[2] \\
\hline
$E_7$&18&$\Delta$&126&\vTm  0,  \,54[2], \,68[4], \,98[4], \,104[5],
\,122[2], \,132[7],
\,150[7], \,152[4],
  \\
&&&&\vTm  \,158, \,166[2], \,172[4], \,186[4],
\,192[10], \,200,  \,202[4], \,204[4],
  \\
  &&&&\vTm \,218[4], \,220[2], \,230[2], \,236[6],
\,246[6], \,248[4], \,254,
   \,260[4],
  \\
  &&&&\vTm \,270[2], \,272[2], \,282[4], \,284[2],
  \,290,  \,296[6], \,314[2], \,316[2],
  \\
  &&&&\vTm  \,324,
  \,334[2], \,336[2], \,342, \,364[2], \,380[2]\\
\hline
$E_8$&30&$\Delta$&240&\vTm 0, \,116[4], \,184[4], \,228[7], \,272[6],
\,300[4], \,336[9], \,344, \,364[9],
\\
&&&& \vTm   \,388[4],  \,412[6], \,440[11], \,452[4], \,456[4], \,480[6],
\,500[6], \,520[6],
\\
&&&&\vTm  \,540[14],
      \,548, \,556[6], \,564,  \,572[2], \,592[8], \,608[6], \,624[6],
\\
&&&&\vTm \,636[6],
      \,65[8],
     \,668[6],  \,684[2],\,700[8], \,712[4], \,724[8], \,740[2],
\\
&&&&\vTm \,752[2], \,768[10],
      \,776,
       \,792[2], \,804[6], \,812,  \,816[2], \,828[2],
\\
&&&&\vTm  \,840[2], \,852[2],
\,864[2], \,876[8],
      \,896[2],       \,904, \,920[2], \,940[2],
\\
&&&&\vTm \,952[2], \,972[2], \,980,
\,992[2], \,1032[2],
        \,1060[2], \,1076[2]\\
\hline
\end{tabular}
\label{ErrootMei}
\end{equation}
The relation between Tr($\widetilde{W}$) and Tr($\bar{M}$),
(\ref{trigMWrel}) are satisfied, since
\(
F^{\Delta}(E_6)=42\),
\(F^{\Delta}(E_7)=66\) and
\(F^{\Delta}(E_8)=114\), see (\ref{fDelta}).

\subsubsection{$F_4$}
\label{f4rtrigWspec}
The eigenvalues of $\widetilde{W}$ are:
\begin{eqnarray}
F_4:\quad \mbox{Spec}(\widetilde{W})&=&-\left\{\vTm 20g_L + 12g_S,
\,36g_L + 24g_S, \,24g_L + 18g_S, \,12g_L + 10g_S\right\}\nonumber\\
&=& -4\varrho\cdot\!\left\{\vTm \lambda_1, \,\lambda_2,
\,\lambda_3,\,\lambda_4,\right\}.
\label{specf4trigW}
\end{eqnarray}
The eigenvalues of the modified Lax matrix $\bar{L}_K$ are simple:
\begin{equation}
\begin{tabular}{|c|c|c|l|}
\hline
$\Delta$&${\cal R}$&$D$&\vT \hspace*{33mm}Spec($\bar{L}_K$)\\
\hline
$F_4$&$\Delta_L$&24&\vTm $\pm$\,(10$g_L$ + 6$g_S$),
\,$\pm$\,(8$g_L$ + 6$g_S$),
\,$\pm$\,(6$g_L$ + 6$g_S$),
  \,$\pm$\,(6$g_L$ + 4$g_S$), \\
&&&\vTm \,$\pm$\,(4$g_L$ + 4$g_S$),
      \,$\pm$\,(6$g_L$ + 2$g_S$), \,$\pm$\,(2$g_L$ + 4$g_S$),
      \,$\pm$\,(4$g_L$ + 2$g_S$), \\
&&&\vTm      \,$\pm$\,4$g_L$,  \,$\pm$\,(2$g_L$ + 2$g_S$),
\,$\pm$\,2$g_L$[2]\\
\hline
$F_4$&$\Delta_S$&24&\vTm       \,(6$g_L$ + 5$g_S$),
\,$\pm$\,(6$g_L$ + 4$g_S$),
\,$\pm$\,(6$g_L$ + 3$g_S$),
  \,$\pm$\,(4$g_L$ + 3$g_S$), \\
&&&\vTm \,$\pm$\,(4$g_L$ +2$g_S$),
      \,$\pm$\,(2$g_L$ + 3$g_S$), \,$\pm$\,(4$g_L$ + $g_S$),
      \,$\pm$\,(2$g_L$ + 2$g_S$), \\
&&&\vTm  \,$\pm$\,(2$g_L$ + $g_S$),    \,$\pm$\,2$g_S$,
\,$\pm$\,$g_S$[2]
\\
\hline
\end{tabular}
\label{F4uniLKei}
\end{equation}
The interpretation of the root type $\bar{L}_K$ eigenvalues in terms
of the ``height" of the roots is also valid for $F_4$, too.
The eigenvalues of $\bar{M}$ can be expressed succinctly in terms of the
fundamental weights $\{\lambda_j\}$, which are listed in (\ref{specf4trigW}).
The
entry
$\lambda_j$ means that the corresponding eigenvalue is
$i4\varrho\cdot\lambda_j$, etc:
\begin{equation}
\begin{tabular}{|c|c|c|l|}
\hline
$\Delta$&${\cal R}$&$D$&\vT \hspace*{13mm}Spec($\bar{M}$)\\
\hline
$F_4$&$\Delta_L$&24&\vTm 0,  \,$\lambda_4$[2], \,$\lambda_1$[4],
\,$\lambda_3$[4],
\,($\lambda_1+ \lambda_4$)[2], \,$\lambda_2$[6],
\\
&&&\vTm \,($\lambda_1 + \lambda_3$)[2],
\,($\lambda_2+ \lambda_4$)[2], \,($\lambda_1 + \lambda_2$) \\
\hline
$F_4$&$\Delta_S$&24&\vTm 0,    \,$\lambda_4$[4], \,$\lambda_1$[2],
\,$\lambda_3$[6],
\,($\lambda_1+ \lambda_4$)[2], \,$\lambda_2$[4],
\\
&&&\vTm \,($\lambda_3 + \lambda_4$),
\,($\lambda_1 + \lambda_3$)[2],  \,($\lambda_2+ \lambda_4$)[2]
\\
\hline
\end{tabular}
\label{F4uniMei}
\end{equation}

\subsubsection{$G_2$}
\label{g2rtrigWspec}
The eigenvalues of $\widetilde{W}$ are:
\begin{eqnarray}
G_2:\quad \mbox{Spec}(\widetilde{W})&=&-\left\{\vTm 4g_L + 8g_S/3,
\,8g_L + 4g_S,\right\}\nonumber\\
&=& -4\varrho\cdot\!\left\{\vTm \lambda_1, \,\lambda_2\right\}.
\label{specg2trigW}
\end{eqnarray}
The eigenvalues of the modified Lax matrix $\bar{L}_K$ are simple:
\begin{equation}
\begin{tabular}{|c|c|c|l|}
\hline
$\Delta$&${\cal R}$&$D$&\vT \hspace*{19mm}Spec($\bar{L}_K$)\\
\hline
$G_2$&$\Delta_L$&6&\vTm  \,$\pm\,(4g_L+2g_S)$, \,$\pm\,(2g_L+2g_S)$,
\,$\pm\,2g_L$\\
\hline
$G_2$&$\Delta_S$&6&\vTm  \,$\pm\,(2g_L+(4g_S/3))$,
\,$\pm\,(2g_L+(2g_S/3))$,
\,$\pm\,2g_S/3$\\
\hline
\end{tabular}
\label{G2uniLKei}
\end{equation}

The eigenvalues of $\bar{M}$ can be expressed succinctly in terms of the
fundamental weights $\{\lambda_j\}$, which are listed in (\ref{specg2trigW}).
The
entry
$\lambda_j$ means that the corresponding eigenvalue is
$i4\varrho\cdot\lambda_j$, etc:
\begin{equation}
\begin{tabular}{|c|c|c|l|}
\hline
$\Delta$&${\cal R}$&$D$&\vT \hspace*{2mm}Spec($\bar{M}$)\\
\hline
$G_2$&$\Delta_L$&6&\vTm 0,  \,$\lambda_1$[2], \,$\lambda_2$[3],\\
\hline
$G_2$&$\Delta_S$&6&\vTm 0,  \,$\lambda_1$[3], \,$\lambda_2$[2],\\
\hline
\end{tabular}
\label{G2uniMei}
\end{equation}

\section{Comments and Discussion}
\label{comm}
\setcounter{equation}{0}
We have shown that the classical Calogero and Sutherland systems at
their equilibrium points have very interesting properties.
The equilibrium point is related to the zeros of classical polynomials
of
Hermite, Laguerre and Jacobi types.
The second derivatives of the potential have ``integer-eigenvalues", and
various Lax matrices also have ``integer-eigenvalues" at
the equilibrium point. Most of these results are obtained by numerical
evaluation and
it remains a real challenge to derive these ``integer-eigenvalues"
analytically.

In this connection, it is interesting to compare with the situation of
another well-known set of integrable multi-particle dynamical systems
based on crystallographic root systems---the Toda  systems.
Since the non-affine Toda molecule systems do not have a finite
equilibrium point, we only consider the affine Toda molecule of the root system
$\Delta$,
\begin{equation}
H={1\over2}p^2+V_{Toda}(q),\quad
V_{Toda}(q)={1\over{\beta^2}}\sum_{j=0}^rn_j\,e^{\beta\alpha_j\cdot q},
\end{equation}
in which $\{\alpha_1,\ldots,\alpha_r\}$ are the  simple roots of $\Delta$
and
\begin{equation}
\alpha_0=-\sum_{j=1}^r n_j\,\alpha_j,\quad (n_0=1),
\end{equation}
is the euclidean part of the additional affine simple root.
The integers $\{n_j\}$, $j=1,\ldots,r$ are called
Coxeter labels and $\beta$ is the real coupling constant.
The above potential is so chosen as to have the equilibrium point
\begin{equation}
\bar{q}=(0,0,\ldots,0).
\end{equation}
The eigenvalues of the second derivatives of the potential
\begin{equation}
V^{\prime\prime}_{Toda}(0)=\sum_{j=0}^rn_j\alpha_j\otimes\alpha_j,
\end{equation}
are not integers but so-called affine Toda masses $\{m_1^2,\ldots,m_r^2\}$,
corresponding to the Perron-Frobenius eigenvector of the incidence
  matrix of the root system $\Delta$, \cite{bcdsa}.
Since the Lax pair of the Toda molecules is expressed in terms of
  the coordinates $q$
and the Lie algebra generators corresponding to $\Delta$ \cite{oltu},
the eigenvalues
of the Lax pair matrices at the equilibrium point
are completely determined by the
chosen representation of the Lie algebra.

\section*{Acknowledgements}
\setcounter{equation}{0}
We thank Kanehisa Takasaki and Toshiaki Shoji for fruitful discussion and
useful comments. This work was supported by the Anglo-Japanese Collaboration
Project  of the Royal Society and the Japan Society for the Promotion of
Science with the title ``Symmetries and Integrability".
         R. S. is partially supported  by the Grant-in-aid from the
Ministry of Education, Culture, Sports, Science and Technology, Japan,
priority area (\#707) ``Supersymmetry and unified theory of elementary
particles".

\bigskip
\appendix
\begin{center}
{\Large \bf Appendix}
\end{center}
\section*{Eigenvalues of $K$ Matrix}
\setcounter{equation}{0}

\renewcommand{\theequation}{A.\arabic{equation}}

Here we  show  that the constant matrix $K$ defined
in (\ref{QLcomm})
\begin{equation}
K\equiv
\sum_{\rho\in\Delta_+}g_{\rho}(\rho\cdot\hat{H})
(\rho^\vee\!\cdot\hat{H})
\hat{s}_\rho
\end{equation}
has a remarkable property that its eigenvalues are
all {\em integer\,\/}$\times$\,{\em
coupling constant\/}.
The $\tilde{K}$ matrix (\ref{Ktilde}) has a similar property.
This matrix plays an important role in the theory
of classical $r$-matrix of Calogero-Moser systems
\cite{rmat}.
First we note that it  is Coxeter invariant and symmetric
\begin{equation}
\hat{s}_\sigma K\hat{s}_\sigma=K,\quad \forall\sigma\in\Delta,
\quad K^{\rm T}=K,
\end{equation}
implying that the eigenvalues  are real, and the eigenvectors
span representation spaces of the Weyl group whose dimensions are the
multiplicities given in the tables below. As simple examples,
we indicate, for the $A_r$ root system, the decomposition of ${\cal R}$ into
the irreducible representations of the Weyl group, which is the symmetric
group.   The diagonal elements of $K$
are all
vanishing:
\begin{equation}
K_{\mu\mu}=\sum_{\rho\in\Delta_+}g_{\rho}
(\rho\cdot\mu)
(\rho^\vee\!\cdot\mu)
\delta_{\mu,s_\rho(\mu)}=0,
\end{equation}
since $\mu-s_\rho(\mu)=(\rho^\vee\!\cdot\mu)\rho=0$ is necessary for
the Kronecker delta to be non-vanishing. Thus it is
traceless
\begin{equation}
\mbox{Tr}K=0,
\label{traceless}
\end{equation}
which is also obvious from the definition as a commutator (\ref{QLcomm}).
Another important property is that it commutes with $M$:
\begin{equation}
[K,M]=0,
\label{kmcomm}
\end{equation}
at the general position $q$ for both Calogero and Sutherland systems.
All the matrix elements of $K$ are non-negative  and the eigenvector
for the highest eigenvalue (the Perron-Frobenius eigenvector) is
in fact ${\bf v}_0$ (\ref{v0def}),
which is a singlet representation of the Weyl
group:
\begin{equation}
K{\bf v}_0=\lambda_{PF}{\bf v}_0,\quad
\lambda_{PF}=2\mu^2(\sum_{\rho\in{\Delta}_+}g_{\rho})/r
=2\mu^2\tilde{\cal E}_0/\omega r.
\label{pFvec}
\end{equation}
Other important eigenvectors of $K$
  are given by
\begin{equation}
Q{\bf v}_0
\end{equation}
in which $Q$ is defined by (\ref{Lpmdef}) and ${\bf v}_0$ is the above
Perron-Frobenius eigenvector introduced in (\ref{v0def}). For all possible
values of the coordinates $q=(q_1,\ldots,q_r)$,
it is always an eigenvector of
$K$:
\begin{equation}
K\,Q{\bf v}_0=\lambda_Q\,Q{\bf v}_0,
\end{equation}
in which the eigenvalue $\lambda_Q$ is expressed by  {\bf boldface fonts}
  in the formulae from (\ref{arvecKei}) to (\ref{I2m}).
This eigenvalue is usually $r$ (rank) fold degenerate and the
corresponding eigenvectors form an ever present $r$  dimensional
irreducible representation of the Weyl group. Exceptional situations of
  additional degeneracies  occur in $A_7$  root type
(\ref{arrootKei}),  $D_4$, $D_6$ and $E_6$ root type (\ref{drrootK}),
(\ref{errootK}) and $H_3$ and
$H_4$ root type (\ref{h34spec2}). For the cases when ${\cal R}$ is the
set of minimal weights, (\ref{arvecKei}), (\ref{drvspec})
and (\ref{e67minK}), $\lambda_Q$ is related to
$\lambda_{PF}$ by the Coxeter number $h$:
\begin{equation}
\lambda_Q=\lambda_{PF}-gh.
\end{equation}
For the crystallographic simply laced root type cases, (\ref{arrootKei}),
(\ref{drrootK}) and
  (\ref{errootK}),
we have
\begin{equation}
\lambda_Q=(h-6)g.
\end{equation}

If the set ${\cal R}$ consists of minimal weights  (\ref{eq:mindef}),
all the matrix elements of $K$ are either
1 or 0 times the coupling constant. If the set ${\cal R}$ coincides with
the set of all roots $\Delta$ for a crystallographic simply laced root
system, the matrix element of $K$ is characterised by the inner products
of the roots (the roots are normalised as $\alpha^2=2$):
\begin{equation}
K_{\alpha \beta}=g\left\{
\begin{array}{ll} 4& \mbox{if}\  \alpha\cdot\beta=-2, \ i.e., \
\alpha=-\beta,\\
  1 & \mbox{if}\ \alpha\cdot\beta=1,\\
0 & \mbox{otherwise},
\end{array}
\right.
\label{rootK}
\end{equation}
and a similar statement holds for  non-simply laced crystallographic
root systems.

\bigskip
We list
  the spectrum of $K$, {\em i.e.,}
set of eigenvalues  with [{multiplicity}] for all $\Delta$ and for typical
choices of the set of single Coxeter (Weyl) orbits  ${\cal R}$ for which the
typical Lax pairs are known. For the simply laced root systems, we omit the
coupling constant $g$ in the spectrum. In these formulae
$h$ denotes the Coxeter
number.

\begin{enumerate}
\item $A_r$ with vector weights embedded in ${\bf R}^{r+1}$,
{\em i.e.,}
\begin{equation}
{\cal R}={\bf V}=\{{\bf e}_j,\
j=1,\ldots,r+1|{\bf e}_j\in{\bf R}^{r+1}, {\bf e}_j\cdot
{\bf e}_k=\delta_{jk}\},
\label{arvecwei}
\end{equation}
\begin{equation}
\begin{tabular}{|c|c|c|c|c|l|}
\hline
$\Delta$&$h$&${\cal R}$&$D$&$\mu^2$&\vT Spec($K$)\\
\hline
$A_r$&$r+1$&{\bf V}&$r+1$&1&\vTm$r$, {\bf -1}[$r$]\\
\hline
\end{tabular}
\label{arvecKei}
\end{equation}
corresponding to the following decomposition into the irreducible
representation of the Weyl group:
\begin{equation}
(1+r)=1\oplus r.
\end{equation}
In this case $K$ has a very simple expression in terms of ${\bf v}_0$:
\begin{equation}
K=g({\bf v}_0{\bf v}_0^T-I),\qquad I:\ \mbox{Identity matrix}.
\label{ArvecK}
\end{equation}
The matrix elements of $K$ are characterised by the inner
products, too:
\begin{equation}
K_{\mu \nu}=g\left\{
\begin{array}{ll}
  1 & \mbox{if}\ \mu\cdot\nu=0,\\
0 & \mbox{otherwise},
\end{array}
\right.\quad \mu,\nu\in{\bf V}.
\label{archar}
\end{equation}

\item
$A_r$ with roots ${\cal R}={\Delta}$:
\begin{equation}
\begin{tabular}{|c|c|c|c|c|l|}
\hline
$\Delta$&$h$&${\cal R}$&$D$&$\mu^2$&\vT \hspace*{15mm}Spec($K$)\\
\hline
$A_2$&3&$\Delta$&6&2&\vTm6, \,3[{2}], \,{\bf -3}[{2}], \,-6\\
\hline
$A_3$&4&$\Delta$&12&2&\vTm8, \,4[{3}], \,2[{2}], \,{\bf -2}[{3}],
\,-6[{3}]\\
\hline
$A_4$&5&$\Delta$&20&2&\vTm10, \,5[{4}], \,2[{5}], \,{\bf -1}[{4}],
\,-6[{6}]\\
\hline
$A_5$&6&$\Delta$& 30&2&\vTm 12, \,6[5], \,2[9],
\,{\bf 0}[5],\ -6[10]\\
\hline
$A_6$&7& $\Delta$&42&2&\vTm 14, \,7[6], \,2[14],
\,{\bf 1}[6], \,-6[15]\\
\hline
$A_7$&8&$\Delta$&56&2&\vTm 16, \,8[7], \,{\bf
2}[27], \,-6[21]\\
\hline
$A_8$&9&$\Delta$&72&2& 18, \,9[8], \,{\bf 3}[8], \,2[27],
\,-6[28]\\
\hline
$A_9$&10&$\Delta$&90&2& \vTm 20, \,10[9],
\,{\bf 4}[9], \,2[35], \,-6[36]\\
\hline
$A_{10}$&11&$\Delta$&110&2& \vTm 22, \,11[10], \,{\bf 5}[10],
\,2[44], \,-6[45]\\
\hline
$A_{11}$&12&$\Delta$&132&2&\vTm 24,
\,12[11], \,{\bf 6}[11], \,2[54], \,-6[55]\\
\hline
$A_{12}$&13&$\Delta$&156&2&
\vTm 26, \,13[12], \,{\bf 7}[12], \,2[65], \,-6[66]\\
\hline
$A_{r}$&$r+1$&$\Delta$&$r(r+1)$&2&\vTm
$2h$, \,$h$\,[$r$], \,{\bf (h-6)}[$r$],\\
&&&&&\vTm  2\,[$(r+1)(r-2)/2$], \,-6\,[$r(r-1)/2$]\\
\hline
\end{tabular}
\label{arrootKei}
\end{equation}
corresponding to the following decomposition into the irreducible
representations of $A_r$ Weyl group:
\begin{equation}
r(r+1)=1 \oplus r \oplus r^\prime \oplus (r+1)(r-2)/2 \oplus r(r-1)/2,
\end{equation}
in which $r$ and $r^\prime$ are two distinct $r$ dimensional irreducible
representations.

\item
$B_r$ with short roots:
\begin{equation}
{\cal
R}=\Delta_S=\{\pm {\bf e}_j,\ j=1,\ldots,r|{\bf e}_j\in{\bf R}^{r},
{\bf e}_j\cdot {\bf e}_k=\delta_{jk}\},
\label{brshorts}
\end{equation}
\begin{equation}
\begin{tabular}{|c|c|c|c|c|l|}
\hline
$\Delta$&$h$&${\cal R}$&$D$&$\mu^2$&\vT \hspace*{30mm}Spec($K$)\\
\hline
$B_r$&2$r$&$\Delta_S$&$2r$&1&
\vTm $2g_L(r-1)+2g_S$, \, ${\bf -2}{g_S}$[$r$],\,
$-2g_L+2g_S$[$r-1$]\\
\hline
\end{tabular}
    \label{brvspec}
\end{equation}
The $C_r$ in the vector representation ${\cal R}={\bf V}=\{\pm
{\bf e}_j\}$, $j=1,\ldots,r$
  is the same as above if the coupling
constants are interchanged, $g_S\leftrightarrow g_L$.
Similarly to the $A_r$ vector weight case (\ref{ArvecK}), we have
\begin{equation}
K=g_L\left({\bf v}_0{\bf v}_0^T-I-{}_SI\right)+2g_S\,{}_SI,
\label{bcrk}
\end{equation}
in which ${}_SI$ is the second identity matrix. It is 1 for the
elements
$({\bf e}_j,-{\bf e}_j)$, $(-{\bf e}_j,{\bf e}_j)$, $j=1,\ldots,r$
and zero otherwise.

\item $D_r$ with the vector weights:
\begin{equation}
{\cal R}={\bf V}=\{\pm
{\bf e}_j| j=1,\ldots,r\},
\label{drvecs}
\end{equation}
\begin{equation}
\begin{tabular}{|c|c|c|c|c|l|}
\hline
$\Delta$&$h$&${\cal R}$&$D$&$\mu^2$&\vT \hspace*{12mm}Spec($K$)\\
\hline
$D_r$&$2(r-1)$&{\bf V}&$2r$&1&
\vTm $2(r-1)$, \, {\bf 0}
[$r$],\, -2[$r-1$]\\
\hline
\end{tabular}
    \label{drvspec}
\end{equation}
Similarly to the $C_r$ vector weight case (\ref{bcrk}),
  we have an expression
\begin{equation}
K=g\left({\bf v}_0{\bf v}_0^T-I-{}_SI\right).
\label{drk}
\end{equation}
This $K$ matrix is also characterised by the inner product as in
(\ref{archar}).

\item $D_r$ with the (anti) spinor weights:
\begin{equation}
{\cal R}={\bf S}=
{1\over2}\{\pm
{\bf e}_1\pm \cdots\pm{\bf e}_r\}\  \mbox{with even (odd) number of} -
\mbox{signs},
\end{equation}
\begin{equation}
\begin{tabular}{|c|c|c|c|c|l|}
\hline
$\Delta$&$h$&${\cal R}$&$D$&$\mu^2$&\vT \hspace*{25mm}Spec($K$)\\
\hline
$D_4$&6&{\bf S}&8&
1&\vTm 6, \,{\bf 0}[4], \,-2[3]\\
\hline
$D_5$ &8&{\bf S}&16&
5/4&\vTm 10, \,{\bf 2}[5], \,-2[10]\\
\hline
$D_6$ &10&{\bf S}&32&
3/2&\vTm 15, \,{\bf 5}[6], \,-1[15], \,-3[10]
\\
\hline
$D_7$ &12&{\bf S}&64&
7/4& \vTm 21, \,{\bf 9}[7], \,1[21], \,-3[35]
\\
\hline
$D_8$ & 14&{\bf S}&128&
2& \vTm 28, \,{\bf 14}[8], \,4[28], \,-2[56], \,-4[35]
\\
\hline
$D_9$&16&{\bf S}&256&
9/4& \vTm 36, \,{\bf 20}[9], \,8[36], \,0[84], \,-4[126]
\\
\hline
$D_{10}$ & 18&{\bf S}&512&
5/2& \vTm 45, \,{\bf 27}[10], \,13[45], \,3[120], \,-3[210], \,-5[126]
\\
\hline
$D_{11}$ &20&{\bf S}&1024&
11/4& \vTm 55,
\,{\bf 35}[11], \,19[55], \,7[165], \,-1[330], \,-5[462]\\
\hline
\end{tabular}
\label{drrootK}
\end{equation}
The characterisation of the spinor $K$ matrix is a bit different:
\begin{equation}
K_{\mu \nu}=g\left\{
\begin{array}{ll}
  1 & \mbox{if}\ \mu\cdot\nu=(r-4)/4,\\
0 & \mbox{otherwise}.
\end{array}
\right.\quad \mu,\nu\in {\bf S}.
\end{equation}

\item $D_r$ with the  roots {\em i.e.,} ${\cal R}=\Delta$:
\begin{equation}
\begin{tabular}{|c|c|c|c|c|l|}
\hline
$\Delta$&$h$&${\cal R}$&$D$&$\mu^2$&\vT \hspace*{23mm}Spec($K$)\\
\hline
$D_4$&6&$\Delta$&24&2& \vTm 12, \,4[9], \,{\bf 0}[6], \,-6[8]
\\
\hline
$D_5$&8&$\Delta$&40&2& \vTm 16, \,6[4], \,4[10], \,{\bf 2}[5], \,0[5],
  \,-6[15]\\
\hline
$D_6$&10&$\Delta$&60& 2&\vTm 20, \,8[5], \,{\bf 4}[21], \,0[9], \,-6[24]\\
\hline
$D_7$&12&$\Delta$&84&2& \vTm 24, \,10[6], \,{\bf 6}[7],
  \,4[21], \,0[14],
\,-6[35]\\
\hline
$D_8$&14&$\Delta$&112&2& \vTm 28, \,12[7], \,{\bf
8}[8], \,4[28], \,0[20], \,-6[48]\\
\hline
$D_9$&16&$\Delta$&144&2& \vTm 32, \,14[8], \,{\bf
10}[9], \,4[36], \,0[27], \,-6[63]\\
\hline
$D_{10}$&18&$\Delta$& 180& 2& \vTm 36, \,16[9], \,{\bf
12}[10], \,4[45], \,0[35], \,-6[80]\\
\hline
$D_{11}$&20&$\Delta$&220&2& \vTm 40, \,18[10], \,{\bf 14}[11], \,4[55],
\,0[44], \,-6[99]\\
\hline
$D_{r}$&$2(r-1)$&$\Delta$&$2r(r-1)$&2& \vTm $2h$, \,$2(h-1)$\,[$r-1$], \,
${\bf (h-6)}$[$r$],\\
&&&&&\,4\,[$r(r-1)/2$], \,0\,[$r(r-3)/2$], \,-6\,[$r(r-2)$]\\
\hline
\end{tabular}
\end{equation}

\item
$E_r$  with the minimal weights:
\begin{equation}
\begin{tabular}{|c|c|c|c|c|l|}
\hline
$\Delta$&$h$&${\cal R}$&$D$&$\mu^2$&\vT \hspace*{10mm}Spec($K$)\\
\hline
$E_6$&12&{\bf 27}&27&4/3&\vTm
6, \,{\bf 4}[6], \,-2[20]\\
\hline
$E_7$&18&{\bf 56}&56&3/2&\vTm
27, \,{\bf 9}[7], \,-1[27], \,-3[21]\\
\hline
\end{tabular}
\label{e67minK}
\end{equation}
These $K$ matrix are characterised by:
\begin{equation}
K_{\mu \nu}=g\left\{
\begin{array}{ll}
  1 & \mbox{if}\ \mu\cdot\nu=1/3,\\
0 & \mbox{otherwise}.
\end{array}
\right.\quad \mu,\nu\in {\bf 27},\qquad
g\left\{
\begin{array}{ll}
  1 & \mbox{if}\ \mu\cdot\nu=1/2,\\
0 & \mbox{otherwise}.
\end{array}
\right.\quad \mu,\nu\in {\bf 56}.
\end{equation}

\item
$E_r$  with the  roots {\em i.e.,} ${\cal R}=\Delta$:
\begin{equation}
\begin{tabular}{|c|c|c|c|c|l|}
\hline
$\Delta$&$h$&${\cal R}$&$D$&$\mu^2$&\vT \hspace*{17mm}Spec($K$)\\
\hline
$E_6$&12&$\Delta$&72&2&
\vTm 24, \,{\bf 6}[26], \,0[15], \,-6[30]\\
\hline
$E_7$&18&$\Delta$&126&2&
\vTm 36, \,{\bf 12}[7], \,8[27], \,0[35],\,
-6[56]\\
\hline
$E_8$&30&$\Delta$&240&2&
\vTm 60, \,{\bf 24}[8], \,12[35], \,0[84], \,-6[112]\\
\hline
$E_r$&$h$&$\Delta$&$D$&2&
\vTm $2h$, \,{\bf (h-6)}\,[$r$], \,$(h/3+2)\,$[$(r-1)(r+2)/2$],\\
&&&&& \,0\,[$(D-r(r+1))/2$],
\,-6\,[$D/2-r$]\\
\hline
\end{tabular}
\label{errootK}
\end{equation}

\item
$F_4$ with the long roots
{\em i.e.,} ${\cal R}=\Delta_{L}$:
\begin{equation}
\begin{tabular}{|c|c|c|c|c|l|}
\hline
$\Delta$&$h$&${\cal R}$&$D$&$\mu^2$&\vT \hspace*{32mm}Spec($K$)\\
\hline
$F_4$&12&$\Delta_L$&24&2&
\vTm $12(g_L+g_S)$, \,$12g_S$[2],\,
$4(g_L-g_S)$[9], \,{\bf 0}[4], \,$-6g_L$[8]\\
\hline
\end{tabular}
\label{f4spec2}
\end{equation}

\item $G_2$ with the long roots {\em i.e.,} ${\cal
R}=\Delta_L$:
\begin{equation}
\begin{tabular}{|c|c|c|c|c|l|}
\hline
$\Delta$&$h$&${\cal R}$&$D$&$\mu^2$&\vT \hspace*{40mm}Spec($K$)\\
\hline
$G_2$&6&$\Delta_L$&6&2&
\vTm $6(g_L+g_S)$, \,$3(g_L-g_S)$[2], \,{\bf -3}$(g_L+g_S)$[2], \,
$6(-g_L+g_S)$\\
\hline
\end{tabular}
\label{g2spec2}
\end{equation}

\item $H_r$ with the roots  {\em i.e.,} ${\cal R}=\Delta$:
\begin{equation}
\begin{tabular}{|c|c|c|c|c|l|}
\hline
$\Delta$&$h$&${\cal R}$&$D$&$\mu^2$&\vT \hspace*{17mm}Spec($K$)\\
\hline
$H_3$&10&$\Delta$&30&1&
\vTm 10, \,4[5], \,3[3], \,0[9], \,{\bf -2}[7], \,-5[5]\\
\hline
$H_4$&30&$\Delta$&120&1&
\vTm 30, \,15[4], \,10[9], \,{\bf 0}[70], \,-5[36]\\
\hline
\end{tabular}
\label{h34spec2}
\end{equation}
These $K$ matrices for $H_3$ and $H_4$ are characterised by:
\begin{equation}
K_{\alpha \beta}=g\left\{
\begin{array}{ll} 1& \mbox{if}\  \alpha\cdot\beta=-1, \ i.e., \
\alpha=-\beta,\\
  1/2 & \mbox{if}\ \alpha\cdot\beta=1/2,\\
(3\pm\sqrt{5})/4 & \mbox{if}\ \alpha\cdot\beta=(1\mp\sqrt{5})/4,
\\
0 & \mbox{otherwise}.
\end{array}
\right.
\end{equation}

\item $I_2(m)$  in the  $m$ dimensional
representation consisting of the vertices of the regular $m$-gon
${\cal R}=R_{m}$:
\begin{eqnarray}
R_m&=&\{\sqrt{2}(\cos({2k\pi/ m}+t_0), \sin({2k\pi/ m}+t_0) )\in{\bf
R}^2|\ k=1,\ldots,m\},
\label{rmvert}\\
t_0&=&0,\ (\pi/2m), \quad \mbox{for}\ \ m \ \mbox{even}\
(\mbox{odd}).
\nonumber
\end{eqnarray}
\begin{equation}
\begin{tabular}{|c|c|c|c|c|l|}
\hline
$\Delta$&$h$&${\cal R}$&$D$&$\mu^2$&\vT \hspace*{20mm}Spec($K$)\\
\hline
$I_2(2n+1)$&$2n+1$&$R_{2n+1}$&$2n+1$&2&\vT
$2(2n+1)$, \,0[$2n-2$], \,{\bf -(2n+1)}[2]\\
\hline
&&&&&\vTm
$2n(g_0+g_e)$, \,$(-1)^nn(g_0-g_e)$[2],\\
$I_2(2n)$&$2n$&$R_{2n}$&$2n$&2&\vTm \quad 0[$2n-6$], \,{\bf
-n}$(g_o+g_e)$[2],\\
&&&&&\vTm \qquad$-(-1)^n2n(g_0-g_e)$\\
\hline
\end{tabular}
\label{I2m}
\end{equation}

\end{enumerate}


\begin{thebibliography}{99}
\bibitem{Cal}  F.~Calogero, ``Solution of the one-dimensional
\(N\)-body problem with quadratic and/or inversely quadratic pair
potentials", J. Math. Phys. {\bf 12} (1971) 419-436.
\bibitem{Sut}
B.~Sutherland, ``Exact results for a quantum many-body problem in
one-dimension. II'', Phys. Rev. {\bf A5} (1972) 1372-1376.
\bibitem{CalMo}
J.~Moser, ``Three integrable Hamiltonian systems connected with
isospectral deformations'',  Adv. Math. {\bf 16} (1975) 197-220;\
J.~Moser,  ``Integrable systems of non-linear evolution equations",
in {\it Dynamical Systems, Theory and Applications\/};\
J. Moser, ed., Lecture Notes in Physics {\bf 38} (1975),
Springer-Verlag;\
F.~Calogero, C.~Marchioro and O.~Ragnisco, ``Exact solution of the
classical and quantal one-dimensional many body problems with
the two body potential \(V_{a}(x)=g^2a^2/\sinh^2\,ax\)'', Lett. Nuovo
Cim. {\bf 13} (1975) 383-387;\
F.~Calogero,``Exactly solvable one-dimensional many body problems'',
Lett. Nuovo Cim. {\bf 13} (1975) 411-416.



\bibitem{OP1} M.\,A.~Olshanetsky and A.\,M.~Perelomov,
``Completely integrable Hamiltonian systems connected with
  semisimple Lie algebras",
  Inventions Math. {\bf 37} (1976), 93-108;
  ``Classical integrable finite-dimensional systems related to Lie
  algebras'',
  Phys. Rep.  {\bf C71} (1981), 314-400.

  \bibitem{bcs2}  A.\,J.~Bordner, E.~Corrigan and R.~Sasaki,
``Generalized Calogero-Moser models and  universal Lax pair operators'',
  Prog. Theor. Phys. {\bf 102}  (1999)  499-529,
  {\tt  hep-th/9905011}.

  \bibitem{DHoker_Phong}
E.~D'Hoker and D.\,H.~Phong, ``Calogero-Moser
Lax pairs with spectral parameter for general Lie algebras'',
Nucl. Phys. {\bf B530} (1998) 537-610, {\tt hep-th/9804124}.




\bibitem{bcs1}
  A.\,J.~Bordner, E.~Corrigan and R.~Sasaki,
``Calogero-Moser models I: a new formulation'',
Prog. Theor. Phys. {\bf 100} (1998) 1107-1129, {\tt hep-th/9805106};
A.\,J.~Bordner,   R.~Sasaki and K.~Takasaki, ``Calogero-Moser models II:
symmetries and foldings'', Prog. Theor. Phys. {\bf
101} (1999) 487-518, {\tt hep-th/9809068};
A.\,J.~Bordner and R.~Sasaki, ``Calogero-Moser models III: elliptic
potentials and
twisting'', Prog. Theor. Phys. {\bf 101} (1999) 799-829, {\tt
hep-th/9812232};
S.\,P.~Khastgir, R.~Sasaki and K.~Takasaki,
``Calogero-Moser models IV: Limits to Toda theory",
  Prog. Theor. Phys. {\bf 102}  (1999), 749-776, {\tt
hep-th/9907102}.


\bibitem{bms}  A.\,J.~Bordner, N.\,S.~Manton and R.~Sasaki,
``Calogero-Moser models V:  Supersymmetry,
and Quantum Lax Pair", Prog. Theor. Phys. {\bf 103} (2000) 463-487,
{\tt hep-th/9910033}.

\bibitem{Dunk}
C.\,F.~Dunkl, ``Differential-difference operators associated to
reflection groups", Trans. Amer. Math. Soc. {\bf 311} (1989) 167-183.


\bibitem{kps}
S.\, P.~Khastgir, A.\, J.~Pocklington and R.~Sasaki,
``Quantum Calogero-Moser Models: Integrability for all Root Systems'',
J.\ Phys. {\bf A33} (2000) 9033-9064,
{\tt hep-th/0005277}.

\bibitem{Heck}
G.\,J.~Heckman, ``A remark on the Dunkl differential-difference
operators", in W.~Barker and P.~Sally (eds.) ``Harmonic analysis
on reductive groups", Birkh\"auser, Basel (1991).

\bibitem{HeOp}
  G.\,J.~Heckman and E.\,M.~Opdam, ``Root
systems and hypergeometric functions I'', Comp. Math. {\bf
64} (1987), 329--352;
G. J. Heckman, ``Root systems and
hypergeometric functions II'', Comp. Math. {\bf 64}
(1987), 353--373;
E. M.~Opdam, `` Root systems and
hypergeometric functions III'', Comp. Math. {\bf 67}
(1988), 21--49;
``Root systems and
hypergeometric functions IV'', Comp.  Math. {\bf 67}
(1988), 191--209.

\bibitem{calmat}
F.~Calogero, ``On the zeros of the classical polynomials'', Lett. Nuovo
Cim. {\bf 19} (1977) 505-507;
``Equilibrium configuration of one-dimensional many-body problems
with quadratic and inverse quadratic pair potentials",
Lett. Nuovo Cim. {\bf 22} (1977) 251-253;
``Eigenvectors of a matrix related to the zeros of Hermite polynomials",
Lett. Nuovo Cim. {\bf 24} (1979) 601-604;
``Matrices, differential operators and polynomials'',
J. Math. Phys. {\bf 22} (1981) 919-934.

\bibitem{calpere}
F.~Calogero and A.\,M.~Perelomov,
``Properties of certain matrices related to the equilibrium
configuration of one-dimensional many-body problems with pair
potentials $V_{1}=-\log|\sin x|$ and $V_{2}=1/\sin^2x$'', Commun. Math.
Phys. {\bf 59} (1978) 109-116.

\bibitem{ahmcal}
S.~Ahmed, M.~Bruschi, F.~Calogero, M.\,A.~Olshanetsky and A.\,M.~Perelomov,
``Properties of the zeros of the classical polynomials and of Bessel
functions", Nuovo Cim. {\bf 49} (1979) 173-199.




\bibitem{halsha}
F.\, D.\, M.~Haldane, ``Exact Jastrow-Gutzwiller resonating
     valence bond ground state of the spin 1/2 antiferromagnetic
     Heisenberg chain with 1/r**2 exchange",
Phys. Rev. Lett. {\bf 60} (1988) 635-638; B.\,S.~Shastry, ``Exact solution
of $S=1/2$ Heisenberg antiferromagnetic chain with long-ranged
interactions", {\it ibid} {\bf 60} (1988) 639-642.

\bibitem{fmp}
A.\,P.~Polychronakos, ``Exchange operator formalism for integrable systems
of particles", Phys. Rev. Lett. {\bf 69} (1992) 703-705;
M.~Fowler and J.\,A.~Minahan, ``Invariants of the Haldane-Shastry $SU(N)$
chain", Phys. Rev. Lett. {\bf 70} (1993) 2325-2328;
A.\,P.~Polychronakos,
``Lattice integrable systems of Haldane-Shastry type",
{\it ibid} {\bf 70} (1993)
2329-2331.

\bibitem{ber}
D.~Bernard, V.~Pasquier and D.~Serban, ``Exact solution of
  long-range interacting spin chains with boundaries",
Europhys. Lett. {\bf 30} (1995) 301-306.

\bibitem{yam}
T.~Yamamoto, ``Multicomponent Calogero model of $B_N$-type confined in
a harmonic potential", Phys. Lett. {\bf A208} (1995) 293;
T.~Yamamoto and O.~Tsuchiya, ``Integrable $1/r^2$ spin chain with
reflecting end", J. Phys. {\bf A29} (1996) 3977-3984, {\tt
cond-mat/9602105}.


\bibitem{is1}
V.\,I.~Inozemtsev and R.~Sasaki,
``Universal Lax pairs for spin Calogero-Moser models and spin
exchange models", J. Phys. {\bf A34} (2001) 7621-7632, {\tt hep-th/0105164}.



\bibitem{is2}
V.\,I.~Inozemtsev and R.~Sasaki,
``Hierarchies of Spin Models related to Calogero-Moser Models",
Nucl. Phys. {\bf B618} 689-698.
{\tt hep-th/0105197}. 


\bibitem{witten}
E.~Witten, ``Dynamical breaking of supersymmetry",  Nucl. Phys. {\bf B188}
(1981) 513-554.

\bibitem{cfs}
R.~Caseiro, J.-P.~Fran\c{c}oise and R.~Sasaki,
``Algebraic Linearization of Dynamics of Calogero Type for any
Coxeter Group'',
J.\ Math.\ Phys.\ {\bf 41} (2000) 4679-4986,
{\tt hep-th/0001074}.


\bibitem{cfs2}
R.~Caseiro, J.-P.~Fran\c{c}oise and R.~Sasaki,
``Quadratic Algebra associated with Rational Calogero-Moser Models",
J. Math. Phys. {\bf 42}  5329-5340, {\tt hep-th/0102153}.


\bibitem{dyson}
F.\,J.~Dyson, ``Statistical theory of the energy levels of complex
systems, I, II, III'', J. Math. Phys.  {\bf 3}, (1962) 140-156,
157-165, 166-175; ``A Brownian motion model for the eigenvalues of a
random matrix'',  J. Math. Phys.  {\bf 3}, (1962) 1191-1198.


\bibitem{szego}
G.~Szeg\"o, ``Orthogonal polynomials'', Amer. Math. Soc. New York
(1939).


\bibitem{rmat}
J.\, Avan and M.\, Talon, ``Classical
\(R\)-matrix structure for the Calogero model",
Phys. Lett. {\bf B303} (1933) 33-37, {\tt hep-th/9210128};
J.\, Avan, O.\,  Babelon and  M. Talon,
``Construction of the classical R matrices for the Toda and Calogero
models", Alg. Anal. {\bf 6}  (1994) 67--89,
  {\tt hep-th/9306102};
Y.~B.~Suris,
``Why are the Ruijsenaars--Schneider and the Calogero--Moser hierarchies
governed by the same $r$--matrix?,''
  Phys. Lett. {\bf A225} (1997), 253--262, {\tt hep-th/9602160}.

\bibitem{bcdsa}
H.\, W.\, Braden, E.\, Corrigan, P.\, E.\, Dorey and R.\,Sasaki,
``Affine Toda Field Theory and Exact S-Matrices",
      Nucl. Phys.  {\bf B338} (1989) 689-746.
\bibitem{oltu}
See, for example, D.~I.~Olive and N.~Turok,
``Algebraic Structure Of Toda Systems,''
Nucl.\ Phys.\ B {\bf 220} (1983) 491.


\end{thebibliography}
\end{document}